\documentclass[a4paper,fleqn,usenatbib]{mnras}

\usepackage{newtxtext,newtxmath}
\usepackage[T1]{fontenc}
\usepackage{ae,aecompl}

\usepackage{graphicx}	% Including figure files
\usepackage{amsmath}	% Advanced maths commands

\usepackage{amssymb}	% Extra maths symbols

\usepackage{enumitem}
\setlist[itemize]{leftmargin=*}

\usepackage{epsfig}
\usepackage{multirow}
\usepackage{float}
\usepackage{footmisc}
\def\mnras{MNRAS}
\def\apj{ApJ}

\def\apjl{ApJL}
\def\apjs{ApJS}
\def\aap{A\& A}
\def\pasp{PASP}

\def\nat{Nature}

\def\xmm{{\it XMM-Newton}}

\def\nustar{{\it NuSTAR}}
\def\swift{{\it Swift}}
\def\rxte{{\it RXTE}}
\def\rxj04{RX J0439.6-5311}
\def\rej1034{RE J1034+396}
\def\1h07{1H 0707-495}
\def\pg12{PG 1244+026}
\def\grs1915{GRS 1915+105}

\title[Re-observing \rej1034. II. New Insights on the Soft Excess and QPO]{Re-observing the NLS1 Galaxy \rej1034. II. New Insights on the Soft X-ray Excess, QPO and the Analogy with \grs1915}

\author[C. Jin, et al.]{
Chichuan Jin$^{1,2}$\thanks{E-mail: ccjin@nao.cas.cn},
Chris Done$^{3}$,
Martin Ward$^{3}$
\\
$^{1}$National Astronomical Observatories, Chinese Academy of Sciences, 20A Datun Road, Beijing 100101, China\\
$^{2}$School of Astronomy and Space Sciences, University of Chinese Academy of Sciences, 19A Yuquan Road, Beijing 100049, China\\
$^{3}$Centre for Extragalactic Astronomy, Department of Physics, University of Durham, South Road, Durham DH1 3LE, UK\\
}

\date{prepared for MNRAS}

\pubyear{2019}

\begin{document}
\label{firstpage}
\pagerange{\pageref{firstpage}--\pageref{lastpage}}
\maketitle

\begin{abstract}
The active galactic nucleus (AGN) \rej1034\ displays the most significant X-ray Quasi-Periodic Oscillation (QPO) detected so far. We perform a detailed spectral-timing analysis of our recent simultaneous \xmm, \nustar\ and \swift\ observations. We present the energy dependence of the QPO's frequency, rms, coherence and phase lag, and model them together with the time-averaged spectra. Our study shows that four components are required to fit all the spectra. These components include an inner disc component ({\tt diskbb}), two warm corona components ({\tt CompTT-1} and {\tt CompTT-2}), and a hot corona component ({\tt nthComp}). We find that {\tt diskbb}, {\tt CompTT-2} (the hotter but less luminous component) and {\tt nthComp} all contain the QPO signal, while {\tt CompTT-1} only exhibits stochastic variability. By fitting the lag spectrum, we find that the QPO in {\tt diskbb} leads {\tt CompTT-2} by 679 s, and {\tt CompTT-2} leads {\tt nthComp} by 180 s. By only varying the normalizations, these components can also produce good fits to the time-averaged and variability spectra obtained from previous observations when QPOs were present and absent. Our multi-wavelength study shows that the detectability of the QPO does not depend on the contemporaneous mass accretion rate. We do not detect a significant Iron K$\alpha$ emission line, or any significant reflection hump. Finally, we show that the rms and lag spectra in the latest observation are very similar to the 67 Hz QPO observed in the micro-quasar \grs1915. These new results support the physical analogy between these two sources. We speculate that the QPO in both sources is due to the expansion/contraction of the vertical structure in the inner disc.
\end{abstract}

\begin{keywords}
accretion, accretion discs - galaxies: active - galaxies: nuclei.
\end{keywords}

%%%%%%%%%%%%%%%%%%%%%%%%%%%%%%%%%%%%%%%%%%%%%%%%%%

%%%%%%%%%%%%%%%%% BODY OF PAPER %%%%%%%%%%%%%%%%%%

\section{Introduction}
\label{sec-introduction} 

\begin{figure}
\centering
\includegraphics[trim=0.0in 0.3in 0.0in 0.0in, clip=1, scale=0.55]{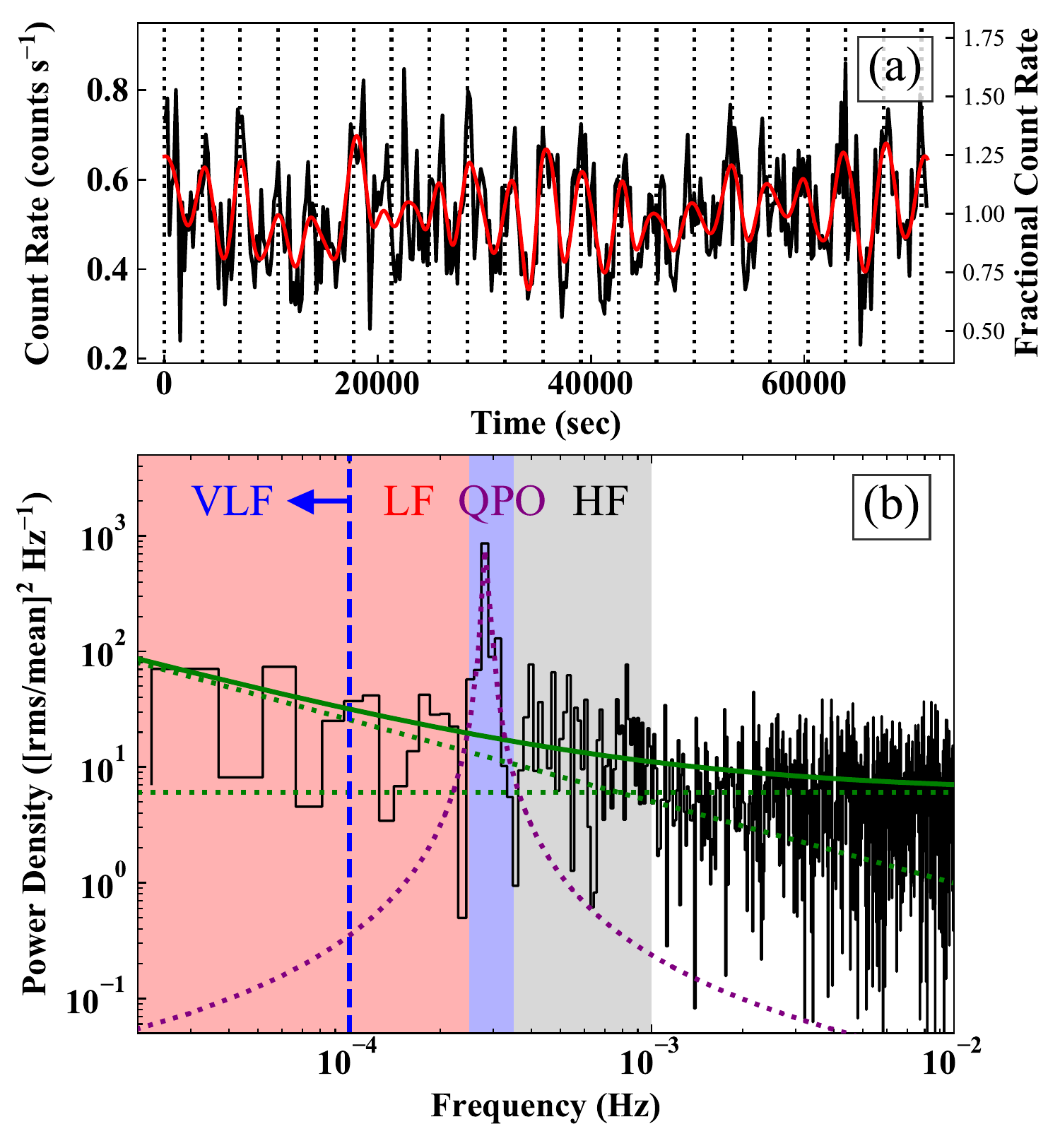} 
\caption{Panel-a: the 1-4 keV light curve of \rej1034\ in the \xmm\ EPIC-pn observation in 2018 (Obs-9). The red solid line shows the QPO component as represented by the intrinsic mode function (see Paper-I). Vertical dotted lines indicate a period of 3550 s. Panel-b: the PSD of the 1-4 keV light curve. The green solid line indicates the best-fit PSD continuum, which consists of a red-noise continuum and Poisson noise (green dotted lines). The purple dotted line indicates the best-fit lorentzian profile of the QPO. See Paper-I for further details. Four frequency bands have been defined, including the high-frequency band (HF): $(0.35-1.0)\times10^{-3}$ Hz, QPO band: $(2.5-3.5)\times10^{-4}$ Hz, low-frequency band (LF): $(0.14-2.5)\times10^{-4}$ Hz, and very-low-frequency (VLF) band: $(0.14-1.0)\times10^{-4}$ Hz. The lowest frequency is determined by the duration of the light curve.}
\label{fig-lc}
\end{figure}

The Quasi-Periodic Oscillation (QPO) in the Narrow Line Seyfert 1 galaxy \rej1034\ ($z=0.0424$) is the most significant and persistent example of an X-ray QPO in Active Galactic Nuclei (AGN: see Figure~\ref{fig-lc}). It has been seen in multiple observations spanning over more than 10 years (Gierli\'{n}ski et al. 2008, Middleton et al. 2009, Alston et al. 2014, Jin, Done \& Ward 2020, hereafter referred to as Paper-I). A key question is what is so special about \rej1034\ that produces such a dramatic QPO? One possible answer is that it is in some way related to the extraordinarily strong soft X-ray excess present in this AGN (Puchnarawicz et al. 1995). A soft excess is defined as additional emission above an extrapolation of the best-fit hard X-ray power law extending to low energies below 2 keV. This phenomenon 
is ubiquitously seen in AGN which lack a significant gas column density (e.g. Arnaud et al. 1985; Turner \& Pounds 1988, Porquet et al. 2004, Gierli\'{n}ski \& Done 2004, Crummy et al. 2006). However, the origin of this component remains an unsolved problem, especially as it has no obvious analog in X-ray binaries (e.g. Kubota, Makishima \& Ebisawa 2001; Kubota \& Done 2004; Page et al. 2004). Currently, there are two most favoured spectral models. The first one is that there is an additional warm, optically thick Comptonisation region as well as the hot, optically thin Comptonisation region which produces the hard X-ray emission (e.g. Porquet et al. 2004, Gierlinski \& Done 2004, Petrucci et al. 2018). This can arise from part of the disc itself where the emission does not completely thermalise (Kubota \& Done 2018). An alternative origin is that the soft excess arises from illumination of the partially ionised accretion disc by harder X-rays. This would imply that the soft X-ray reflected/reprocessed spectrum should be dominated by atomic features, but their presence is smeared out by strong relativistic effects (Ross \& Fabian 1993; Ballantyne, Ross \& Fabian 2001; Crummy et al. 2006; Kara et al. 2016). Despite these models being physically very different, they still cannot be easily distinguished by spectral fitting alone, especially over the limited 0.3-10~keV bandpass of the X-ray data.  

The soft excess is observed to be the strongest in the subset of AGN classified as Narrow Line Seyfert 1 galaxies (NLS1s, e.g. Boller, Brandt \& Fink 1996).  NLS1s typically have low 
black hole masses and high mass accretion rates (e.g. Boroson 1992), so they are also the accretion systems with the highest predicted intrinsic disc temperatures. In \rej1034\ the inner disc is predicted to emit mostly in the soft X-ray bandpass, so part of its very steep soft X-ray spectrum can arise from the disc itself (Jin et al. 2012a, Done et al. 2012). Hence, there are potentially four components in the 
X-ray spectrum, namely the intrinsic disc emission, the soft excess, the hard X-ray power law and its reflection/reprocessed emission from the illuminated disc. In this paper we will explore how the QPO affects each of these components in order to better constrain its physical origin.

Since the results from the spectral fitting procedure are degenerate, we must use additional information contained in the variability characteristics to identify these components. There are a variety of spectral-timing techniques available (Uttley et al. 2014). 
Some of these explore the level of variability for a given timescale as a function of energy (e.g. rms spectra and energy-resolved power spectra), while more powerful techniques search for correlated variability across different energy bands. Such correlations are not only technically useful (e.g. they can enhance the signal-to-noise: S/N), but also physically important by providing clues to causality.
For example, if the soft X-ray excess results solely from illumination of the disc by the variable hard X-ray corona, then the soft X-ray variability should be similar to that seen in the harder X-ray emission, but smoothed and lagged by the light travel timescale, giving a soft X-ray lag from reverberation (Fabian et al. 2009). Conversely, if the soft excess is a separate additional component, it will be likely to vary independently of the hard X-rays. However, if this variability is due to mass accretion rate fluctuations then it should propagate inwards on the viscous timescale, and then modulate the mass accretion rate in the hard X-ray emitting region, giving a soft X-ray lead from the propagation time (Kotov, Churazov \& Gilfanov 2010; Ar\'{e}valo \& Uttley 2006). 

Previous applications of these correlation techniques to NLS1s have shown that both reflection and propagation are important. The correlated variability shows a soft lead for long timescale fluctuations, and a soft lag on the shortest timescales
(Fabian et al. 2009; Emmanoulopoulos, M$^{c}$Hardy \& Papadakis 2011; Zoghbi et al. 2013; Kara et al. 2017; Parker, Miller \& Fabian 2018). This implies that the soft X-ray excess contains both slow variability of intrinsic soft X-ray emission which propagates inwards, and fast variability of the intrinsic hard X-ray component which reverberates. A complete spectral-timing model of this physical picture can fit the full range of spectral/timing properties of the NLS1 PG 1244+026 (Gardner \& Done 2014, hereafter GD14), including the spectrum of the fastest variability (Jin et al. 2013).

In Paper-I we showed that the QPO in a recent \xmm\ observation of \rej1034\ in 2018 (Obs-9) is strong and highly coherent. 
There is a clear soft X-ray lead at the QPO frequency, which is the opposite to the result from a previous observation in 2007 (Obs-2) as reported in Zoghbi \& Fabian (2011). This new soft X-ray lead supports the analogy between this QPO in \rej1034\ and the 67~Hz QPO seen the stellar mass black hole \grs1915\ (Middleton et al. 2009; Middleton, Uttley \& Done 2011; M{\'e}ndez et al. 2013). 

Here we use a range of spectral-timing techniques to explore both the standard AGN stochastic variability and the QPO component present in these \xmm\ 
data. Further, we show how this can be used to break degeneracies in the spectral modelling of the soft X-ray excess.  
In addition, we have \nustar\ data which extends the available spectral bandpass with relatively good S/N up to $\sim$40 keV, which strongly constrains any reflection component present in the data (Sections 3 and 4). The best-fit model is then applied to previous observations as a further test and a consistency check (Section 5). We also study the broadband spectral energy distribution of \rej1034, as well as its long term UV variability measured simultaneously with the X-ray data by the \xmm\ OM (Section 6). In Section 7 we make a detailed comparison with the 67~Hz QPO observed in \grs1915. We discuss possible origins of the QPO in Section 8. The final section of the paper summarizes our main results and
conclusions.

Throughout this paper we adopt a flat universe model with the Hubble constant H$_{0} = 72$ km s$^{-1}$ Mpc$^{-1}$, $\Omega_{\Lambda} = 0.73$ and $\Omega_{\rm M} = 0.27$. All the spectral fittings are performed with the {\sc xspec} software (v12.11.0l, Arnaud 1996).

\section{Observations and Data Reduction}
\label{sec-obs}
\subsection{\xmm}
\xmm\ (Jansen et al. 2001) has observed \rej1034\ nine times so far (see Paper-I). In this paper we make use of data from Obs-2, 3, 6 and 9. Obs-2 and 9 are of much higher quality than the other observations both in terms of exposure time and low background, and so the QPO in these two observations can be studied in considerable detail. Obs-3 and 6 are the two observations when the QPO was absent. As described in Paper-I, we downloaded the data from the \xmm\ Science Archive (XSA), and re-reduced it using the \xmm\ Science Analysis System (SAS v18.0.0) with the most recent calibration files. The same data reduction procedure is applied to all the observations. We define a circular source extraction region of 80 arcsec radius. Obs-2 suffers from significant pile-up and so we exclude a circular region of 10 arcsec radius around the source position for this dataset (see Appendix~\ref{sec-pileup}). The {\sc epproc}, {\sc emproc} and {\sc evselect} scripts were used to reduce the data from the European Photon Imaging Camera (EPIC), and extract the spectra and light curves. The {\sc rgsproc} script was used to reprocess the data from the Reflection Grating Spectrometer (RGS) and to extract the RGS spectra. The {\sc omfchain} script was used to reprocess the data from the Optical Monitor (OM), and to extract the photometric flux from the {\tt Imaging}-mode exposures and a light curve from the {\tt Fast}-mode exposures taken through the UVW1 filter.

\subsection{\nustar}
\nustar\ (Harrison et al. 2013) performed an observation of \rej1034\ on 2018-10-30 (Coordinated Universal Time: 10:29:38) for 100 ks exposure time, which overlapped with the entire observing window of the Obs-9 of \xmm. We used the {\sc nupipeline} script in {\sc HEASOFT} (v6.26.1, Blackburn 1995) to reprocess the data together with the most recent calibration database. To identify the South Atlantic Anomaly (SAA) passages and then remove contaminated data, we chose the {\tt optimized} mode of the SAA calculation with the default algorithm option. The {\sc nuscreen} script was used to produce the cleaned event files with the GRADE range of 0-4. The source extraction region was chosen to be a circle with a radius of 1 arcmin centered on \rej1034. The background was extracted from an aperture of the same radius located on a nearby source-free region. The {\sc nuproducts} script was used to extract the source and background spectra, and produce the response and auxiliary files.

\subsection{\swift}
\swift\ performed a target-of-opportunity (ToO) observation of \rej1034\ on 2018-10-30 with 1.7 ks exposure time, which was simultaneous with the \xmm\ and \nustar\ campaign. We downloaded the data from the High Energy Astrophysics Science Archive (HEASARC), and reprocessed it with {\sc HEASOFT} (v6.26.1) and the most recent calibration files. We ran the {\sc xrtpipeline} to reprocess the X-ray Telescope (XRT) data and used the {\sc xselect} tool to extract the spectrum from a circular region of 30 arcsec radius. 

Six optical/UV filters (UVW2, UVM2, UVW1, U, B and V)
were used by the ultraviolet-optical telescope (UVOT) during this observation.
The host galaxy of \rej1034\ is apparent in both the optical and UV bands, so we used a circular aperture of 10 arcsec radius to include both the AGN and host galaxy emission. The background flux was determined from a nearby source-free region with a circular radius of 40 arcsec. The {\sc uvotimsum} and {\sc uvotsource} scripts were run to extract the integrated sky image and source flux in every UVOT filter used during the observation. We also checked that our data are not affected by small scale changes in sensitivity within the detector
(Edelson et al. 2015)\footnote{https://heasarc.gsfc.nasa.gov/docs/heasarc/caldb/swift/docs/ uvot/uvotcaldb\_sss\_01.pdf}.

\begin{figure*}
\centering
\begin{tabular}{cc}
\includegraphics[trim=0.0in 0.1in 0.0in 0.1in, clip=1, scale=0.48]{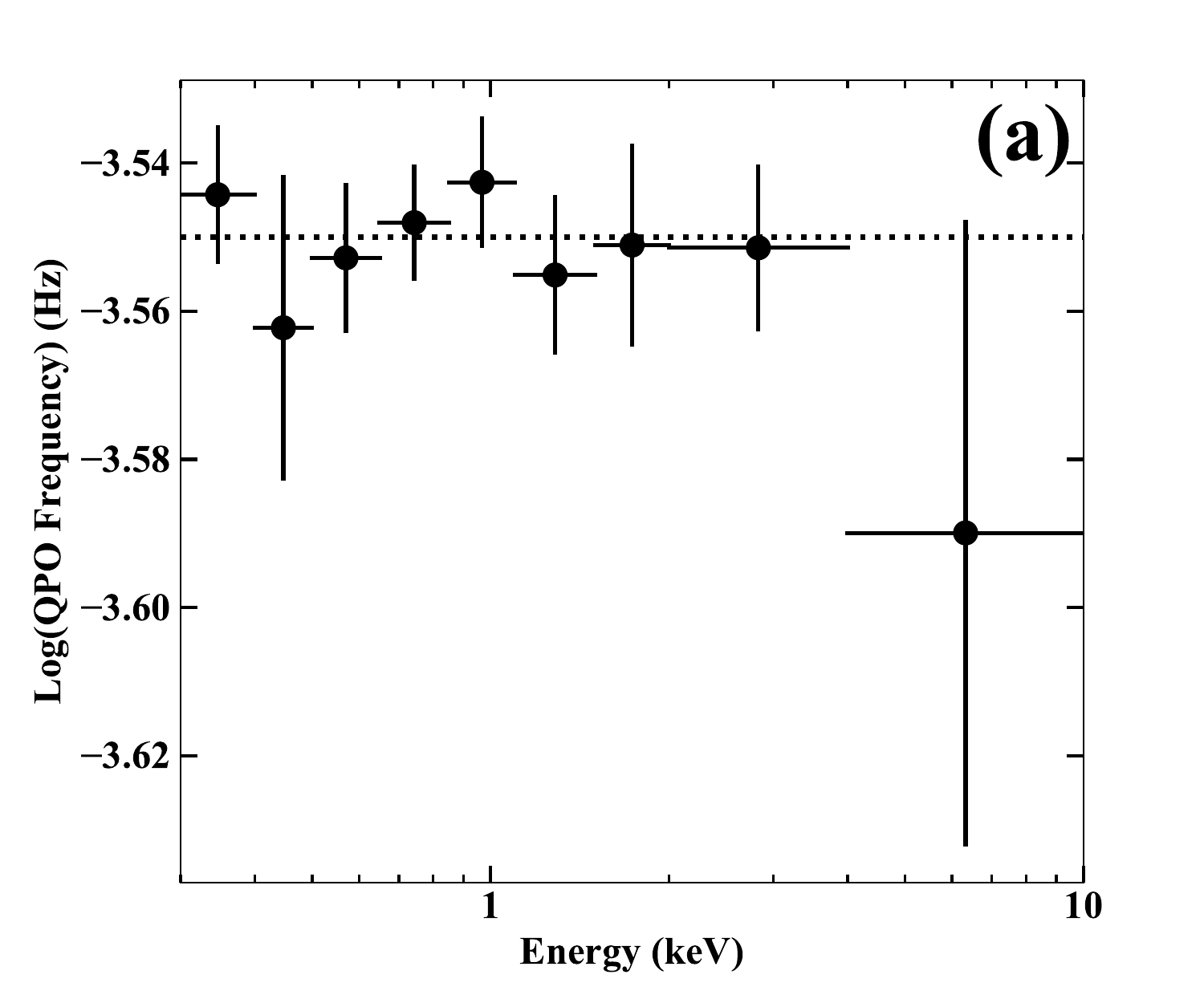} &
\includegraphics[trim=0.2in 0.1in 0.0in 0.1in, clip=1, scale=0.48]{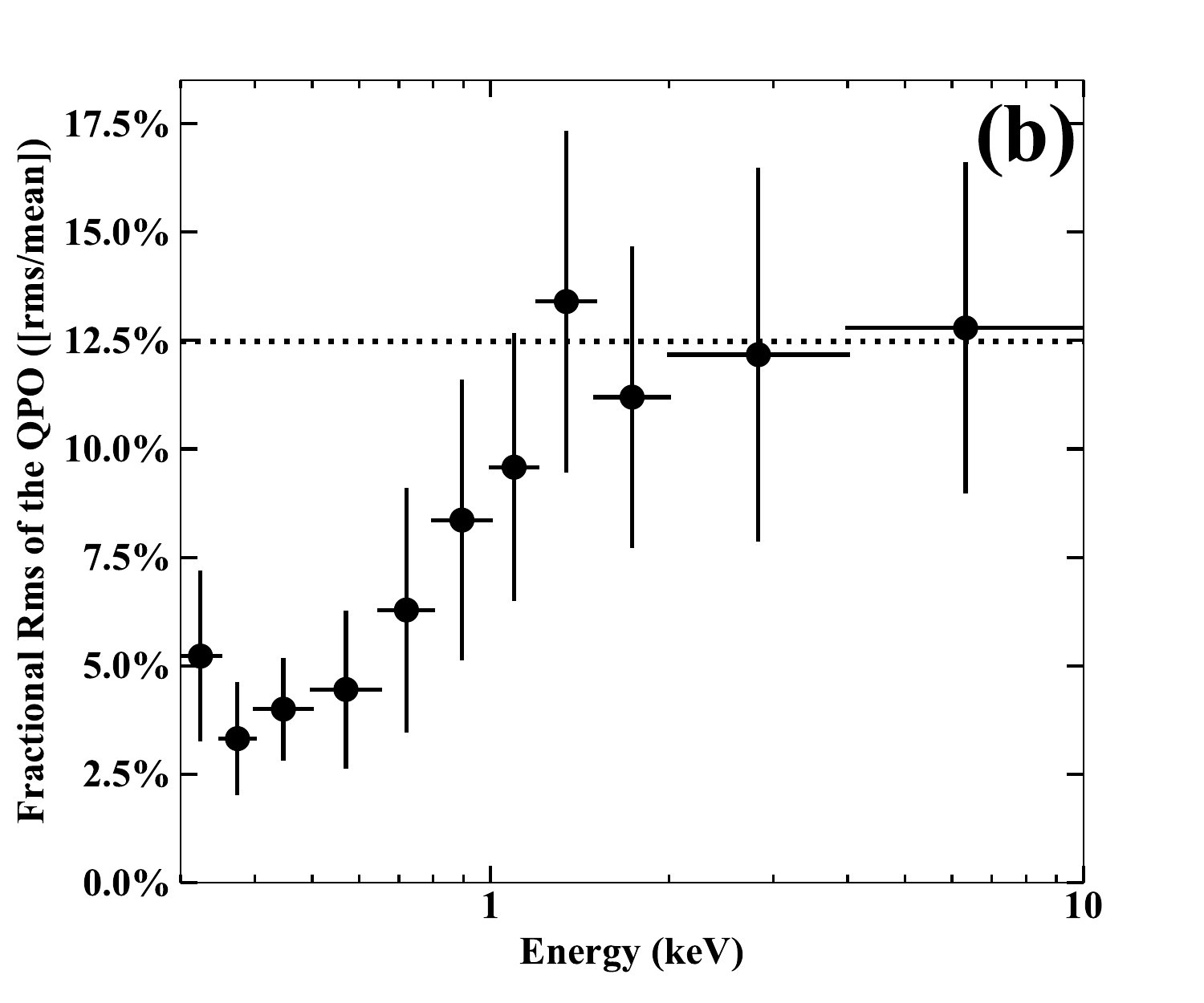} \\
\includegraphics[trim=0.0in 0.1in 0.0in 0.1in, clip=1, scale=0.48]{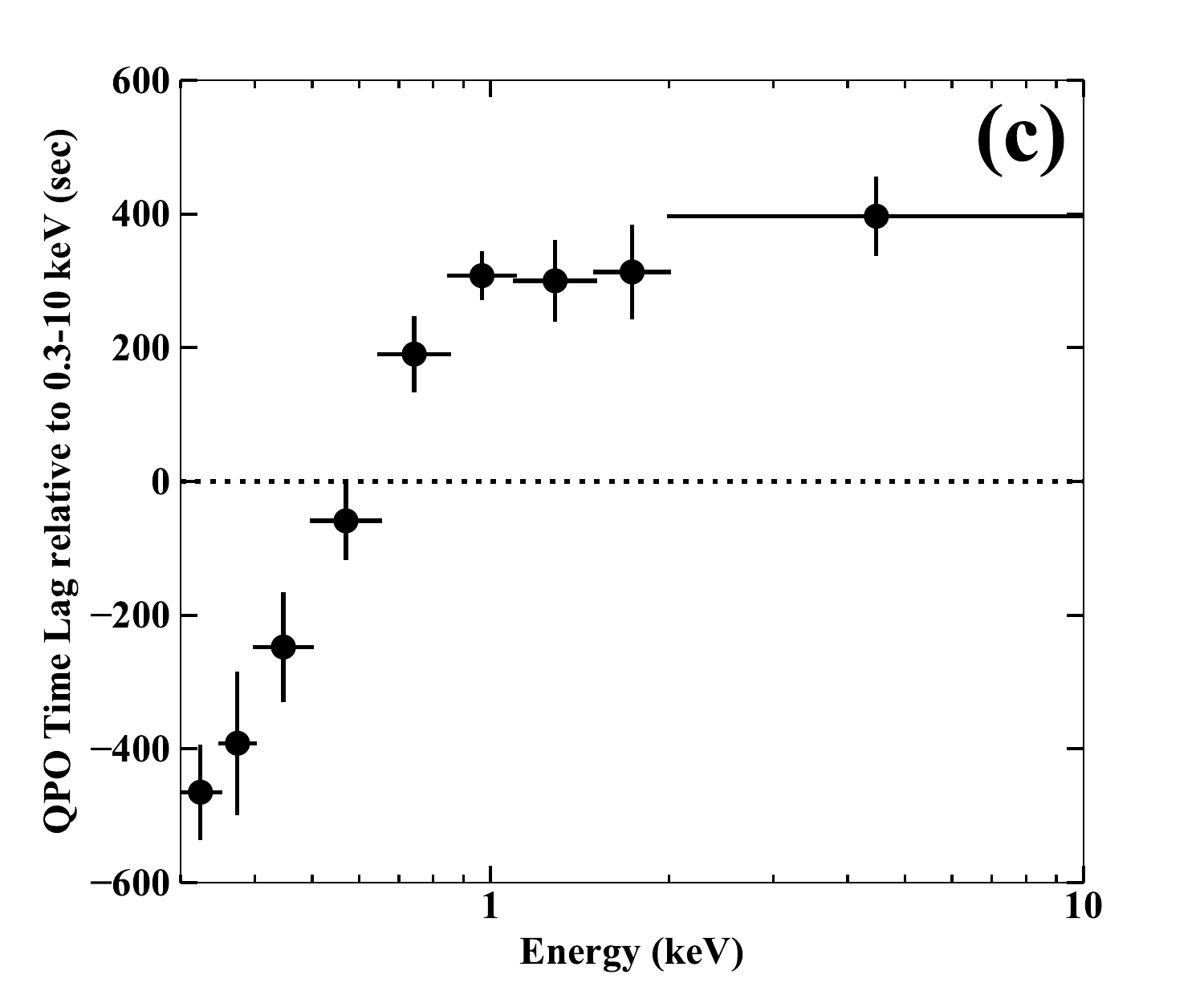} &
\includegraphics[trim=0.2in 0.1in 0.0in 0.1in, clip=1, scale=0.48]{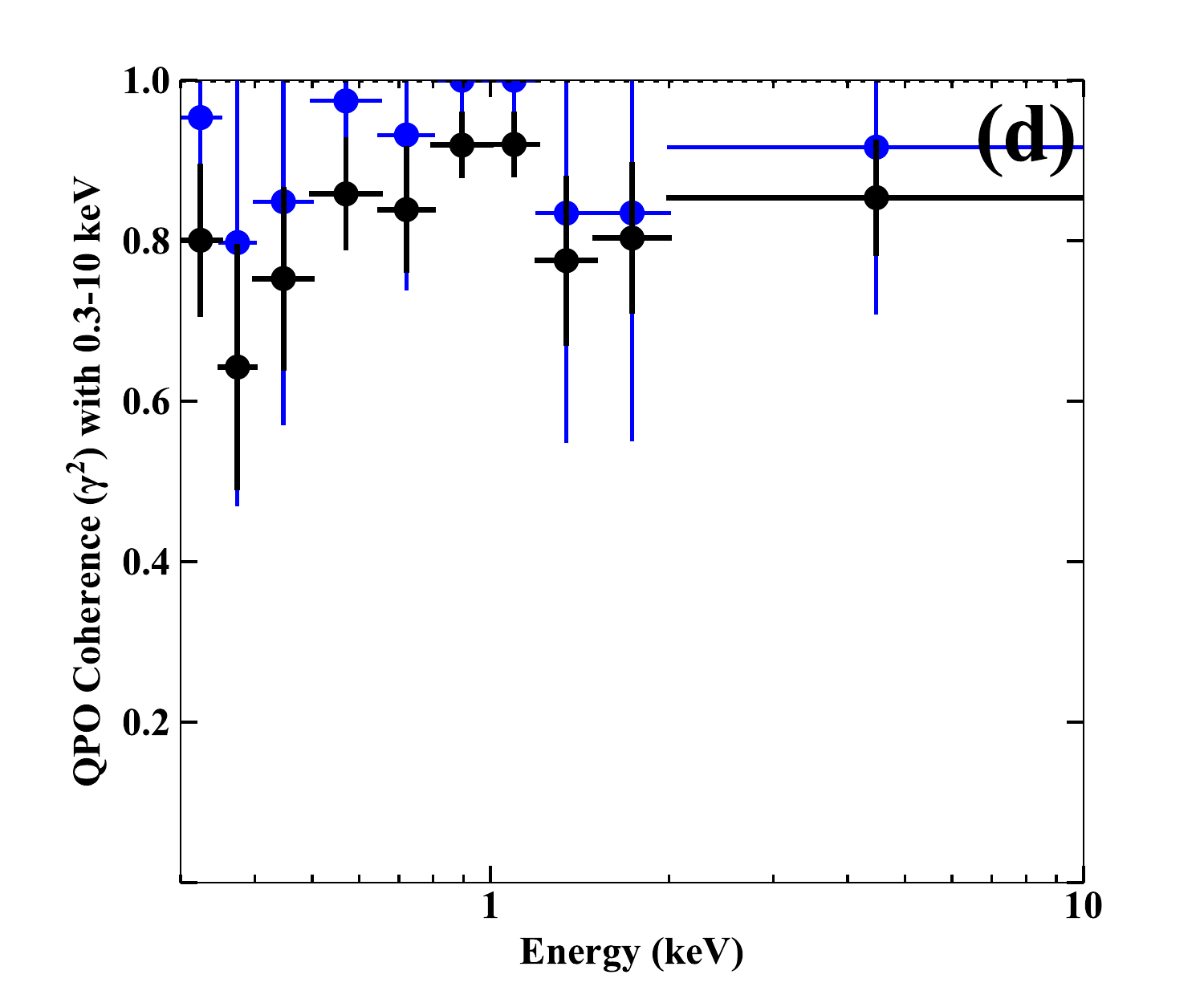} \\
\end{tabular}
\caption{The energy dependences for selected QPO properties, including the QPO's peak frequency (Panel-a), fractional rms amplitude (Panel-b), time lag relative to the QPO in 0.3-10 keV (Panel-c), and the coherence at the QPO frequency also relative to 0.3-10 keV (Panel-d). In Panel-a, the black dotted line indicates the QPO frequency of $2.8\times10^{-4}$ Hz in 0.3-10 keV. In Panel-b, the black dotted line indicates the mean rms of 12.5 per cent above 2 keV. In Panel-c, positive values indicate a lag behind the reference band of 0.3-10 keV. In Panel-d, the black points are the raw coherences, and the blue points are the Poisson-noise-corrected coherences. All the error bars represent 1$\sigma$ uncertainties.}
\label{fig-qpopar-ene}
\end{figure*}

\section{X-ray Spectral-timing Properties}
\label{sec-spectraltiming}
\subsection{Energy Dependence of the QPO Properties}
\label{sec-ene-qpo}
In Paper-I we compared the QPO period and rms variability among 0.3-1, 1-4 and 2-10 keV bands. We found no significant change in the QPO period, but the fractional rms amplitude increases from 4\% in 0.3-1 keV to 12.4\% in 1-4 keV. In this paper, we divide the 0.3-10 keV band into smaller energy bins, and carry out  a more comprehensive study into the energy dependences of the QPO's peak frequency, rms, time lag and coherence.

In order to determine the peak frequency of the QPO feature and its fractional rms amplitude, it is necessary to perform a careful modelling of the PSD. Following the same procedures of Paper-I, we use a power law to fit the red noise continuum, and a free constant to account for the Poisson noise. The QPO feature is modelled with a Gaussian profile. The maximum likelihood estimate (MLE) method is used to derive the best-fit parameters of the PSD model, and the rms is derived by integrating the best-fit QPO profile under the Belloni-Hasinger normalization (Belloni \& Hasinger 1990). Based on the best-fit PSD model, we simulate $10^{5}$ periodograms, and perform the same PSD fitting to each of them. Then the probability distributions of model parameters are derived, from which their uncertainties are measured.

Figure~\ref{fig-qpopar-ene}a shows the QPO frequency plotted against energy. We confirm our result in Paper-I with a higher energy resolution, showing that the QPO's peak frequency does not depend on the energy. Therefore, it is possible that the QPO signal present in different energy bands may have the same physical origin.

The fractional rms spectrum is shown in Figure~\ref{fig-qpopar-ene}b. This spectrum displays a typical shape for some super-Eddington NLS1s (e.g. Jin et al. 2009, 2013; Jin, Done \& Ward 2016, 2017a). It rises from $\sim$3\% at 0.5 keV to 12\% at 2 keV, and flattens towards the hard X-rays. Therefore, we can infer that the hard X-ray is dominated by a single component, and the soft excess contains a separate and less variable component. However, the ratio of rms between 2 keV and 0.3 keV is $\sim$4, while the ratio of flux at these energies in the time-averaged spectrum is $\sim$16, so there must be at least one component present in the soft X-ray excess which also contains the QPO.

\begin{figure*}
\centering
\begin{tabular}{ccc}
\includegraphics[trim=0.5in 0.2in 0.0in 0.1in, clip=1, scale=0.4]{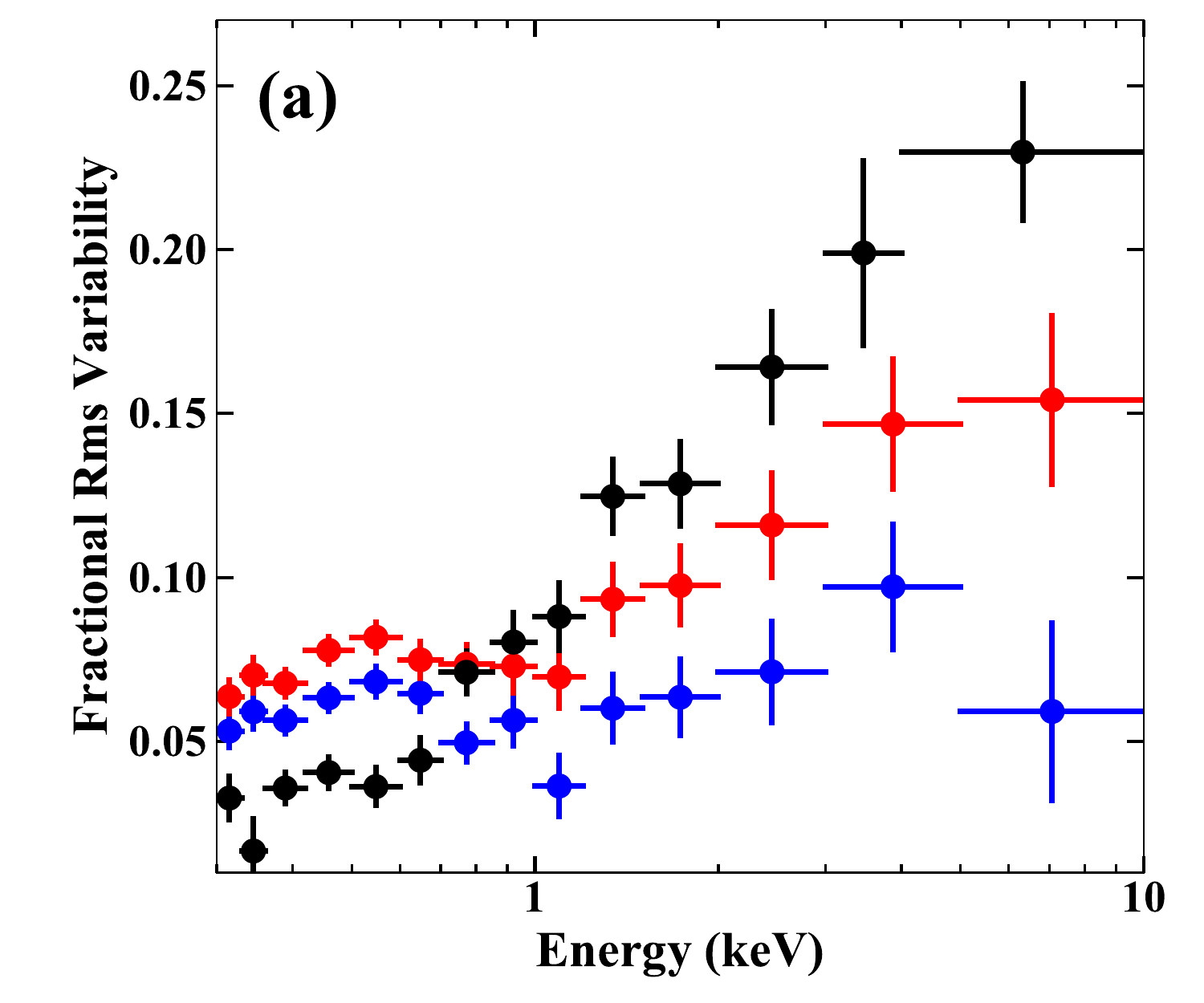} &
\includegraphics[trim=0.6in 0.2in 0.0in 0.1in, clip=1, scale=0.4]{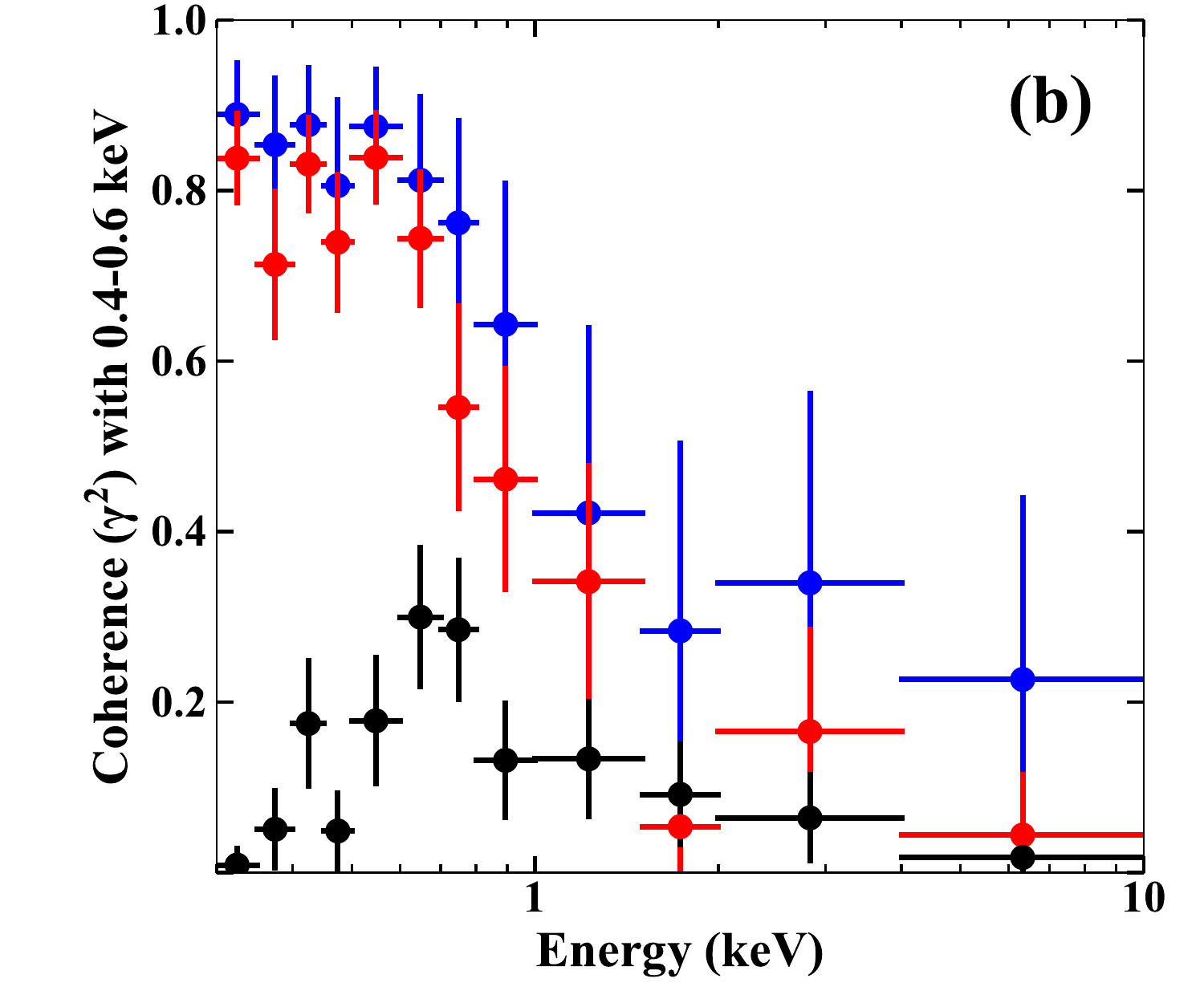} &
\includegraphics[trim=0.4in 0.2in 0.0in 0.1in, clip=1, scale=0.4]{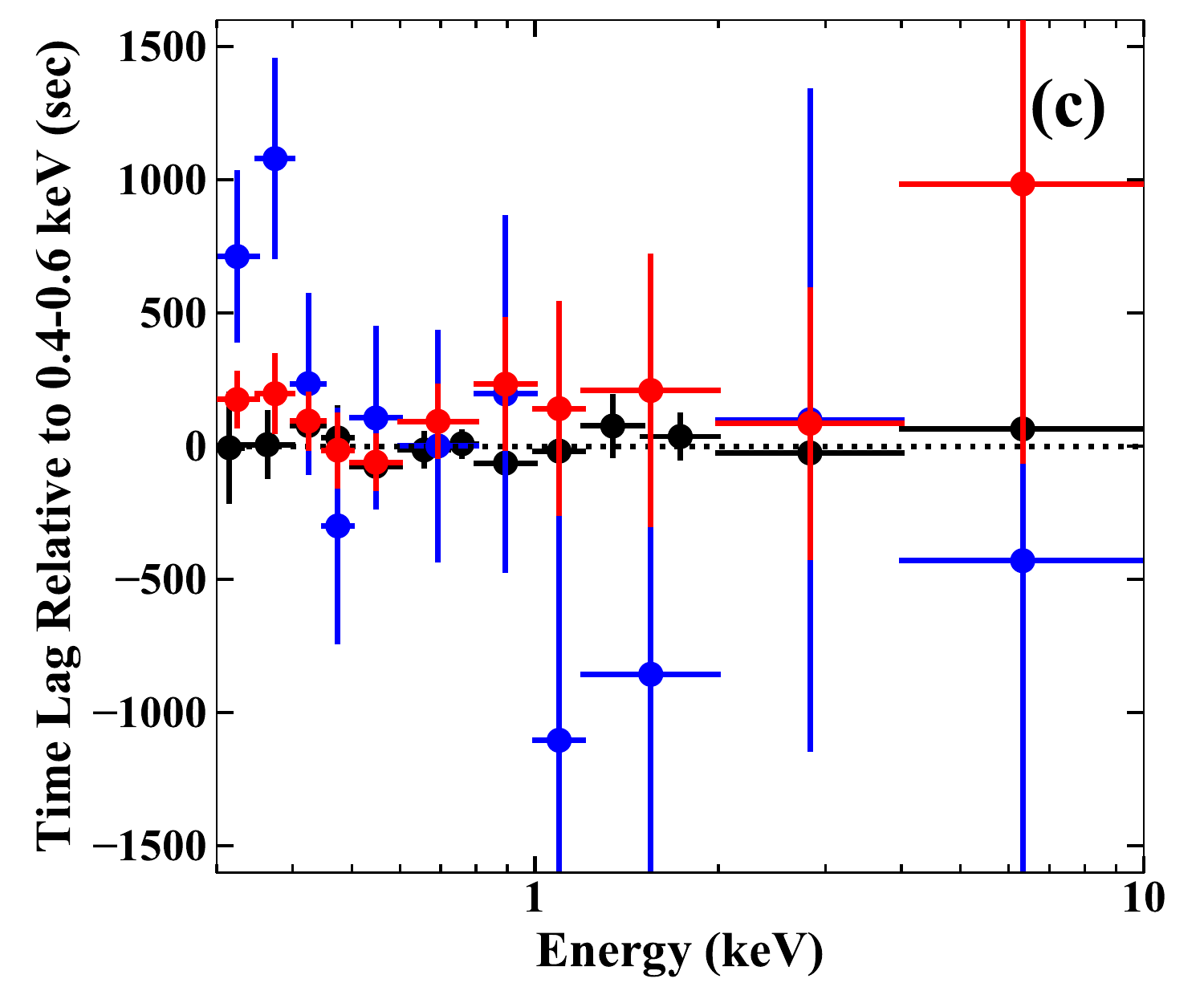} \\
\end{tabular}
\caption{Panel-a: fractional rms spectra for the low-frequency band (red points), high-frequency band (black points) and very-low-frequency band (blue). Panel-b: coherence spectra relative to the 0.4-0.6 keV band for the same three frequency bands. Panel-c: lag spectra relative to the 0.4-0.6 keV band for the same three frequency bands. The definition of various frequency bands is given in Figure~\ref{fig-lc}.}
\label{fig-cohlag}
\end{figure*}

In Paper-I we reported the discovery that the QPO in 0.3-1 keV leads that in 1-4 keV by 430 s. We now investigate this result in more detail by producing the lag spectrum. The QPO's frequency bin is chosen to be $(2.5-3.5)\times10^{-4}$ Hz, which includes 7 data points in the PSD. The 0.3-0.4 keV band is selected to be the reference band. Figure~\ref{fig-qpopar-ene}c shows how the time lag changes with energy, where a positive lag indicates that the QPO in 0.3-0.4 keV leads. It is clear that the lag increases monotonically from 0.3 to 1 keV, and then flattens towards harder X-rays in a similar way to that of the rms spectrum. The maximum time lag between the soft and hard bands is $\sim$ 800 s. This shape of the QPO lag spectrum also supports the suggestion that the soft X-ray excess is dominated by a separate component from that of the hard X-rays.

Since a time lag between two time series is meaningful only if there is a strong coherence between them, we also calculate the coherence of the QPO between 0.3-0.4 keV and other energy bands. The raw and Poisson-noise-corrected coherences are calculated in the Fourier domain following the prescriptions described in Vaughan \& Nowak (1997) and Nowak et al. (1999) (also see the review by Uttley et al. 2014). Figure~\ref{fig-qpopar-ene}d shows the coherences at different energies.
It is clear that the QPO's coherence is close to unity across the 0.3-10 keV range, which indicates that the QPO signals in different energy bands are highly coherent. Therefore, the phase lag of the QPO should represent the actual time lag between different energy bands. This further supports our point in Paper-I that the positive QPO lag seen in Obs-9 is more robust than the negative lag observed in Obs-2, because the QPO in Obs-2 has a much lower coherence.

\subsection{Frequency-differentiated Non-QPO Variability Spectra}
\label{sec-varspec}
During Obs-9, \rej1034\ displays a strong stochastic variability as well as the presence of the QPO. 
Here we explore the spectral-timing properties of the stochastic variability, and compare them with those of the QPO. We define a low-frequency (LF) range as $(0.14-2.5)\times10^{-4}$ Hz, where the lower frequency limit is determined by the duration of the clean light curve. The high-frequency (HF) range is set to be $(0.35-1.0)\times10^{-3}$ Hz. These frequency ranges are specially selected to exclude the QPO signal. We also explore the very low frequency band (VLF), defined as $(0.14-1.0)\times 10^{-4}$~Hz (see the shaded regions in Figure~\ref{fig-lc}).

We note that due to the relatively low count rate above 2 keV, there can exist a significant number of zero-count bins in the hard X-ray light curves if the binning time is too small (e.g. 50 s). These zero-count bins can severely bias the calculation of variability if not treated corrected. A possible effect is that the high-frequency power can deviate significantly from Poisson due to the existence of these zero-count bins. Therefore, we choose a large binning time of 500 s to ensure that there are no zero-count bins contained in the light curves. This requirement imposes the upper limit of $10^{-3}$ Hz on the high frequency band.

First we produce the fractional rms spectra at LF and HF, which are shown in 
Figure~\ref{fig-cohlag}a as the red and black spectra.
The HF rms spectrum appears similar to the QPO's rms spectrum below 1~keV (both in shape and normalization), but above 1~keV it continues to rise to a maximum above 20\% at 10 keV (although with relatively low S/N), whereas the QPO's rms spectrum flattens at $\sim 12$\% from 1~keV onwards. 
The HF rms spectrum is similar to those seen in some other super-Eddington NLS1s which also have a smooth and steep soft X-ray excess (e.g. Jin et al. 2009; Jin, Done \& Ward 2013), although so far no QPOs of comparable significance to that in  \rej1034 have been detected from these NLS1s. 

The rms spectrum of the LF stochastic variability is different from both the HF and the QPO. It has a bump below 1 keV, and then a gradual increase towards hard X-rays. Its rms is higher than the HF in the soft X-rays, but lower than the HF in the hard X-rays. This shows that the soft X-rays vary more at lower frequencies than the hard X-rays, suggesting that they originate from a more extended region (e.g. Gardner \& Done 2014, Jin et al. 2017b). The rms of the VLF variability (blue) continues this trend, but with even less variability observed above 2~keV. 

We select the high S/N 0.4-0.6 keV band light curve as the reference band to produce the coherence and lag spectra (Panels b and c in Figure~\ref{fig-cohlag}) at HF (black), LF (red) and VLF (blue). There is almost no coherence at HF and the HF lags are not statistically significant. In contrast, the slower stochastic variability at LF varies coherently across the 0.3-0.7~keV energy range, and then drops 
rapidly to zero towards 1 keV and at higher energies. Therefore, the LF variability of the soft X-rays is not correlated with the hard X-rays, and there is no significant LF time lag across the 0.3-10 keV band. This is a rather different situation from that observed in some other super-Eddington NLS1s (e.g. \pg12, \rxj04), in which it is found that the soft X-rays lead the hard X-rays at LF with a high coherence (Jin et al. 2013, 2017a). We point out that the X-ray flux of \pg12\ is higher than of \rej1034\ which, in turn, is higher than \rxj04, and so the difference of LF time lag is not simply a result of the level of Poisson noise present in the hard X-ray band.

Considering even slower variability (VLF: blue) there is possibly some
correlated variability in the 2-10~keV emission, although none of the lags in this energy band are statistically significant. However, the VLF lags are marginally significant at lower energies, with the 
0.3-0.35 keV band lagging behind 0.4-0.6 keV by 713 $\pm$ 313 s with a coherence of 0.89 $\pm$ 0.06, while the 0.35-0.4 keV band lags by 1080 $\pm$ 367 s with a coherence of 0.85 $\pm$ 0.08.

We also explored the results by using the light curve in 2-10~keV as the reference light curve, as this energy
band has high rms variability (see Figure~\ref{fig-cohlag}a). The HF variability in 2-10 keV is also likely to be most sensitive to any soft X-ray reflection/reverberation processes arising from hard X-ray illumination of the inner disc, as well as any influence from the shape of the variable hard X-ray component (see e.g. Fig. 8 for the NLS1 \pg12: Jin et al. 2013). However, the S/N of the data is insufficient to determine the shape of the 2-10~keV emission component when this band is divided up to provide higher energy resolution, so there are no significant results from this analysis.

\begin{figure*}
\centering
\begin{tabular}{ccc}
\includegraphics[trim=0.2in 0.2in 0.0in 0.1in, clip=1, scale=0.41]{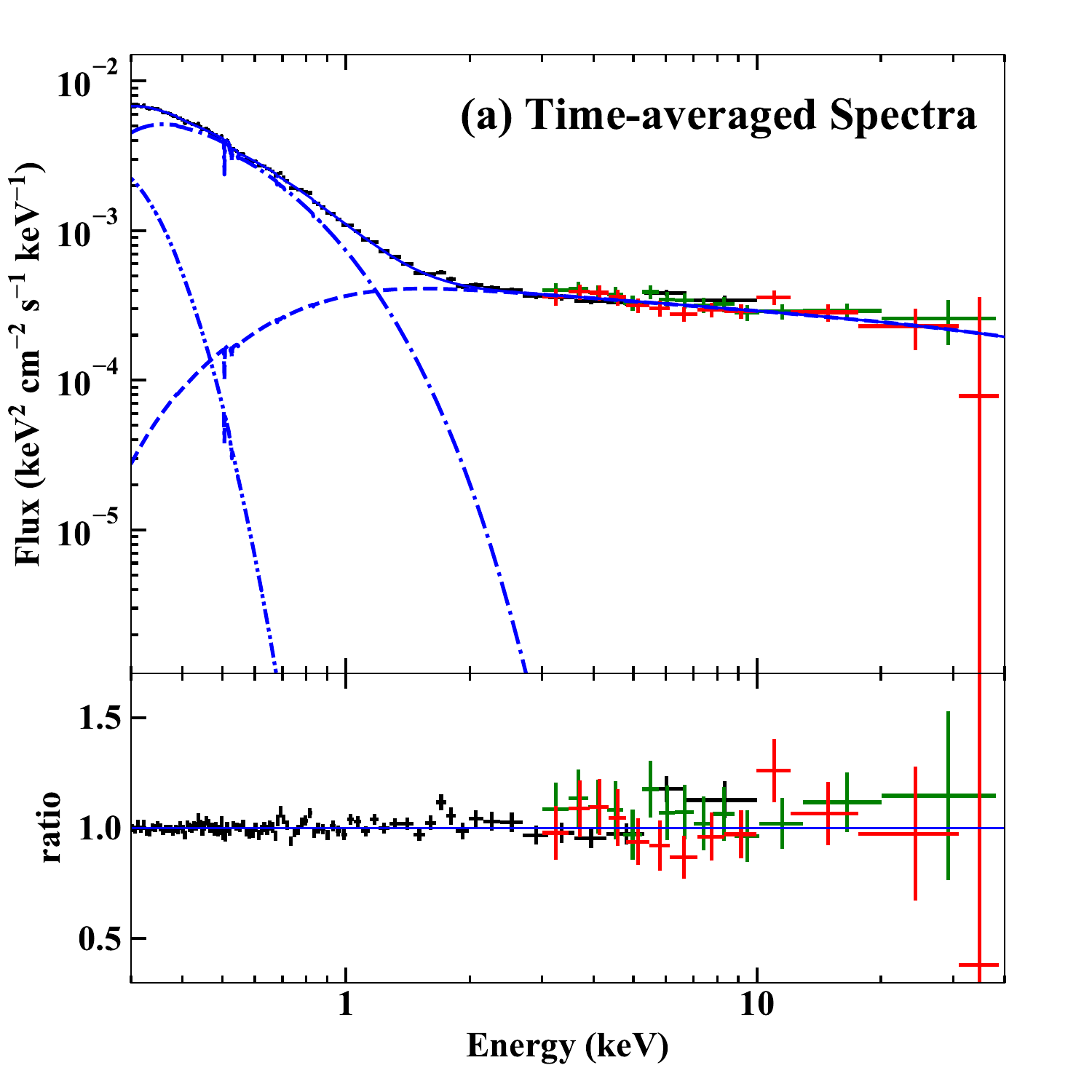} &
\includegraphics[trim=0.7in 0.2in 0.0in 0.1in, clip=1, scale=0.41]{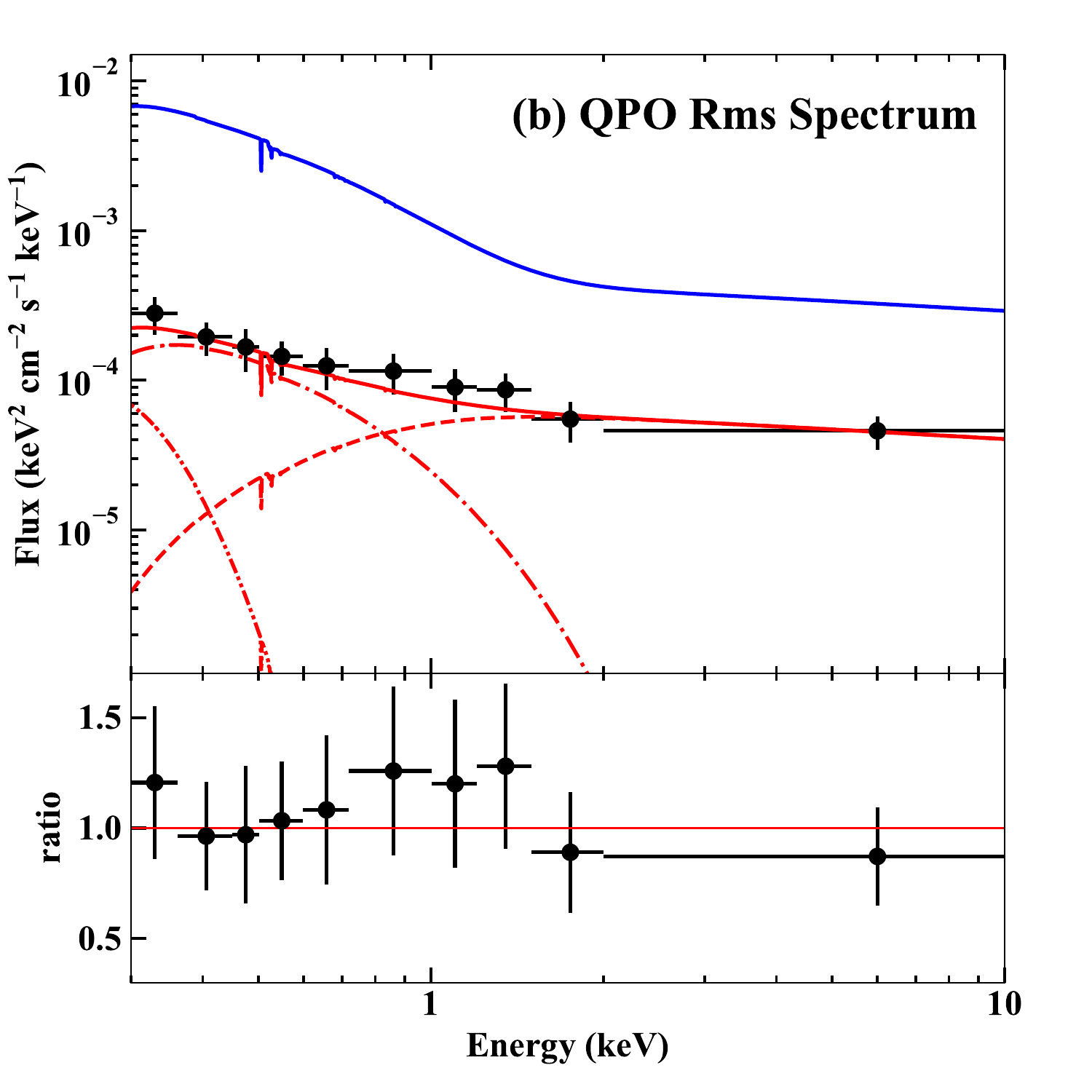} &
\includegraphics[trim=0.75in 0.2in 0.0in 0.1in, clip=1, scale=0.41]{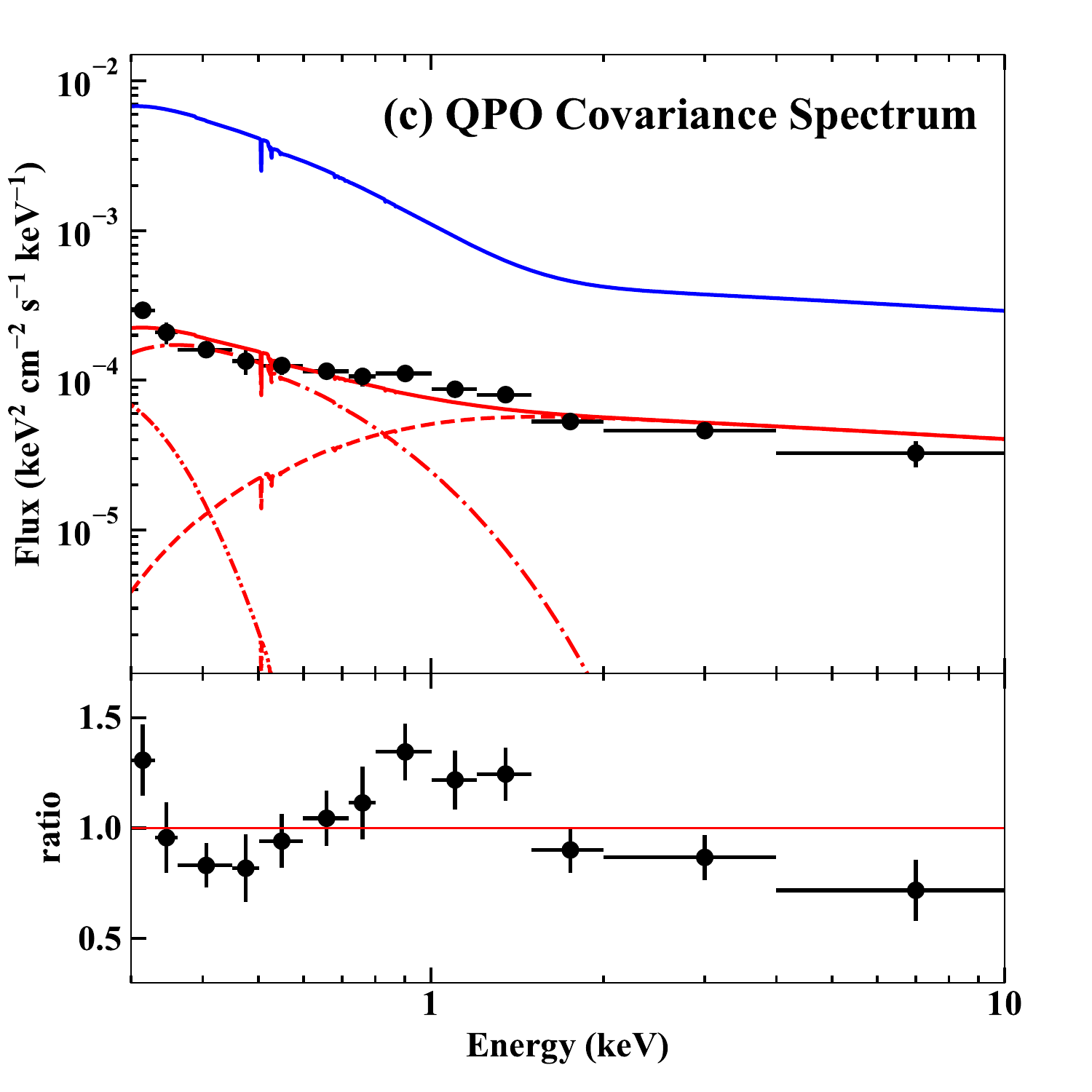} \\
\includegraphics[trim=0.2in 0.2in 0.0in 0.1in, clip=1, scale=0.41]{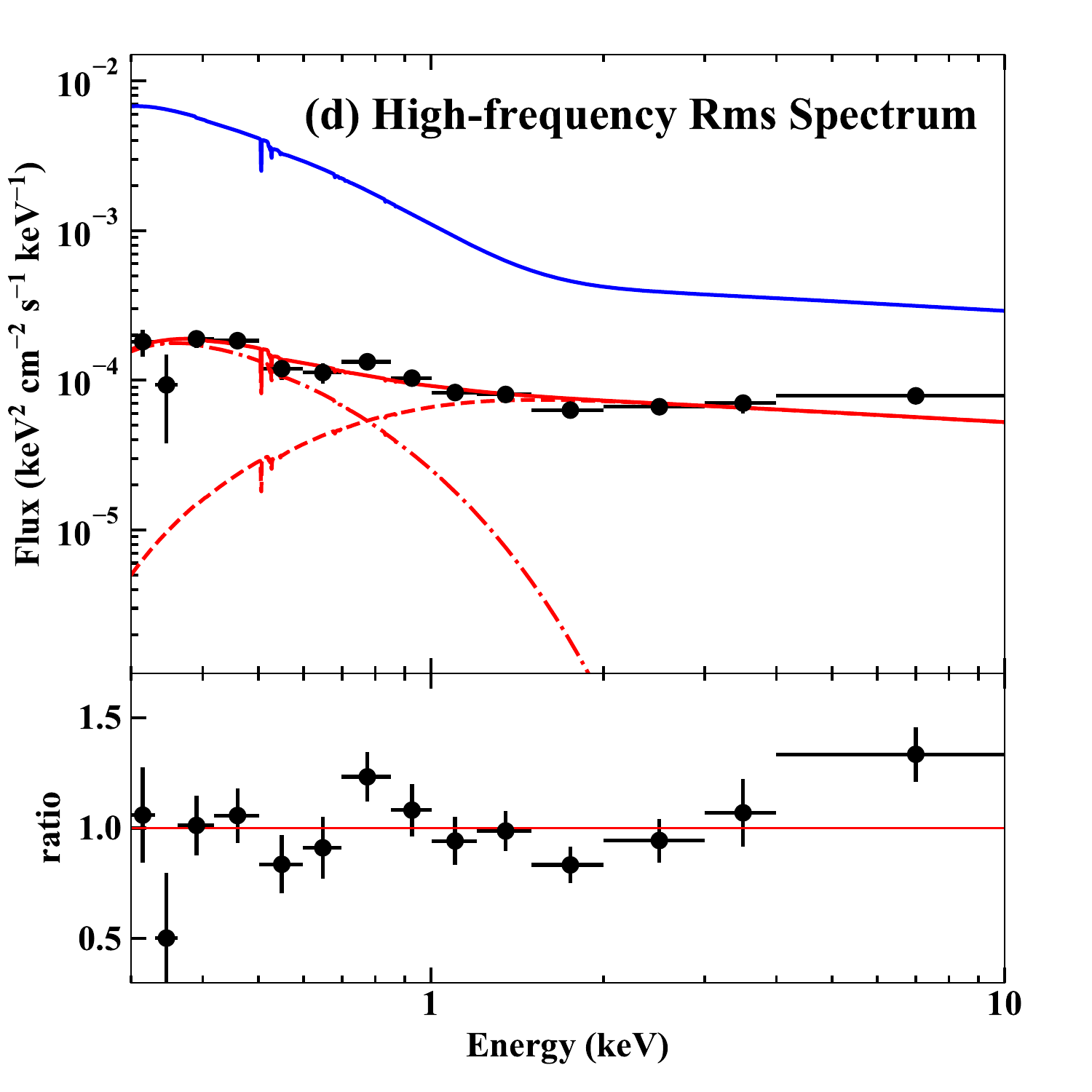} &
\includegraphics[trim=0.7in 0.2in 0.0in 0.1in, clip=1, scale=0.41]{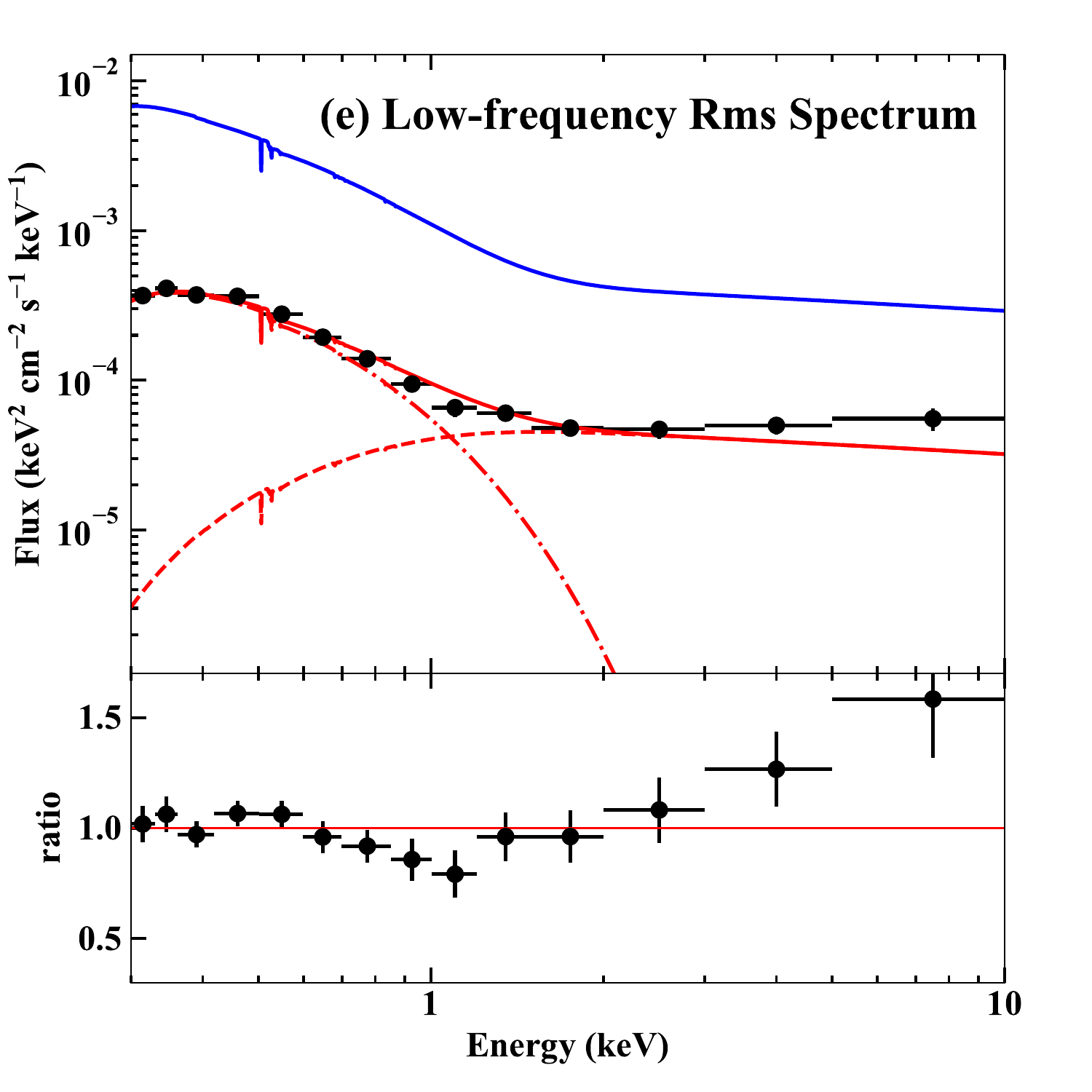} &
\includegraphics[trim=0.75in 0.2in 0.0in 0.1in, clip=1, scale=0.41]{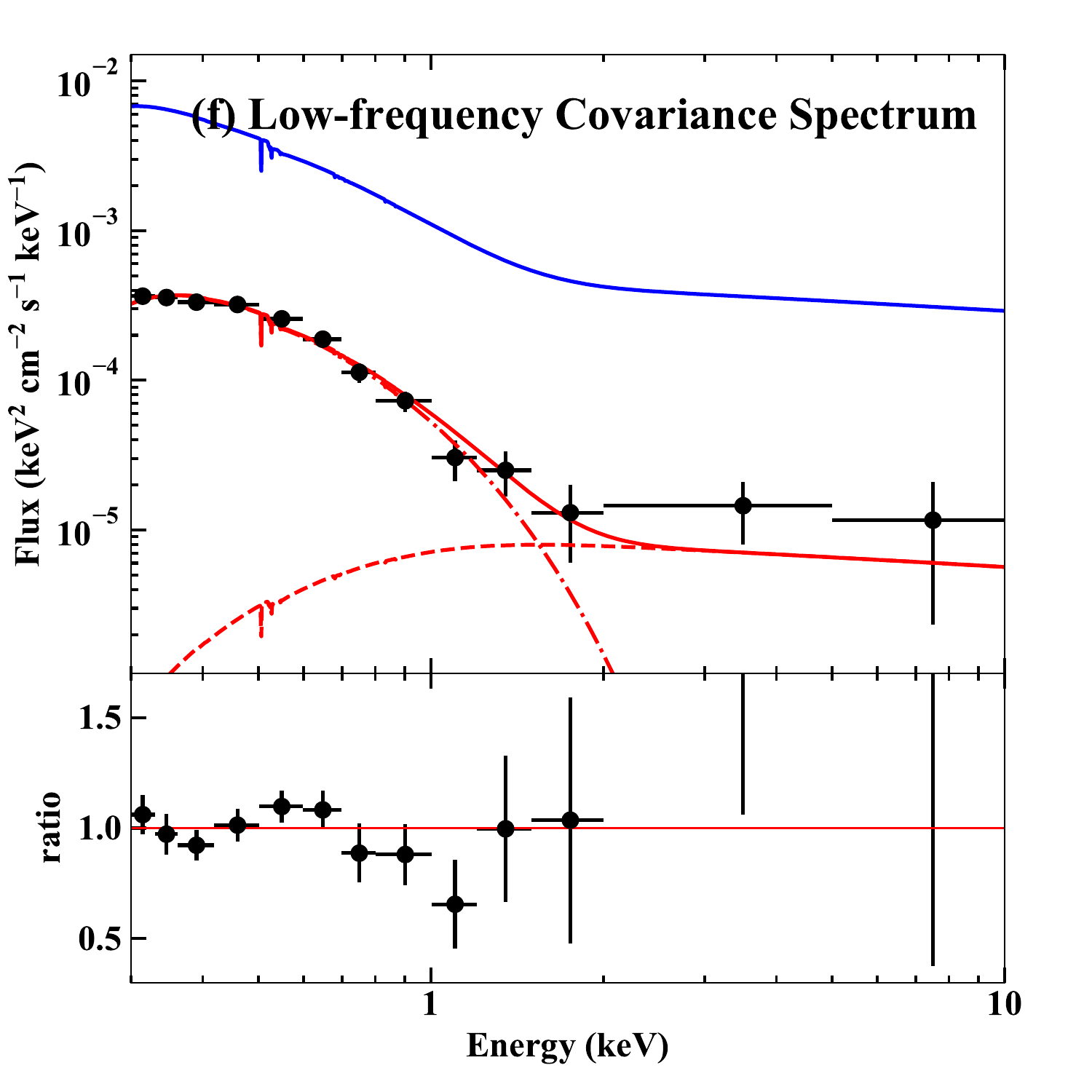} \\
\end{tabular}
\caption{Simultaneous fits of the time-averaged spectra (Panel-a), rms spectrum (Panel-b) and covariance spectrum (Panel-c) at the QPO frequency, high-frequency rms spectrum (Panel-d), low-frequency rms spectrum (Panel-e), and low-frequency covariance spectrum (Panel-f) from Obs-9. The Model-1 configuration is used. In panel-a, the black, green and red spectra are from \xmm\ EPIC-pn, \nustar/FPMA, \nustar/FPMB, respectively. The blue solid, dash, dash-dotted and dotted lines indicate the best-fit {\tt nthComp}, {\tt CompTT} and {\tt diskbb} components, respectively. In Panels b-f, the red solid, dash, dash-dotted and dotted lines indicate their individual best-fit models and separate components, whose shapes are the same as in Panel-a, but with different normalizations. For comparison, the best-fit model of the time-averaged spectrum in Panel-a is also shown in the other panels as the blue solid line.}
\label{fig-spec-rmscov1}
\end{figure*}

\begin{figure*}
\centering
\begin{tabular}{ccc}
\includegraphics[trim=0.2in 0.2in 0.0in 0.1in, clip=1, scale=0.41]{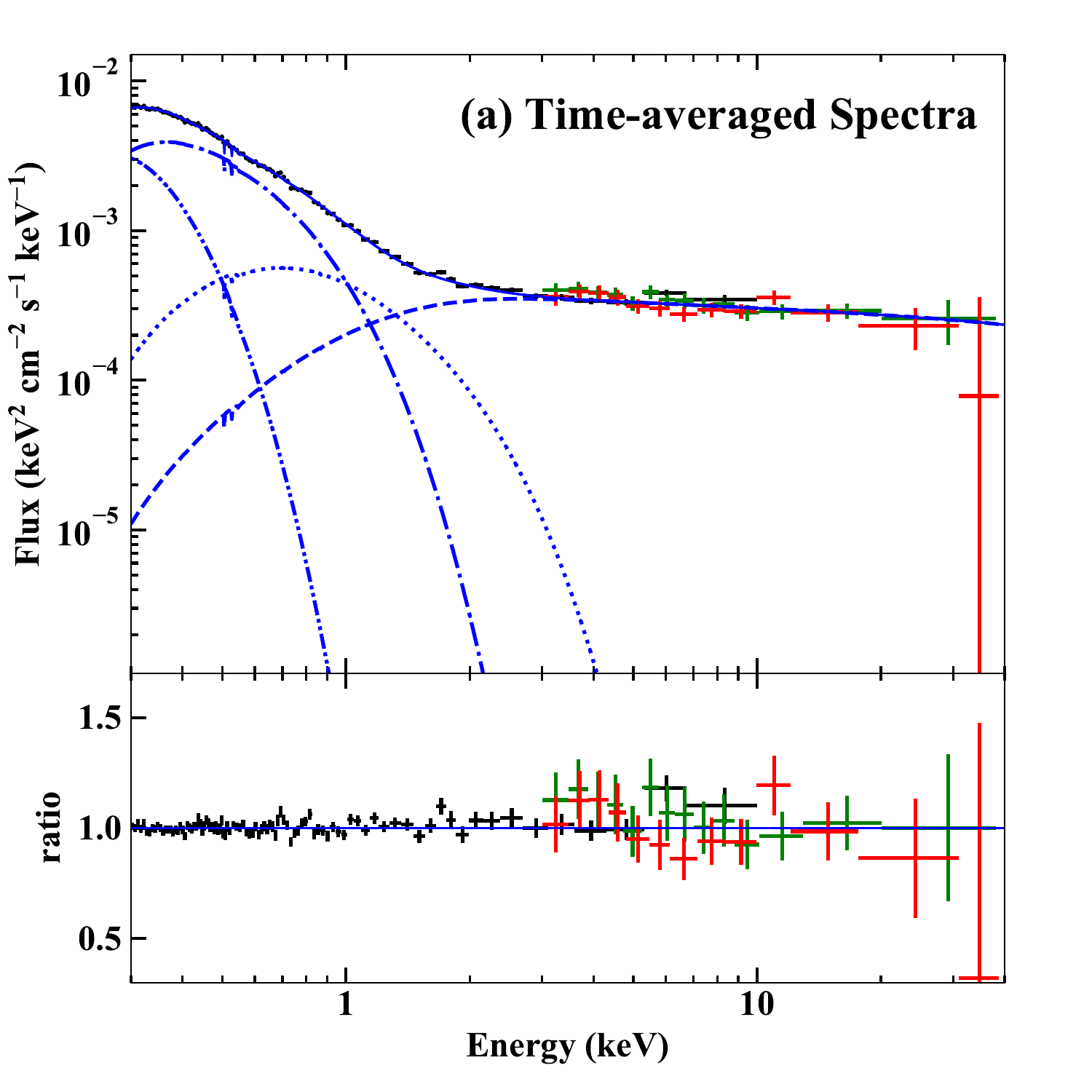} &
\includegraphics[trim=0.7in 0.2in 0.0in 0.1in, clip=1, scale=0.41]{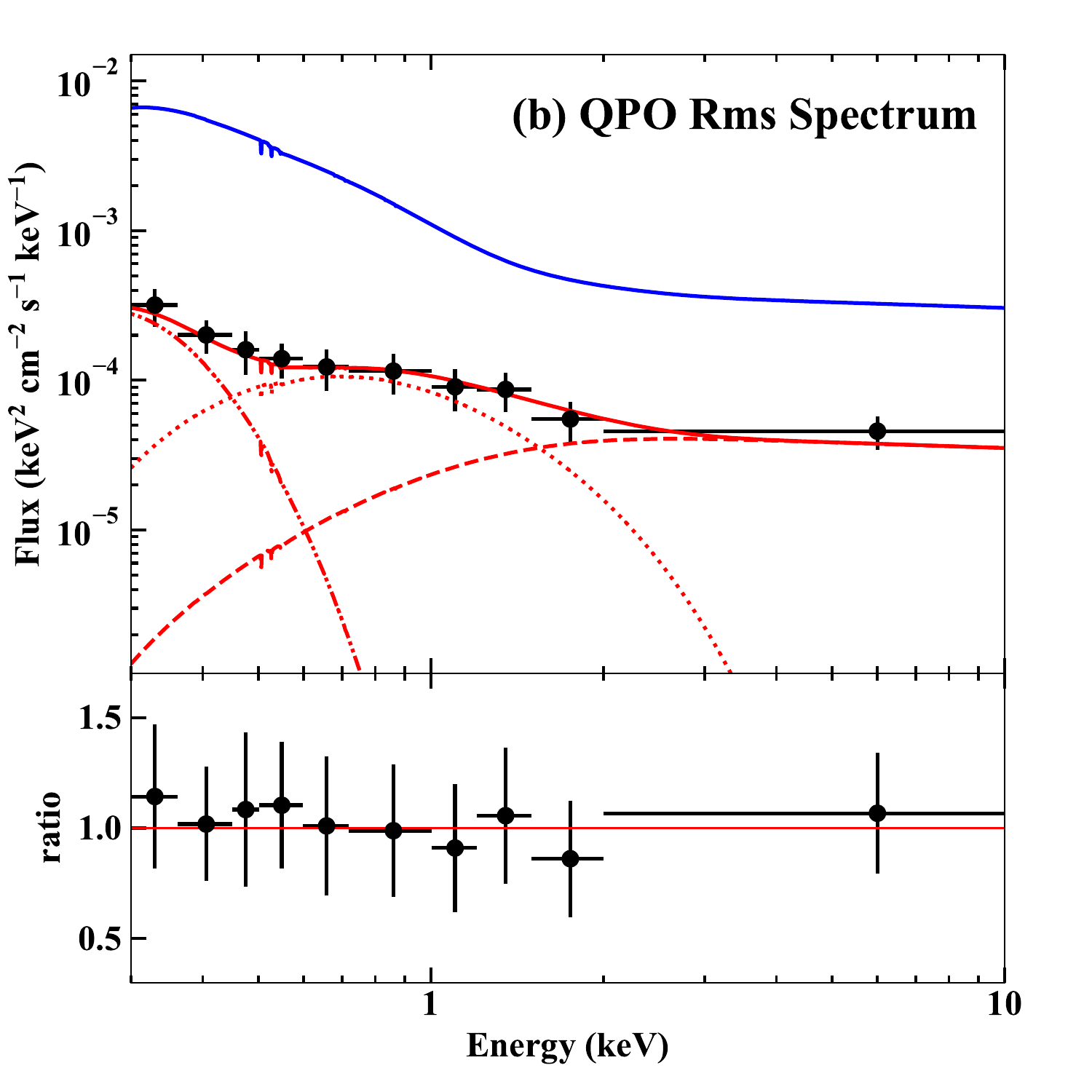} &
\includegraphics[trim=0.75in 0.2in 0.0in 0.1in, clip=1, scale=0.41]{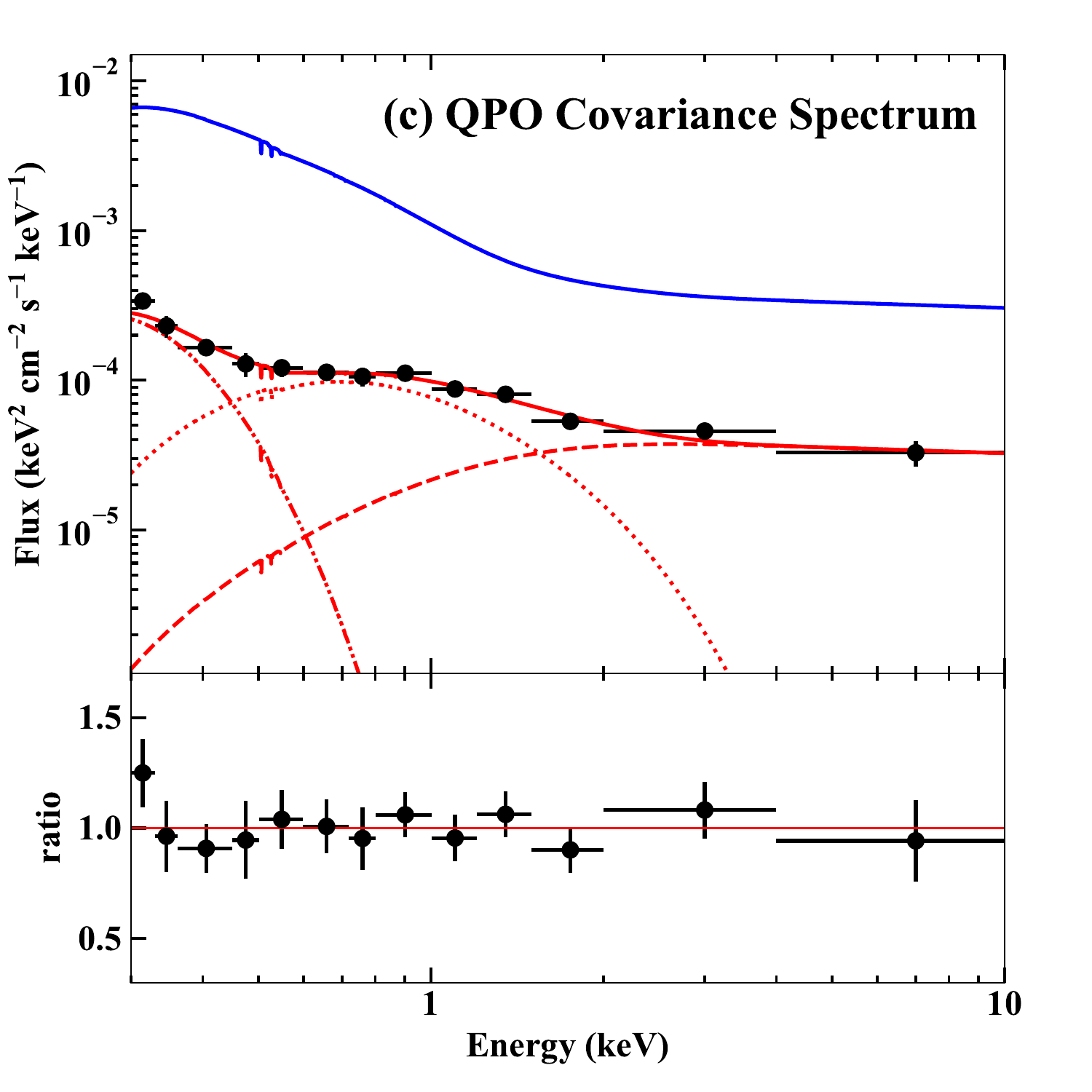} \\
\includegraphics[trim=0.2in 0.2in 0.0in 0.1in, clip=1, scale=0.41]{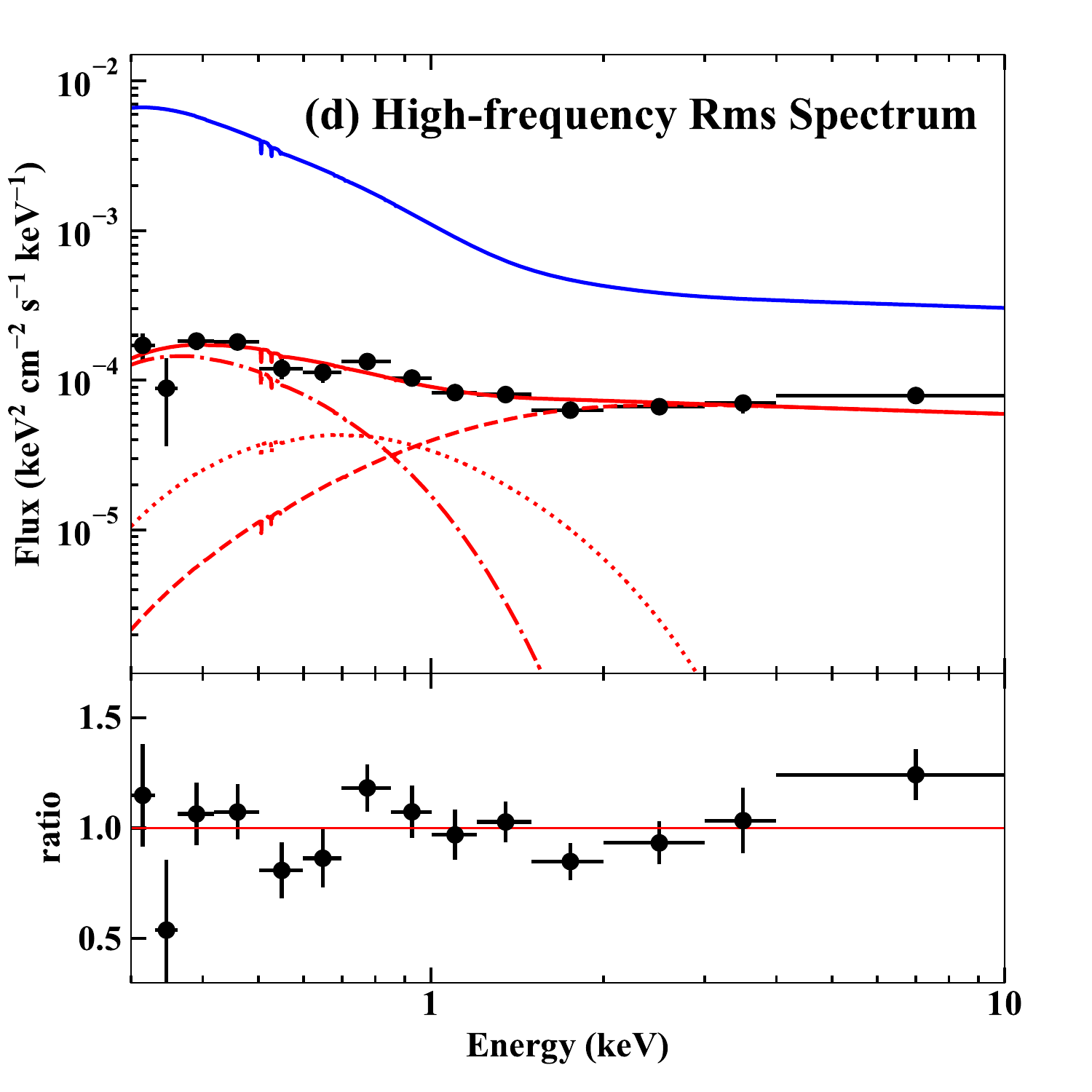} &
\includegraphics[trim=0.7in 0.2in 0.0in 0.1in, clip=1, scale=0.41]{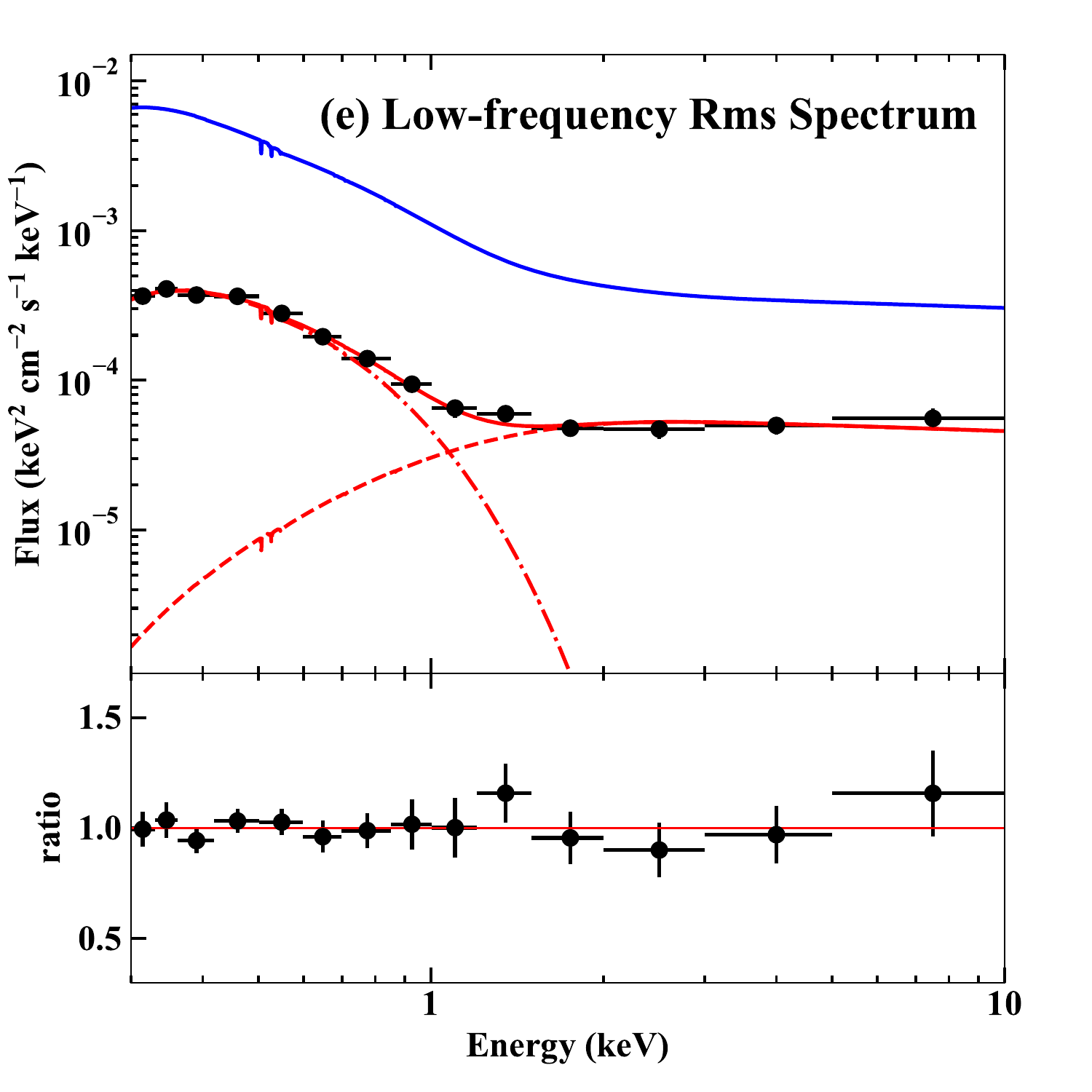} &
\includegraphics[trim=0.75in 0.2in 0.0in 0.1in, clip=1, scale=0.41]{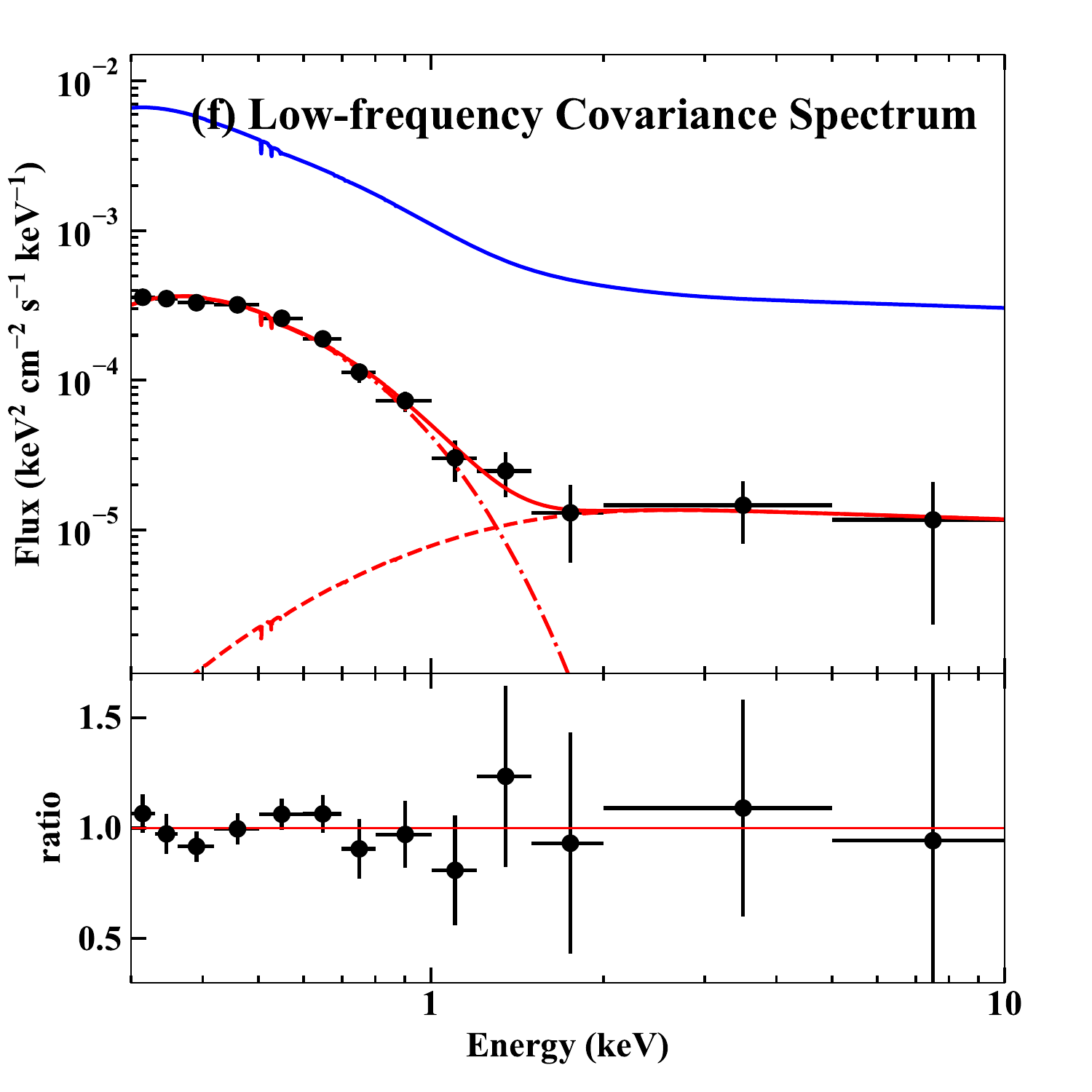} \\
\end{tabular}
\caption{Similar to Figure~\ref{fig-spec-rmscov1}, but for the Model-2 configuration, which produces good fits to all the spectra from Obs-9. The dash-dot-dot, dash-dot, dotted and dash lines indicate the {\tt diskbb}, {\tt CompTT-1},  {\tt CompTT-2} and {\tt nthComp} components, respectively.}
\label{fig-spec-rmscov2}
\end{figure*}

\section{X-ray Spectral-timing modelling}
\label{sec-xray-modelling}
\subsection{Modelling the Time-averaged and Variability Spectra}
\label{sec-varspec-model}
The time-averaged spectra of AGN are often degenerated to different models, meaning that very different physical models can give equally statistically good fits. Hence it is important to include additional information obtaining from the variability. The rms and coherence spectra reported here can be used together with the time-averaged spectra to provide some additional constraints. 
For \rej1034, previous analysis on the time-averaged spectrum, QPO rms and covariance spectra from Obs-2 have shown that the soft X-rays are better explained by a warm, optically thick Comptonisation model 
than by relativistic reflection or absorption dominated models (Middleton et al. 2009; Middleton, Uttley \& Done 2011). Such a warm corona model is also favoured for other NLS1s (e.g. Jin, Done \& Ward 2013, 2016, 2017a; Kara et al. 2017; Parker, Miller \& Fabian 2018). 

There are six types of spectra which can be derived from the spectral-timing analysis described in the previous section. First, the time-averaged spectrum itself, then the absolute rms spectrum,
constructed by multiplying the fractional rms spectra by the time-averaged spectrum.
We derive these for the QPO itself, and also for the HF and LF stochastic variability. 
The correlated variability can be clearly shown by the covariance spectrum, formed by multiplying the absolute rms spectra by the square root of the coherence (Uttley et al. 2014).
Again we construct these for the QPO itself, and for the HF and LF stochastic variability separately. But since there is no significant coherence for the HF stochastic variability, we do not produce the HF covariance spectrum. In addition, we extend the time-averaged spectrum to higher energies using the \nustar\ spectra observed simultaneously with \xmm\ Obs-9.

\subsubsection{Model-1: disc, soft X-ray \& hard X-ray Comptonisation}
\label{sec-model1}
The first model configuration (hereafter: Model-1) includes an accretion disc component, a soft X-ray Comptonisation component and a hard X-ray Comptonisation component. These components are modeled by {\tt diskbb}, {\tt CompTT} (Titarchuk 1994) and {\tt nthComp} (Zdziarski, Johnson \& Magdziarz 1996) in {\sc xspec}, respectively. We further assume that the inner disk emission provides the seed photons for the soft X-ray Comptonisation which, in turn, provides seed photons for the hard X-ray Comptonisation. Thus the seed photon temperature of {\tt CompTT} is tied to the temperature of {\tt diskbb}, and the seed photon temperature of {\tt nthCompt} is tied to the electron temperature of {\tt CompTT}. The electron temperature of {\tt nthCompt} is fixed at 200 keV. The Galactic X-ray absorption{\footnote{https://www.swift.ac.uk/analysis/nhtot/index.php}} ($N_{\rm H, gal}$) towards \rej1034\ is 1.36$\times10^{20}$ cm$^{-2}$ (Willingale et al. 2013), modeled using {\tt TBabs} with the cross-sections taken from Verner et al. (1996) and abundances from Wilms, Allen \& McCray (2000). Any additional intrinsic neutral absorption within \rej1034\ ($N_{\rm H, host}$) is modelled by {\tt zTBabs}, with the column being a free parameter. When fitting the variability spectra we assume that they have the same model components as the time-averaged spectra, but with different normalizations. Furthermore, as a result of the QPO's high coherence, the shapes of the two QPO spectra are consistent with each other, except that the covariance spectrum has smaller errors than the rms spectrum (Wilkinson \& Uttley 2009). 

We fit all the six types of spectra simultaneously assuming that the model components differ only in their normalizations, but not in their shapes across all of the different types of spectra. We use this combined fit to obtain better constraints on the model parameters, with the best-fit results listed in Table~\ref{tab-varspecfit}, and the best-fit spectral decompositions shown in Figure~\ref{fig-spec-rmscov1}.

We find that Model-1 fits the time-averaged spectra very well. The best-fit $N_{\rm H, host}$ is $2.95^{+1.49}_{-0.97}\times10^{20}$ cm$^{-2}$, indicating a low amount of host galaxy absorption. The \nustar\ spectra show that the hard X-ray emission of \rej1034\ has the form of a single power law up to 40 keV, with a photon index $\Gamma$ of 2.20 $\pm$ 0.04. There is no indication of either a reflection hump or high-energy cut-off. However, the S/N is not sufficient to provide any useful spectral constraints above 40 keV. The best-fit disc emission has an inner temperature of 34.6$^{+10.4}_{-4.9}$ eV. The electron temperature of {\tt CompTT} is 0.20 $\pm$ 0.02 keV, similar to the broadly constant value of $\sim$0.2 keV found for many other AGN (Gierli\'{n}ski \& Done 2004; Crummy et al. 2006; Jin et al. 2012a). The optical depth $\tau$ is 12.4$^{+2.1}_{-1.4}$, indicating an optically thick Comptonisation medium.

These components can also be used to fit the variability spectra, although there are some problems associated with this modelling, which are discussed in the next section.
Figure~\ref{fig-spec-rmscov1} Panels-b and c show that the QPO rms and covariance spectra require contributions from all three components. The ratio of the overall normalization between these two variability spectra is 1.05$\pm$ 0.17, confirming that they are fully consistent with each other. 

The HF and LF rms spectra only require contributions from {\tt CompTT} and {\tt nthCompt} i.e. there is no {\tt diskbb} present. The LF covariance spectrum is almost entirely dominated by the {\tt CompTT} component, suggesting that although there is significant LF variability in the hard X-ray Comptonisation component, it is not significantly correlated with the LF variability in the soft X-ray Comptonisation component.

However, as suggested by the overall $\chi^2$ of 704.4 for 653 dof, it is still possible to improve the fits. Indeed, as shown in Figure~\ref{fig-spec-rmscov1} Panels-b and c, the QPO's variability spectra exhibit an extra hump at $\sim$ 1 keV, implying that the best-fit {\tt CompTT} may have a too-low electron temperature or too-small optical depth. This problem was also noted by Middleton, Uttley \& Done (2011), who analyzed the QPO covariance spectrum from Obs-2. These authors suggested that the shape of the soft excess in the QPO covariance spectrum could indicate the presence of an additional soft X-ray Comptonisation component, with a temperature higher than that requited in the time-averaged spectrum. 
Changes in the shape of the soft X-ray Comptonisation emission is also suggested by the LF rms spectrum, where the best-fit {\tt CompTT} over-predicts the emission around 1~keV. There are also some difficulties with the hard X-ray Comptonisation component in fitting these spectra. These issues lead us to further explore the effect of an additional soft X-ray Comptonisation component.

\subsubsection{Model-2: disc, soft X-ray \& hard X-ray Comptonisation, plus an extra intermediate Comptonisation}
\label{sec-model2}
In order to improve the fits of Model-1, we explore whether there is an intermediate Comptonisation region 
between the soft and hard X-ray Comptonisation regions. In this new model (hereafter: Model-2), there is still a {\tt diskbb} component, which provides the seed photons for the first Comptonisation component ({\tt CompTT-1}), but this now provides the seed photons for a second Comptonisation component ({\tt CompTT-2}), which in turn provides the seed photons for the hard X-ray Comptonisation component ({\tt nthCompt}). All of the previous model configurations and assumptions remain the same. Similarly, we fit all of the six types of spectra simultaneously, only allowing the normalizations of the spectral components to vary when fitting the variability spectra. The best-fit parameters are listed in Table~\ref{tab-varspecfit}. In comparison with Model-1, the overall $\chi^2$ of Model-2 decreases by 51.2 for 7 extra free parameters, indicating 5.8 $\sigma$ improvement of the fitting. Therefore, Model-2 provides a significant improvement to the fit, and more importantly it allows the fitting issues of Model-1 highlighted above to be addressed.

The Model-2 fitting of the time-averaged spectra is similarly good as Model-1, although the best-fit parameters are very different because of the inclusion of an additional component. In Model-2, the best-fit $N_{\rm H, host}$ decreases to $1.20^{+3.83}_{-1.20}\times10^{20}$ cm$^{-2}$. The temperature of {\tt diskbb} increases to 52.2$^{+7.9}_{-19.1}$ eV. {\tt CompTT-1} dominates the soft excess below 1 keV, with a lower electron temperature of 0.14$^{+0.01}_{-0.02}$, and a larger optical depth of 20.2$^{+4.0}_{-4.9}$. {\tt CompTT-2} dominates the flux within 1-2 keV, with a higher electron temperature of 0.33$^{+0.10}_{-0.11}$ keV and an optical depth of 12.5$^{+11.7}_{-2.5}$. {\tt nthCompt} still dominates the hard X-rays above 2 keV, with a slightly smaller photon index of 2.11 $\pm$ 0.05.

The main improvement of Model-2 lies in the fits to the variability spectra, as shown in Figure~\ref{fig-spec-rmscov2}. Especially, the QPO and stochastic LF variability split the spectral components in the soft excess. The soft X-ray Comptonisation in the QPO is consistent with only the hotter soft X-ray Comptonisation component ({\tt CompTT-2}), whereas that in the LF variability is consistent with only the 
cooler soft X-ray Comptonisation component ({\tt CompTT-1}).
Only the HF rms spectrum (and time-averaged spectra) require both {\tt CompTT-1} and {\tt CompTT-2}. The QPO spectrum also includes a contribution from the disc, which is not seen in either HF or LF stochastic variability spectra. All spectra also exhibit some contribution from {\tt nthCompt}, but this has a higher seed photon temperature than in Model-1, which is now set by the electron temperature of {\tt CompTT-2}.
Table~\ref{tab-varspecfit-ratio} lists the fractional normalizations of spectral components in every variability spectrum for the best-fit Model-2. 

We can allow a greater range of freedom for the seed photon temperature between the 
various components as it is possible that both {\tt diskbb} and {\tt CompTT-1} can 
provide seed photons for {\tt CompTT-2}, and that {\tt diskbb}, {\tt CompTT-1} and {\tt CompTT-2} can provide seed photons for {\tt nthComp}. Therefore, we set the seed photon temperature of {\tt CompTT-2} and {\tt nthComp} as free parameters (hereafter: Model-2b), and then check if the fitting can be improved. Table~\ref{tab-varspecfit} shows the new fitting results. In this case, the new $\chi^2$ decreases by 5.4 for 2 dof, equivalent to a 1.8 $\sigma$ significance. Thus the fitting is not improved significantly. The seed photon temperature of {\tt CompTT-2} is found to be 0.14$^{+0.03}_{-0.04}$ KeV, which is still consistent with the electron temperature of {\tt CompTT-1}. But now the electron temperature of {\tt CompTT-2} increases significantly to 1.98 keV (although poorly constrained), which is much larger than the best-fit seed photon temperature of {\tt nthComp}. So it is indeed possible that {\tt nthComp} can receive seed photons from each of the soft X-ray components. The temperature of {\tt diskbb} decreases to 30.5$^{+8.1}_{-3.9}$ eV, so its contribution in the soft excess is smaller. Despite these detailed differences in the best-fit values, the spectral decompositions of Model-2b are generally similar to those of Model-2, so we do not plot the fitting results of Model-2b separately.

\begin{table*}
\centering
\caption{The best-fit parameters of Model-1, Model-2 and Model-2b. $N_{\rm H, gal}$ and $N_{\rm H, ins}$ are the Galactic absorption column and intrinsic absorption column of \rej1034, separately. `f' indicates that this parameter is fixed. For linked parameters, we put the parameter names in the table instead of values. The error bars indicate 90 percent confidence limits.}
\begin{tabular}{lccccccc} 
\hline
Comp. & Par. & Model-1 & Model-2 & Model-2 & Model-2b & Model-2b  & Unit \\
\hline
 & & Obs-9 & Obs-9 & Obs-2 & Obs-9 & Obs-2  \\
{\tt TBabs} & $N_{\rm H, gal}$ & 1.36 (f) & 1.36 (f) & 1.36 (f) & 1.36 (f) & 1.36 (f) & $10^{20}$ cm$^{-2}$ \\
{\tt zTBabs} & $N_{\rm H, host}$ & 2.95 $^{+1.49}_{-0.97}$ & 1.20 $^{+3.83}_{-1.20}$ & 2.59 $^{+0.63}_{-0.65}$ & 5.34 $^{+2.27}_{-2.08}$ & 5.88 $^{+1.07}_{-1.08}$ & $10^{20}$ cm$^{-2}$ \\
{\tt diskbb} & $T_{\rm in}$ & 34.6 $^{+10.4}_{-4.9}$ & 52.2 $^{+7.9}_{-19.1}$ & 52.2 (f) & 30.5 $^{+8.1}_{-3.9}$ & 30.5 (f) & eV  \\
{\tt diskbb} & {\it norm} & 1.24 $^{+9.76}_{-1.19}\times10^{7}$ & 3.60 $^{+533.37}_{-2.57}\times10^{5}$ & 4.31 $^{+1.00}_{-0.94}\times10^{5}$ & 0.13 $^{+1.23}_{-0.12}\times10^{9}$ & 1.06 $^{+0.58}_{-0.46}\times10^{8}$ & \\
{\tt compTT-1} & $T_{\rm 0}$ & {\tt diskbb}-$T_{\rm in}$ & {\tt diskbb}-$T_{\rm in}$ & {\tt diskbb}-$T_{\rm in}$ & {\tt diskbb}-$T_{\rm in}$ & {\tt diskbb}-$T_{\rm in}$ & eV  \\
{\tt compTT-1} & {\it kT} & 0.20 $^{+0.02}_{-0.02}$ & 0.14 $^{+0.01}_{-0.02}$ & 0.14 (f) & 0.15 $^{+0.06}_{-0.01}$ & 0.15 (f) & keV  \\
{\tt compTT-1} & $\tau$ & 12.4 $^{+2.1}_{-1.4}$ & 20.2 $^{+4.0}_{-4.9}$ & 20.2 (f) & 13.8 $^{+3.4}_{-3.8}$ & 13.8 (f) &  \\
{\tt compTT-1} & {\it norm} & 1.67 $^{+1.22}_{-0.94}$ & 0.52 $^{+2.31}_{-0.22}$ & 0.56 $^{+0.02}_{-0.02}$ & 4.51 $^{+4.61}_{-2.40}$ & 4.69 $^{+0.34}_{-0.34}$ &  \\
{\tt compTT-2} & $T_{\rm 0}$ & -- & {\tt compTT-1}-{\it kT} & {\tt compTT-1}-{\it kT} & 0.14 $^{+0.03}_{-0.04}$ &  0.14 (f) & KeV  \\
{\tt compTT-2} & {\it kT} & -- & 0.33 $^{+0.10}_{-0.11}$ & 0.33 (f) & 1.98 $^{+\infty}_{-1.65}$ & 1.98 (f) & keV \\
{\tt compTT-2} & $\tau$ & -- & 12.5 $^{+11.7}_{-2.5}$ & 12.5 (f) & 2.71 $^{+9.69}_{-2.52}$ & 2.71 (f) &  \\
{\tt compTT-2} & {\it norm} & -- & 7.67 $^{+7.82}_{-2.88}\times10^{-3}$ & 8.42 $^{+0.41}_{-0.41}\times10^{-3}$ & 0.88 $^{+7.78}_{-0.78}\times10^{-3}$ & 1.01 $^{+0.06}_{-0.06}\times10^{-3}$ &  \\
{\tt nthComp} & $kT_{\rm e}$ & 200 (f) & 200 (f) & 200 (f) & 200 (f) & 200 (f) & keV \\
{\tt nthComp} & $kT_{\rm bb}$ & {\tt compTT-1}-{\it kT} & {\tt compTT-2}-{\it kT} & {\tt compTT-2}-{\it kT} & 0.12 $^{+0.24}_{-0.12}$ & 0.12 (f) & keV  \\
{\tt nthComp} & $\Gamma$ & 2.20 $^{+0.04}_{-0.04}$ & 2.11 $^{+0.05}_{-0.05}$ & 2.11 (f) & 2.05 $^{+0.08}_{-0.05}$ & 2.05 (f) &  \\
{\tt nthComp} & {\it norm} & 3.93 $^{+0.40}_{-0.39}\times10^{-4}$ & 2.11 $^{+1.34}_{-0.72}\times10^{-4}$ & 2.04 $^{+0.05}_{-0.05}\times10^{-4}$ & 3.45 $^{+0.68}_{-1.84}\times10^{-4}$ & 3.29 $^{+0.10}_{-0.10}\times10^{-4}$ & \\
\hline
$\chi^2/dof$ &  & 704.4/653 & 653.2/646 & 585.3/573 & 647.8/644 & 579.5/571 \\
\hline
\end{tabular}
\label{tab-varspecfit}
\end{table*}

\begin{table}
\centering
\caption{The contributions of various spectral components to the variability spectra, as indicated by the percentage of normalization relative to the components in the time-averaged spectra for the best-fit Model-2.}
\begin{tabular}{@{}lccccc@{}} 
\hline
 & QPO Rms & QPO Cov & HF Rms & LF Rms & LF Cov \\
 & (\%) & (\%) & (\%) & (\%) & (\%) \\
\hline
\multicolumn{6}{@{}l}{\it Figure~\ref{fig-spec-rmscov2}: Obs-9 Spectra} \\
{\tt diskbb} & 9.1 $^{+2.4}_{-2.3}$ & 8.4 $^{+1.7}_{-1.7}$ & 0.0 $^{+4.9}_{-0.0}$ & 0.0 $^{+2.4}_{-0.0}$ & 0.0 $^{+2.5}_{-0.0}$ \\
{\tt compTT-1} & 0.0 $^{+1.2}_{-0.0}$ & 0.0 $^{+1.1}_{-0.0}$ & 3.7 $^{+0.8}_{-0.8}$ & 10.2 $^{+0.4}_{-0.4}$ & 9.3 $^{+0.4}_{-0.5}$ \\
{\tt compTT-2} & 18.8 $^{+3.6}_{-3.6}$ & 17.3 $^{+1.9}_{-1.9}$ & 7.6 $^{+2.9}_{-2.9}$ & 0.0 $^{+3.4}_{-0.0}$ & 0.0 $^{+2.8}_{-0.0}$ \\
{\tt nthComp} & 11.6 $^{+2.6}_{-2.6}$ & 10.7 $^{+1.7}_{-1.6}$ & 19.6 $^{+2.0}_{-2.0}$ & 15.0 $^{+1.5}_{-1.5}$ & 3.9 $^{+1.8}_{-1.8}$ \\
\hline
\multicolumn{6}{@{}l}{\it Figure~\ref{fig-spec-rmscov2-obs2}: Obs-2 Spectra} \\
{\tt diskbb} & 0.0 $^{+0.8}_{-0.0}$ & 0.0 $^{+0.8}_{-0.0}$ & 0.0 $^{+3.6}_{-0.0}$ & 0.0 $^{+3.7}_{-0.0}$ & 0.0 $^{+3.2}_{-0.0}$ \\
{\tt compTT-1} & 0.0 $^{+0.4}_{-0.0}$ & 0.0 $^{+0.4}_{-0.0}$ & 2.9 $^{+0.7}_{-0.7}$ & 8.5 $^{+0.5}_{-0.5}$ & 7.4 $^{+0.6}_{-0.6}$ \\
{\tt compTT-2} & 17.0 $^{+3.3}_{-3.3}$ & 15.9 $^{+1.4}_{-1.4}$ & 4.2 $^{+2.4}_{-2.4}$ & 3.9 $^{+1.9}_{-1.9}$ & 5.3 $^{+2.2}_{-2.2}$ \\
{\tt nthComp} & 10.4 $^{+2.3}_{-2.3}$ & 9.8 $^{+1.4}_{-1.4}$ & 10.7 $^{+1.8}_{-1.8}$ & 13.0 $^{+1.5}_{-1.5}$ & 5.8 $^{+1.8}_{-1.8}$ \\
\hline
\end{tabular}
\label{tab-varspecfit-ratio}
\end{table}

\subsection{Modelling the QPO's Lag Spectrum}
\label{sec-lagspec-model}
We can explore further the appropriateness of our best-fit spectral decompositions by fitting them to the QPO lag spectrum (Figure~\ref{fig-qpopar-ene}c). We assume that the shape of the 
lag spectrum is produced by the relative contributions of the various spectral components in different energy bins, and fit for the lags between these various components in order to fit the normalization and shape of the lag spectrum.

First we test the spectral decomposition of the best-fit Model-1. As shown in Figure~\ref{fig-spec-rmscov1}a, the {\tt diskbb} component has a very small flux contribution in the time-averaged spectrum, and so its effect on the lag spectrum can be ignored. We only consider the lag between {\tt compTT-1} and {\tt nthComp}, which is signified by $t_{\rm lag,1}$. The intrinsic light curve of {\tt compTT} is assumed to be the light curve in 0.4-0.6 keV, because this band is dominated by {\tt compTT} in Model-1. Likewise, the 2-10 keV light curve is used to represent {\tt nthComp}. The adoption of these observed light curves avoids potential problems from simulating light curves. Then it is necessary to force the two light curves to have the required lag of $t_{\rm lag,1}$ in the QPO frequency bin of $(2.5-3.5)\times10^{-4}$ Hz. To achieve this we perform a Fourier transform on the light curve of {\tt compTT}, and then modify the phase in the QPO frequency bin to introduce the required lag of $t_{\rm lag,1}$ relative to {\tt nthComp}. Then we perform an inverse Fourier transform to derive a new light curve for {\tt compTT}, whose QPO phase leads {\tt nthComp} by $t_{\rm lag,1}$ exactly.

\begin{figure}
\centering
\includegraphics[trim=0.0in 0.0in 0.0in 0.0in, clip=1, scale=0.48]{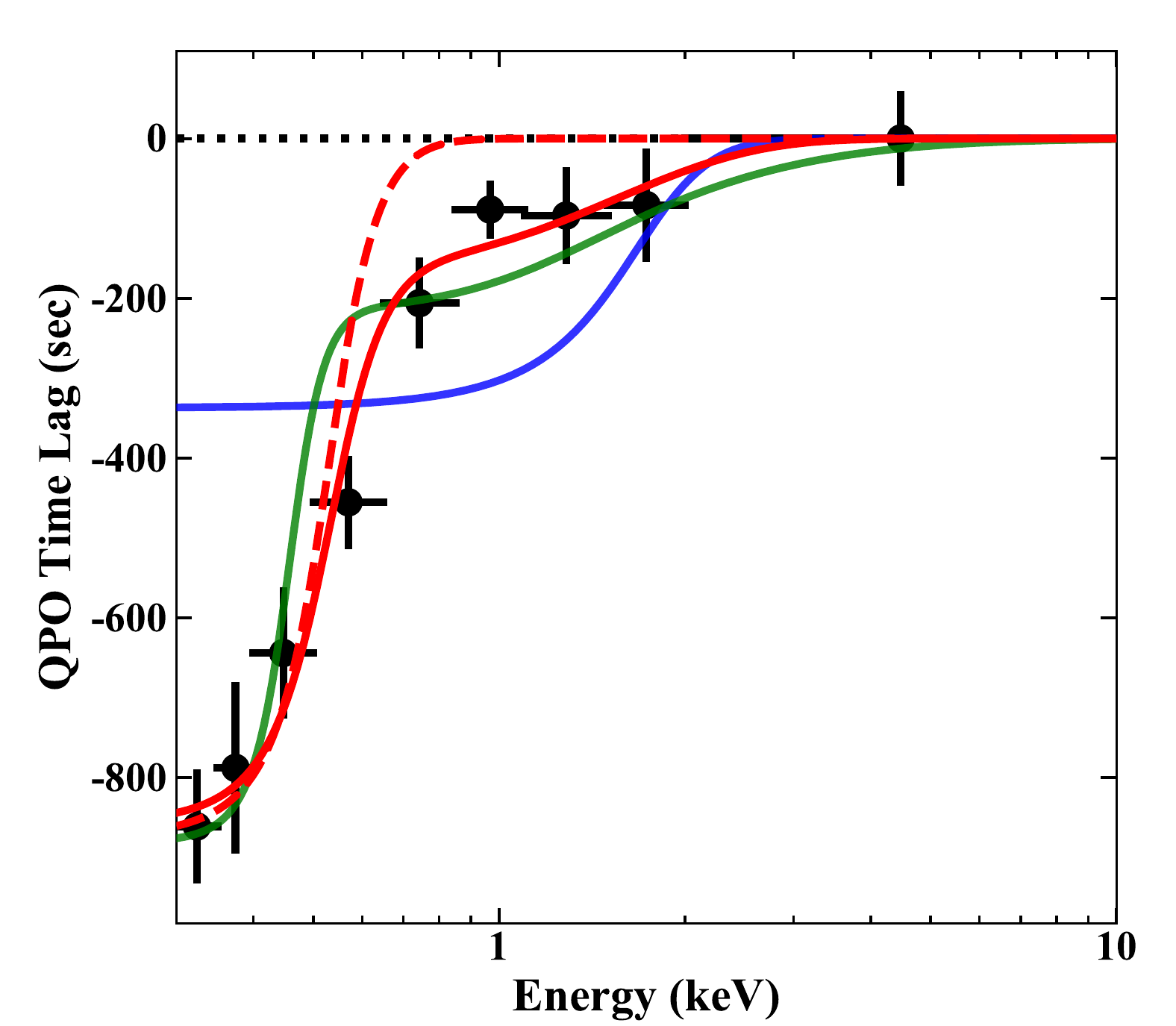}
\caption{Fitting the QPO's lag spectrum. The data are shifted down by 421 s to allow the data point in the 2-10 keV band to be at zero. The blue solid, red dash, red solid and green solid lines represent the best-fit results for the Model-1-T2, Model-2-T1, Model-2-T2, Model-2b-T2 as listed in Table~\ref{tab-lagspec-model}, respectively.}
\label{fig-lagspec-model}
\end{figure}

As the first attempt (Model-1-T1), we fix $t_{\rm lag,1}$ at -861 s, which is the observed lag between 0.3-0.35 keV and 2-10 keV, as shown in Figure~\ref{fig-qpopar-ene}c. The minus sign indicates the soft X-ray lead. Then the lag is calculated for different energy bins between 0.3 and 10 keV.
However, this gives an extremely poor fit to the 
observed lag spectrum, with $\chi^2$ is 666.4 for 9 dof.
Then we set $t_{\rm lag,1}$ as a free parameter (Model-1-T2), and fit the lag spectrum again. Now the best-fit $t_{\rm lag,1}$ is -337 $\pm$ 23 s, with the $\chi^2$ being 137.0 for 8 dof, so it remains as a poor fit, as shown by the blue solid line in Figure~\ref{fig-lagspec-model}. Therefore, we conclude that the best-fit Model-1 cannot reproduce the observed lag spectrum.

We now examine the spectral decomposition of Model-2.
The QPO's rms and covariance spectra can be fit by using {\tt diskbb}, {\tt compTT-2} and {\tt nthComp}.
Thus we introduce two lag variables. $t_{\rm lag,1}$ signifies the lag between {\tt diskbb} and {\tt nthComp}, and $t_{\rm lag,2}$ signifies the lag between {\tt compTT-2} and {\tt nthComp}. Following the spectral decomposition of Model-2, we adopt the light curve in the 0.3-0.35 keV band for {\tt diskbb}, 0.85-1 keV for {\tt compTT-2}, and 2-10 keV for {\tt nthComp}. These light curves are manipulated in the same way so as to provide the required lags of $t_{\rm lag,1}$ and $t_{\rm lag,2}$.

We first attempt to fit a single lag (Model-2-T1), with $t_{\rm lag,2}$ being fixed at 0 s, i.e. assuming that there is no lag time between {\tt compTT-2} and {\tt nthComp}. The resultant best-fit $t_{\rm lag,1}$ is found to be -877 $\pm$ 47 s, and the $\chi^2$ is 33.8 for 8 dof, indicating a much better fit than Model-1 (red dash line in Figure~\ref{fig-lagspec-model}). 
However, this lag model does not reproduce the lag spectrum in 0.6-2 keV. 

Then we allow both $t_{\rm lag,1}$ and $t_{\rm lag,2}$ to be free parameters (Model-2-T2). This produces a very good fit to the lag spectrum, with the $\chi^2$ being 4.9 for 7 dof, as shown by the red solid line in Figure~\ref{fig-lagspec-model}. The best-fit $t_{\rm lag,1}$ is -859 $\pm$ 50 s, and $t_{\rm lag,2}$ is -180 $\pm$ 34 s. In summary, the QPO's lag spectrum requires {\tt diskbb} to lead {\tt compTT-2} by 679 s, and requires {\tt compTT-2} to lead {\tt nthComp} by 180 s. Based on the simplest constraint of causality in time, we can speculate that the QPO is created in {\tt diskbb}. It then propagates to {\tt compTT-2} and {\tt nthComp}, and it is also amplified during this process.

Finally, we check if the spectral decomposition of Model-2b (where the seed photon assumptions were changed) can also fit the QPO's lag spectrum. In this spectral decomposition, if we fix $t_{\rm lag,2}$ at 0 s (Model-2b-T1), the best-fit $t_{\rm lag,1}$ is found to be 905 $\pm$ 54 s, but this is clearly a bad fit, with $\chi^2$ = 82.1 for 8 dof (see Table~\ref{tab-lagspec-model}). If we allow both time lags to be free parameters (Model-2b-T2), then $t_{\rm lag,1}$ and $t_{\rm lag,2}$ are found to be -882 $\pm$ 56 s and -307 $\pm$ 40 s. The resultant $\chi^2$ is 22.9 for 7 dof, which is worse than Model-2-T2. As shown by the green solid line in Figure~\ref{fig-lagspec-model}, the main problem with Model-2b-T2 is that its {\tt diskbb} has a too low temperature.

Consequently, we conclude that the QPO's lag spectrum indicates that Model-2 provides a better spectral decomposition than Model-2b and Model-1. The best-fit spectral components derived from the combined spectral-timing analysis (Model-2) can also fit the independent data from the QPO's lag spectrum. This demonstrates that this model can capture all the spectral variability information, and that the complexity of an additional soft Compton component ({\tt CompTT-2}) is required by the data.

\begin{table}
\centering
\caption{Results of fitting the QPO's lag spectrum. Model-1-T1 and Model-1-T2 are based on the best-fit Model-1, where $t_{\rm lag,1}$ is the time lag between {\tt compTT} and {\tt nthComp}. Model-2-T1 and Model-2-T2 are based on the best-fit Model-2, where $t_{\rm lag,1}$ is the time lag between {\tt diskbb} and {\tt nthComp}, and $t_{\rm lag,2}$ is the time lag between {\tt compTT-2} and {\tt nthComp}. Model-2b-T1 and Model-2b-T2 are based on the best-fit Model-2b. `f' indicates that this parameter is fixed. All the error bars indicate 1 $\sigma$ uncertainties.}
\begin{tabular}{lccr}
\hline
Model & $t_{\rm lag,1}$ & $t_{\rm lag,2}$ & $\chi^{2}/dof$ \\
& (sec) & (sec) & \\
\hline
Model-1-T1 & -861 (f) & -- & 666.4/9 \\
Model-1-T2 & -337 $\pm$ 23 & -- & 137.0/8 \\ 
Model-2-T1 & -877 $\pm$ 47 & -0 (f) & 33.8/8 \\
Model-2-T2 & -859 $\pm$ 50 & -180 $\pm$ 34 & 4.9/7 \\
Model-2b-T1 & -905 $\pm$ 54 & -0 (f) & 82.1/8 \\
Model-2b-T2 & -882 $\pm$ 56 & -307 $\pm$ 40 & 22.9/7 \\
\hline
\end{tabular}
\label{tab-lagspec-model}
\end{table}

\begin{figure*}
\centering
\begin{tabular}{ccc}
\includegraphics[trim=0.2in 0.2in 0.0in 0.1in, clip=1, scale=0.41]{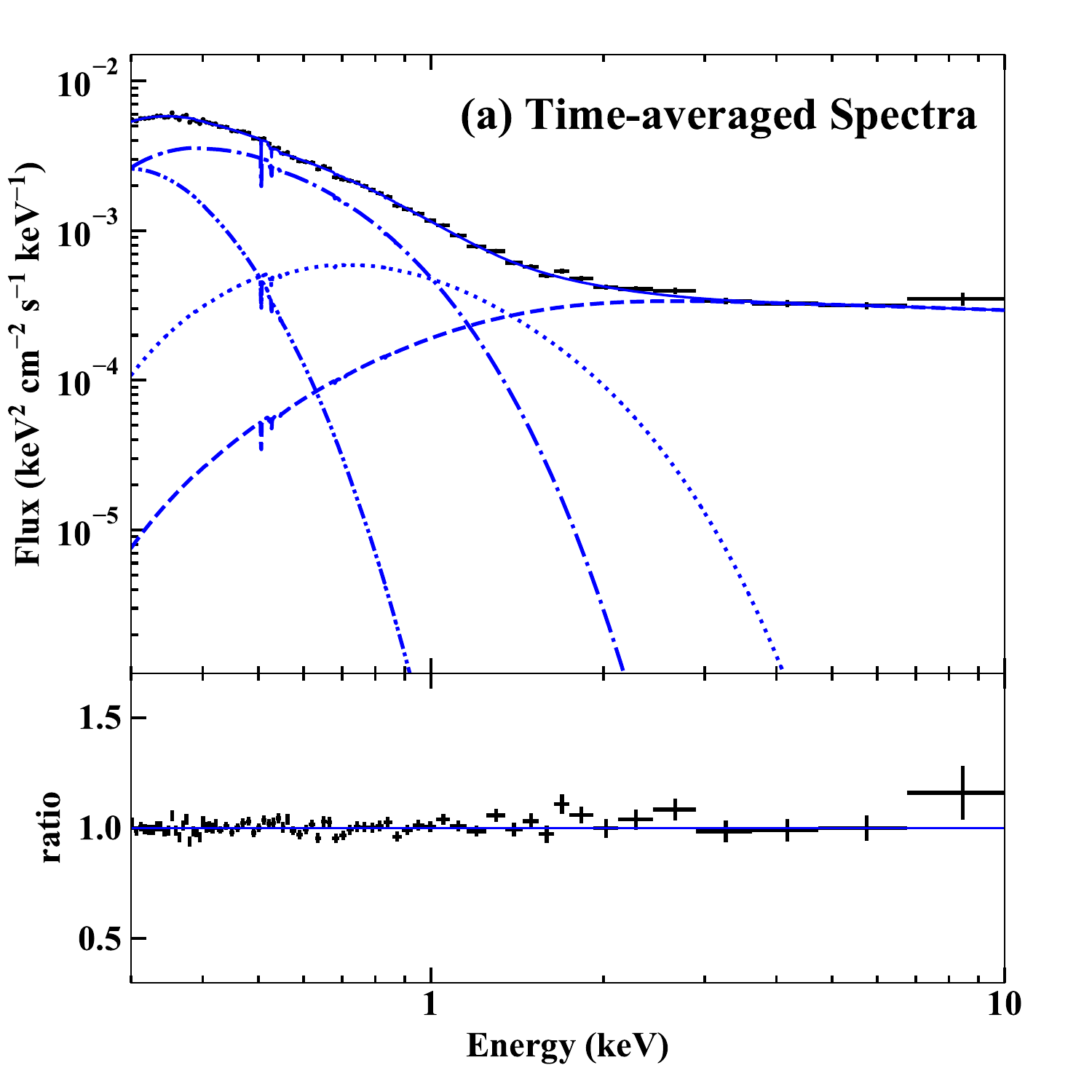} &
\includegraphics[trim=0.7in 0.2in 0.0in 0.1in, clip=1, scale=0.41]{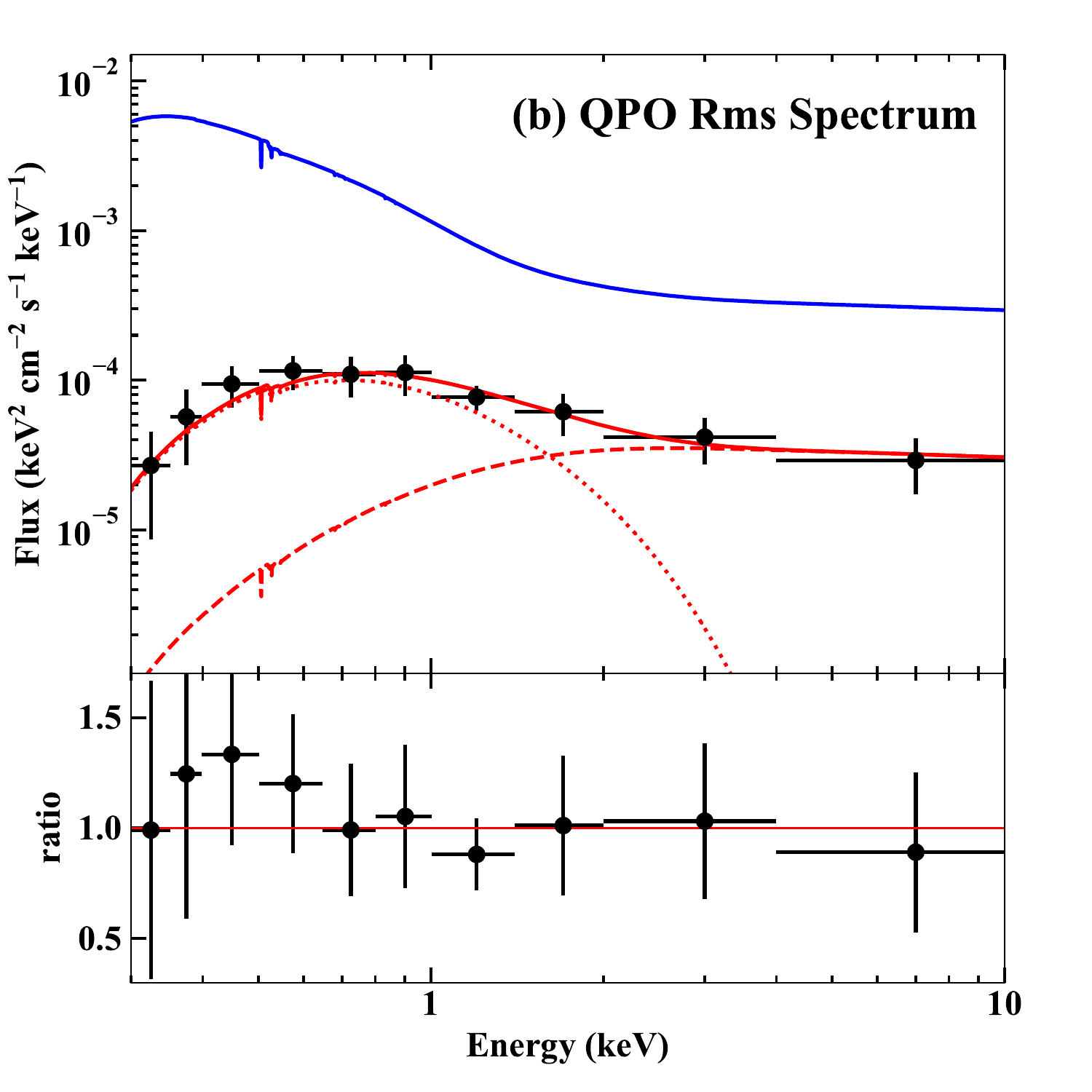} &
\includegraphics[trim=0.75in 0.2in 0.0in 0.1in, clip=1, scale=0.41]{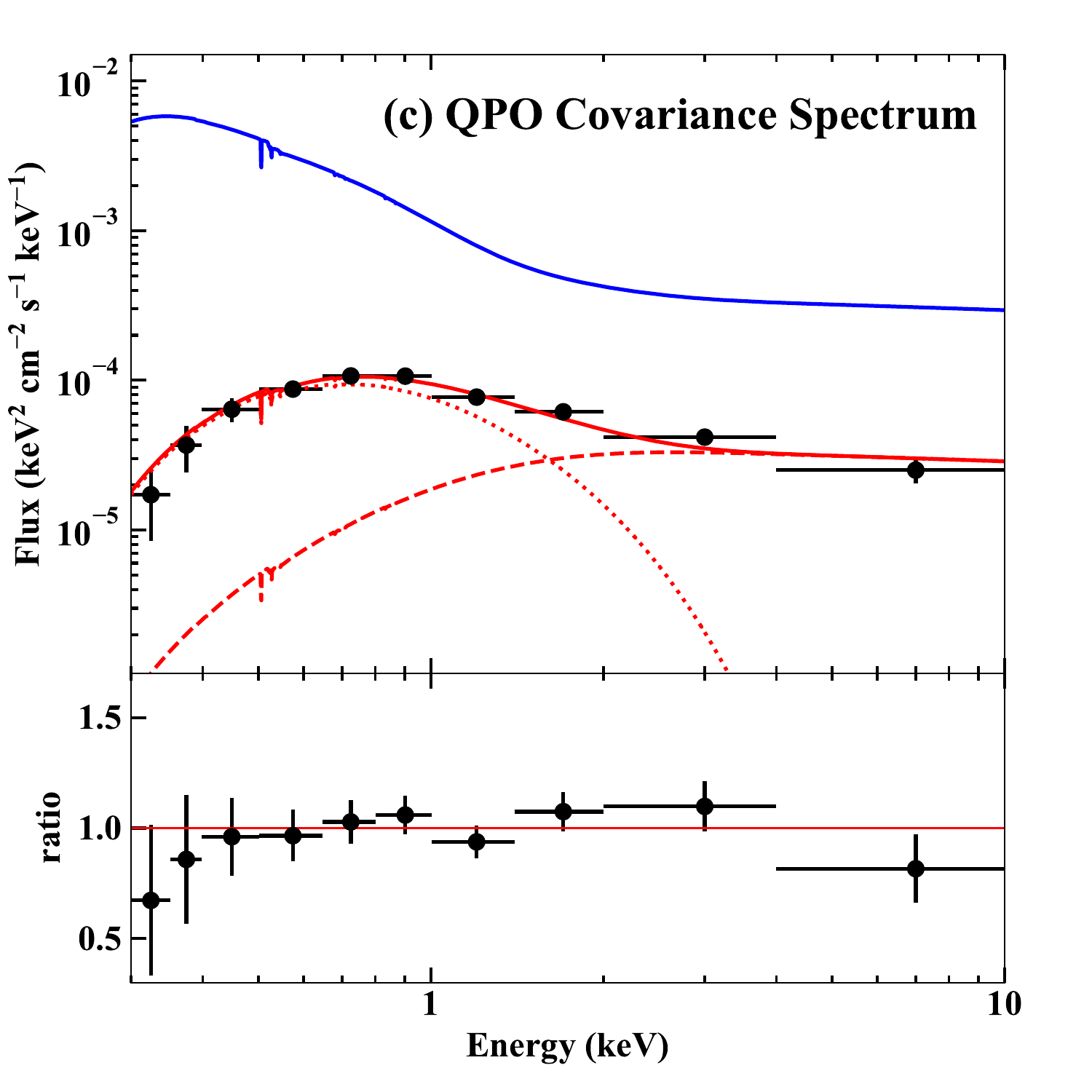} \\
\includegraphics[trim=0.2in 0.2in 0.0in 0.1in, clip=1, scale=0.41]{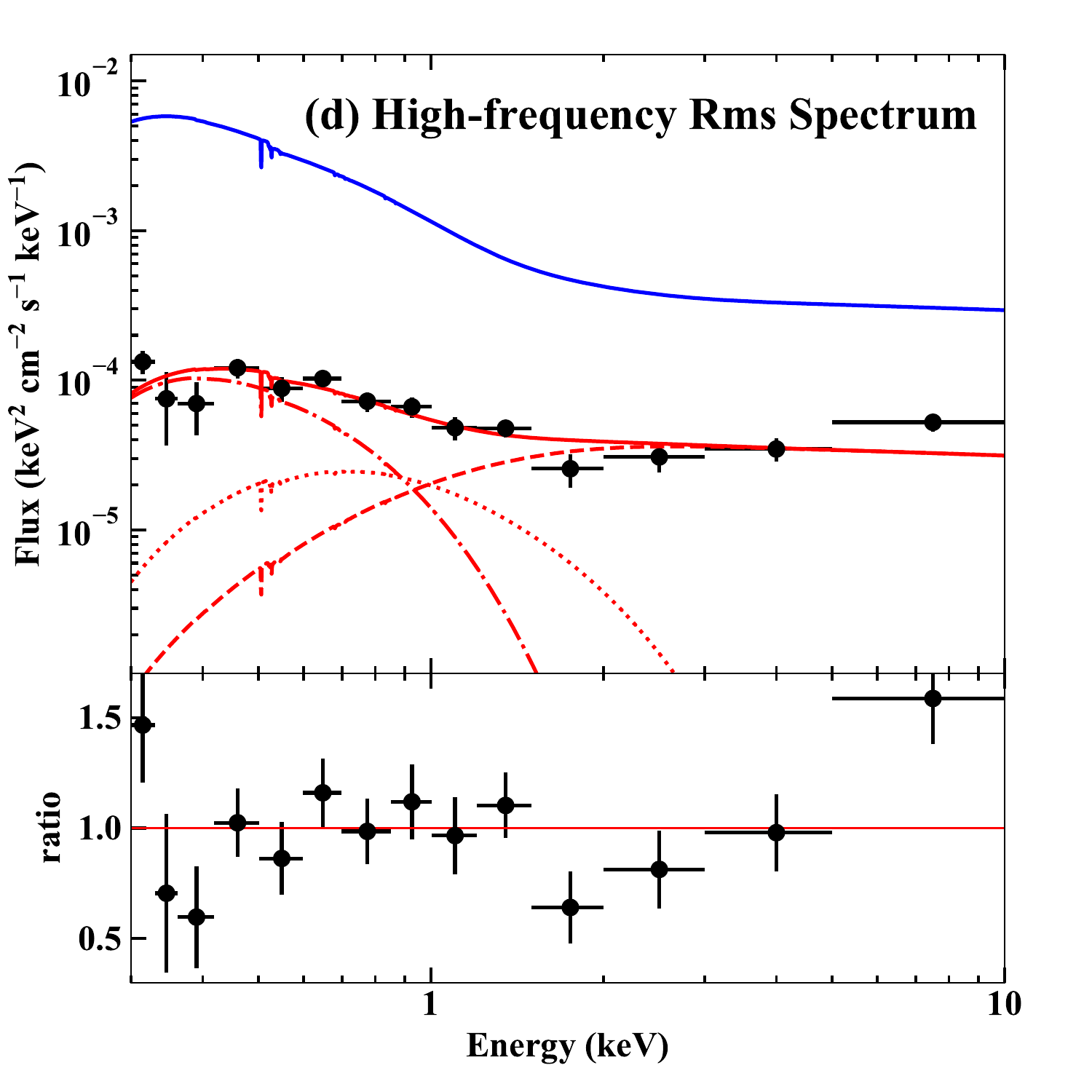} &
\includegraphics[trim=0.7in 0.2in 0.0in 0.1in, clip=1, scale=0.41]{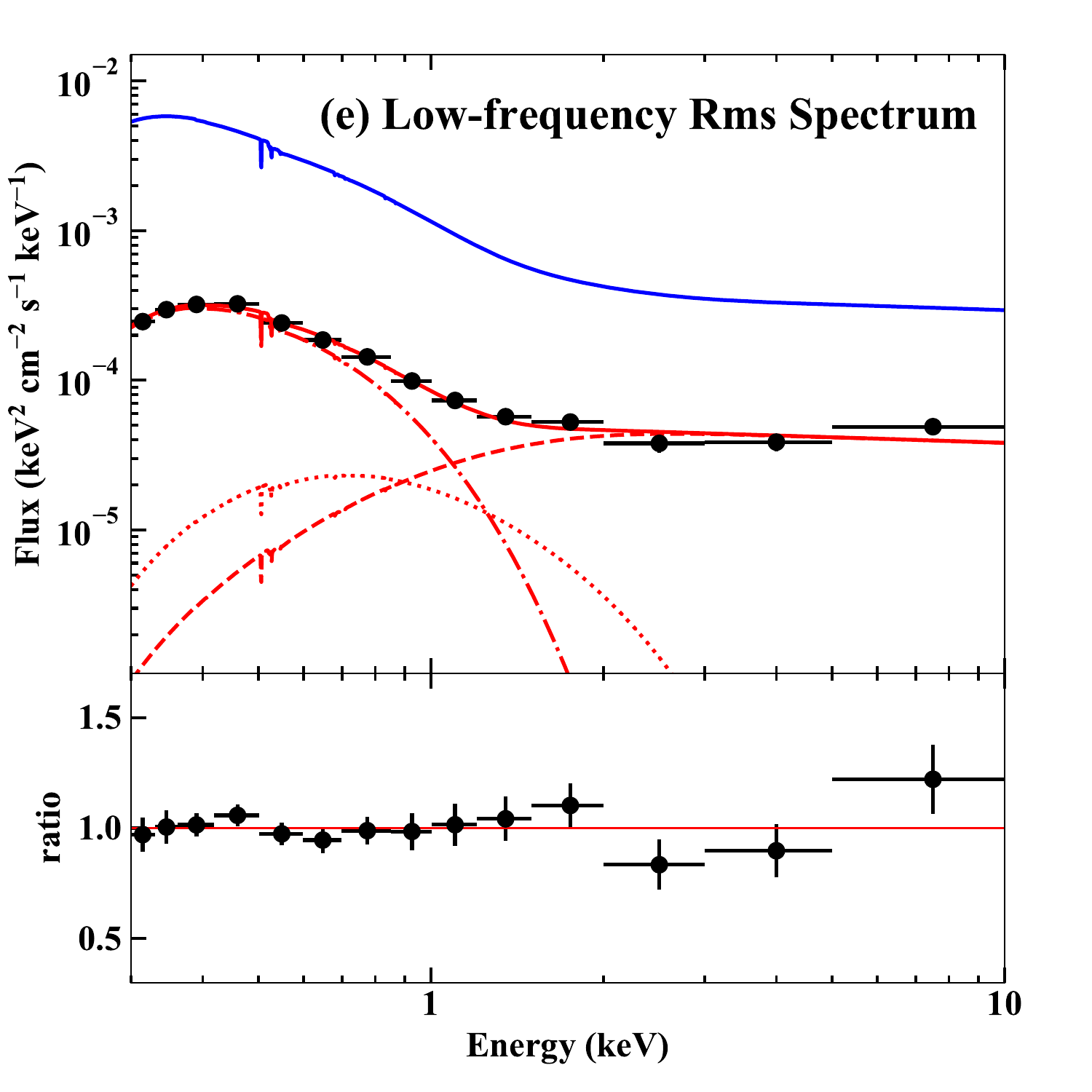} &
\includegraphics[trim=0.75in 0.2in 0.0in 0.1in, clip=1, scale=0.41]{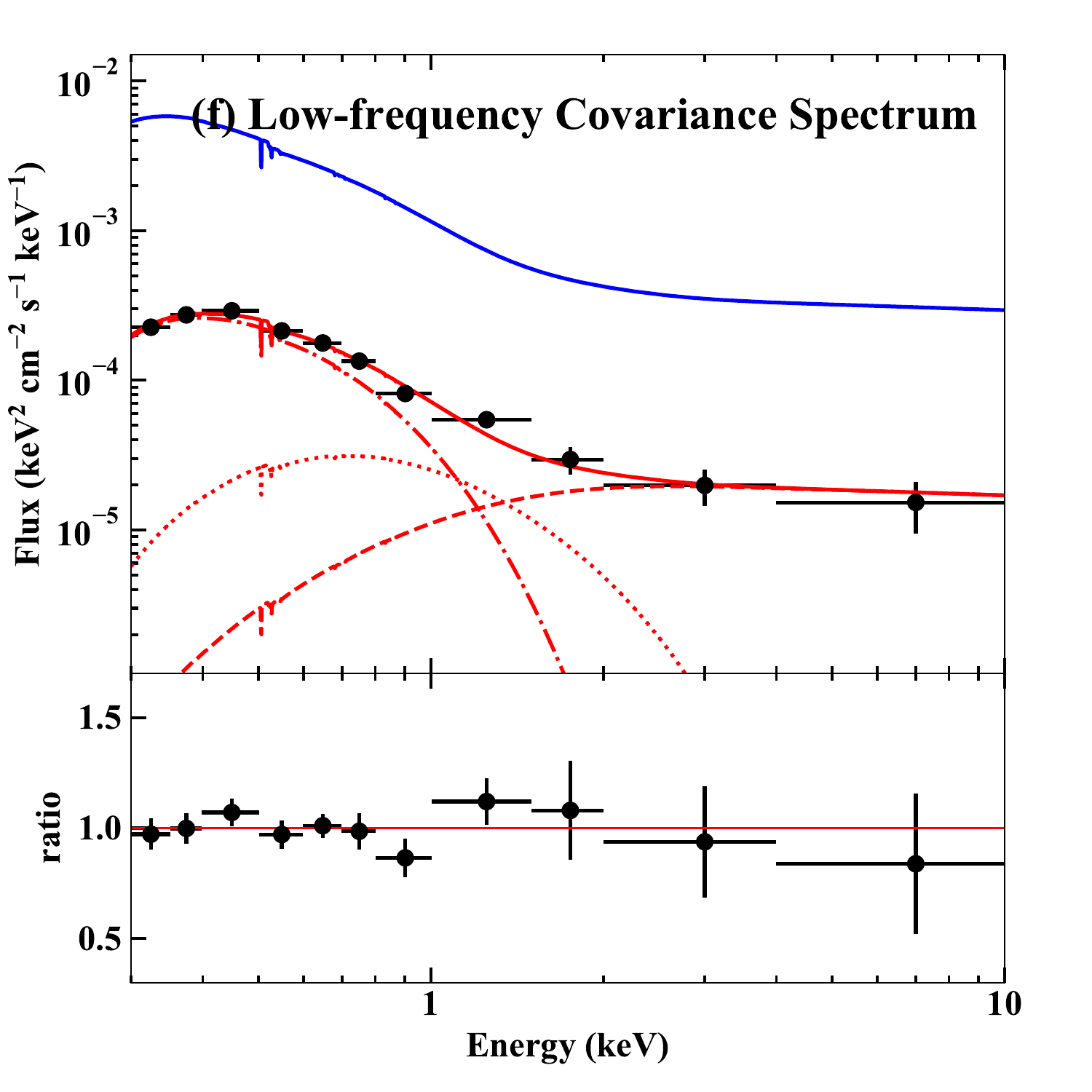} \\
\end{tabular}
\caption{Similar to Figure~\ref{fig-spec-rmscov2}, but now applying the best-fit Model-2 of Obs-9 to the same set of spectra from Obs-2. Only the intrinsic absorption and the normalizations of spectral components are allowed to vary, while the shape of every component remains the same. This model also produces good fits to Obs-2.}
\label{fig-spec-rmscov2-obs2}
\end{figure*}

\subsection{Constraining the Potential Presence of a Reflection Component}
\label{sec-reflect}
The iron K$\alpha$ line is a typical feature present in an X-ray reflection spectrum. This line is generally weak or undetected in X-ray `simple' super-Eddington NLS1s (Gallo 2006), while the X-ray `complex' NLS1s have much
stronger features around Fe K$\alpha$ 
(Gallo 2006, Done \& Jin 2016). There has been a long and continuing debate about the cause of the Fe K$\alpha$ weakness in X-ray `simple' NLS1s, with suggestions including  
an intrinsically weak reflection component, or reflection which is so highly smeared that the line is lost into the continuum. However, recent studies are pointing to the former (a weak reflection) possibility (e.g. Jin, Done \& Ward 2016; Kara et al. 2017; Parker, Miller \& Fabian 2018). 
An Iron K$\alpha$ line has never been detected in \rej1034, but the limits were not very constraining due to the low S/N above 4 keV. Our 72 ks \xmm\ observation in Obs-9 provides a similar spectral quality as previous observations, but for the first time the simultaneous 100 ks \nustar\ exposure provides much better spectral quality above 3 keV. Thus now we can combine the \xmm\ and \nustar\ data to constrain a potential underlying reflection component. In order to maximize the spectral constraints, we also add the two MOS spectra, as well as the entire set of variability spectra. 

We now consider Model-2 as the continuum model, and add a reflection component to it by convolving {\tt nthComp} with the reflection model {\tt rfxconv} (Done \& Gierli\'{n}ski 2006; Kolehmainen, Done \& D{\'\i}az Trigo 2011). The key parameters of {\tt rfxconv} include the relative reflection normalization\footnote{In the {\tt rfxconv} model, $R_{\rm refl}$ is actually a negative value in order to show the reflection component only.} ($R_{\rm refl}$).
The other free parameter is the logarithm of the ionization parameter (log $\xi$). The iron abundance ($A_{\rm iron}$)
is fixed at the Solar value, and the inclination angle is fixed at 30$\degr$.
This reflection spectrum is convolved with the {\tt kdblur} model (Laor 1991) to account for the relativistic smearing. The emissivity index of {\tt kdblur} is fixed at its default value of 3. The outer radius fixed at 100 $R_{\rm g}$, and the inclination angle is fixed at 30$\degr$. The inner radius ($R_{\rm in}$) remains as a free parameter. 

\begin{figure}
\centering
\includegraphics[trim=0.0in 0.3in 0.0in 0.0in, clip=1, scale=0.48]{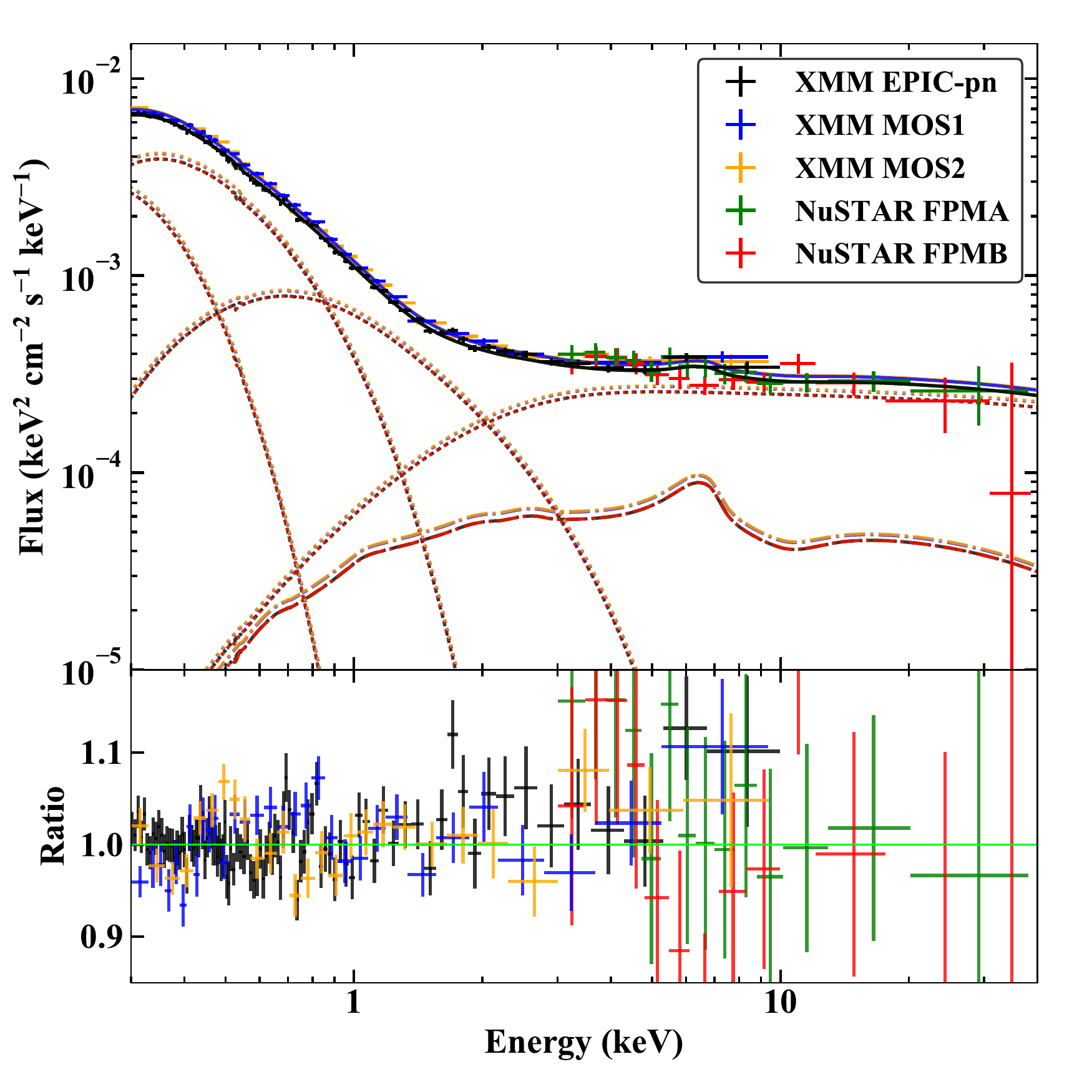}
\caption{Constraining the potential reflection component using all the spectra available in Obs-9. The dotted lines correspond to the four continual components in Model-2 as shown in Figure~\ref{fig-spec-rmscov2}a. Dash-dotted lines indicate the best-fit reflection component in every spectrum.}
\label{fig-reflect}
\end{figure}

Figure~\ref{fig-reflect} shows the best-fit result, with $\chi^2$ = 1074.5 for 1032 dof. The reflection component is found to be very weak. Comparing this to the best-fit Model-2 without reflection, the improvement of $\chi^2$ is only 8.6 for 7 additional free parameters, and so there is no statistically significant requirement for this component to be present. We note that the main contributor to the $\chi^2$ improvement is the high S/N time-averaged spectrum from the EPIC-pn alone, which is 8.5 for 3 dof, corresponding to a slightly higher significance of 2.1$\sigma$.
The best-fit $R_{\rm refl}$ is 0.20$^{+0.16}_{-0.11}$, indicating that 
the reflecting material occupies about 20\% of the sky as seen from the X-ray source. 
The constraint placed on $R_{\rm refl}$ comes mainly from the spectra above 5 keV, driven both by the 
lack of the K$\alpha$ emission line and the Compton hump. The best-fit log$\xi$ is 3.06$^{+0.21}_{-0.21}$,
so the reflecting gas is very highly ionized, and the line is intrinsically broadened by the gas temperature. The best-fit $R_{\rm in}$ is 4.18$^{+8.42}_{-2.57}$. Thus although the spin cannot be directly constrained, nevertheless the result is consistent with a low-spin black hole. 

\begin{figure}
\centering
\includegraphics[trim=0.0in 0.5in 0.0in 0.0in, clip=1, scale=0.45]{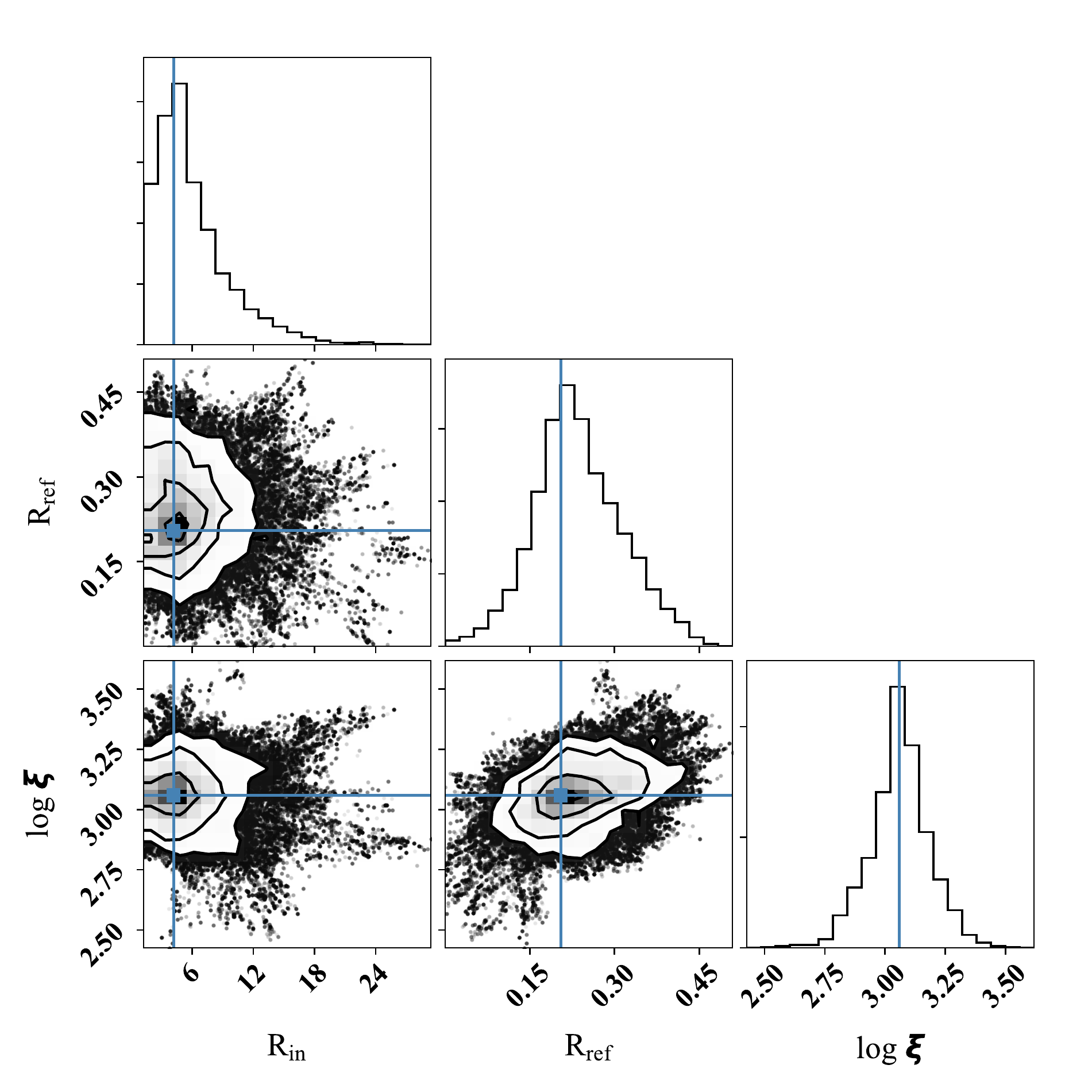}
\caption{Posterior probability distributions of the inner radius $R_{\rm in}$ of the relativistic blurring ({\tt kdblur}), the relative reflection normalization $R_{\rm refl}$ and the ionization parameter log$\xi$ of the reflection component, and their covariances derived from the MCMC sampling. Their best-fit values are indicated by the blue solid lines.}
\label{fig-reflect-mcmc}
\end{figure}

In order to examine the degeneracy of model parameters, we perform Markov Chain Monte Carlo (MCMC) sampling of the parameter space using the {\sc xspec-emcee} program\footnote{The {\sc xspec-emcee} program is developed by Jeremy Sanders, which makes use of the {\sc emcee} package (Foreman-Mackey et al. 2013).}. Figure~\ref{fig-reflect-mcmc} shows the level of degeneracy among $R_{\rm in}$, $R_{\rm refl}$ and log $\xi$. It is clear that $R_{\rm refl}$ and log $\xi$ are somewhat correlated, with a smaller $R_{\rm refl}$ allowed for smaller log $\xi$. This is understandable because if the material is less ionized, the emission lines in the reflection component will appear stronger and narrower, so the reflection fraction $R_{\rm refl}$ must decrease in order to maintain a good fit to the smooth spectra. Actually, if we examine Figure~\ref{fig-reflect} closely, it can be noticed that the \nustar/FPMB spectrum does not show any line feature around Iron K$\alpha$, although the spectral resolution and S/N are not sufficient to provide a statistically significant constraint.

The iron abundance ($A_{\rm iron}$) is also a sensitive parameter because it strongly affects the intensity of the Iron K$\alpha$ feature in the model. We also try to set $A_{\rm iron}$ as a free parameter, and find that the $\chi^2$ only decreases by 1.2 for 1 more free parameter, and $A_{\rm iron}$ is found to be 2.84$^{+0.16}_{-0.44}$, indicating a super-Solar iron abundance. In addition, $R_{\rm refl}$ decreases to 0.15$^{+0.08}_{-0.08}$, and log $\xi$ remains at a large value of 3.30$^{+0.39}_{-0.16}$. Since relaxing $A_{\rm iron}$ does not result in a significant improvement to the fit, but does increase the parameter degeneracy, we no longer consider it in this work. 

We note that we are not able to explore the sort of reflection models
derived for the most extreme NLS1 spectra, where the emissivity is highly centrally peaked, with strong iron overabundance as the upper limit on iron in {\tt rfxconv} is only 3. However, our spectra do not give any indication of a Compton hump, so the limits on the reflected continuum as well as the line seem robust. 
Future observations with deeper exposures are required to place stronger constraints on the underlying reflection and K$\alpha$ emission feature in  \rej1034.

\section{Comparisons with Previous Observations}
\subsection{Fitting the Spectra from Obs-2}
\label{sec-lagspec-obs2}
We now return to the previous data and see whether the new spectral decomposition of Model-2
can provide us with any additional insights into the longer-term behaviour of the QPO in this source.
Obs-2 is the first \xmm\ observation of \rej1034\ showing a significant QPO (Gierli\'{n}ski et al. 2008). It is also the only data set whose quality allows us to extract a similar set of variability spectra. We perform careful analysis of the pile-up effect in Obs-2, and ensure that it does not affect the main results of this paper\footnote{see Appendix~\ref{sec-pileup}.}. Then we use the spectra from Obs-2 to verify the best-fit Model-2 and Model-2b from Obs-9. Since there is no simultaneous \nustar\ observation for Obs-2, the time-averaged spectrum is only available within 0.3-10 keV from \xmm. We assume that the spectra from Obs-2 have the same components as in Obs-9, and the shape of each component is kept the same, only their normalizations are allowed to vary. The intrinsic absorption ($N_{\rm H, host}$) is also treated as a free parameter. Then we fit all the time-averaged and variability spectra of Obs-2, simultaneously.

For Model-2 the overall minimal $\chi^2$ is found to be 585.3 for 573 dof, indicating reasonably good fits to all the spectra. The best-fit $N_{\rm H, host}$ is found to be 2.59$^{+0.63}_{-0.65}\times10^{20}$ cm$^{-2}$, which is slightly higher than that in Obs-9. The spectral decomposition of the time-averaged spectrum is shown in Figure~\ref{fig-spec-rmscov2-obs2}a, which is very similar to Obs-9. The QPO's rms and covariance spectra are well fitted by {\tt compTT-2} and {\tt nthComp}, as shown in Panels-b and c. The HF rms, LF rms and covariance spectra all require contributions from {\tt compTT-1}, {\tt compTT-2} and {\tt nthComp}. We also apply the best-fit Model-2b to the spectra from Obs-2, and the resultant $\chi^2$ is 579.5 for 571 dof, which represents an improvement of 1.9 $\sigma$ significance comparing to Model-2. Hence the results are very similar between Model-2b and Model-2.

Comparing the fitting results between Obs-2 and 9, we find that there
are subtle differences in the rms and covariance spectra of the QPO frequency and the LF band.
In terms of the spectral components, the
main difference in the QPO is that the {\tt diskbb} component is present in Obs-9 but not in Obs-2.
Otherwise the rest of the QPO is similarly well fitted by the combination of 
{\tt compTT-2} and {\tt nthComp}. Besides, {\tt comptt-2}
is present in the LF rms and covariance spectra in Obs-2 but not in Obs-9.
This means that {\tt comptt-2} contained more stochastic variability in Obs-2, which might have affected its intrinsic QPO more severely, and so the QPO is less coherent in Obs-2 than in Obs-9.

\subsection{Fitting the Spectra from the non-QPO Obs-3 and Obs-6}
\label{sec-lagspec-obs36}
Amongst all of the 9 \xmm\ observations of \rej1034, Obs-3 and 6 are the only two observations in which the QPO signal was not detected even though the data quality is sufficiently good (Alston et al. 2014). These two observations were carried out by \xmm\ on 2009-05-31 and 2011-05-07 separately. Interestingly, \rej1034\ showed a much stronger soft excess during these two observations than during the other observations, and so it appears that there is an anti-correlation between the intensity of the soft excess and the QPO's detectability (Paper-I). We test if this different spectral state in Obs-3 and 6 can also be fitted by the same spectral components in the best-fit Model-2. Similarly, we only allow $N_{\rm H, ins}$ and the normalization of every spectral component to be free parameters.

\begin{figure}
\centering
\includegraphics[trim=0.1in 0.4in 0.0in 0.0in, clip=1, scale=0.48]{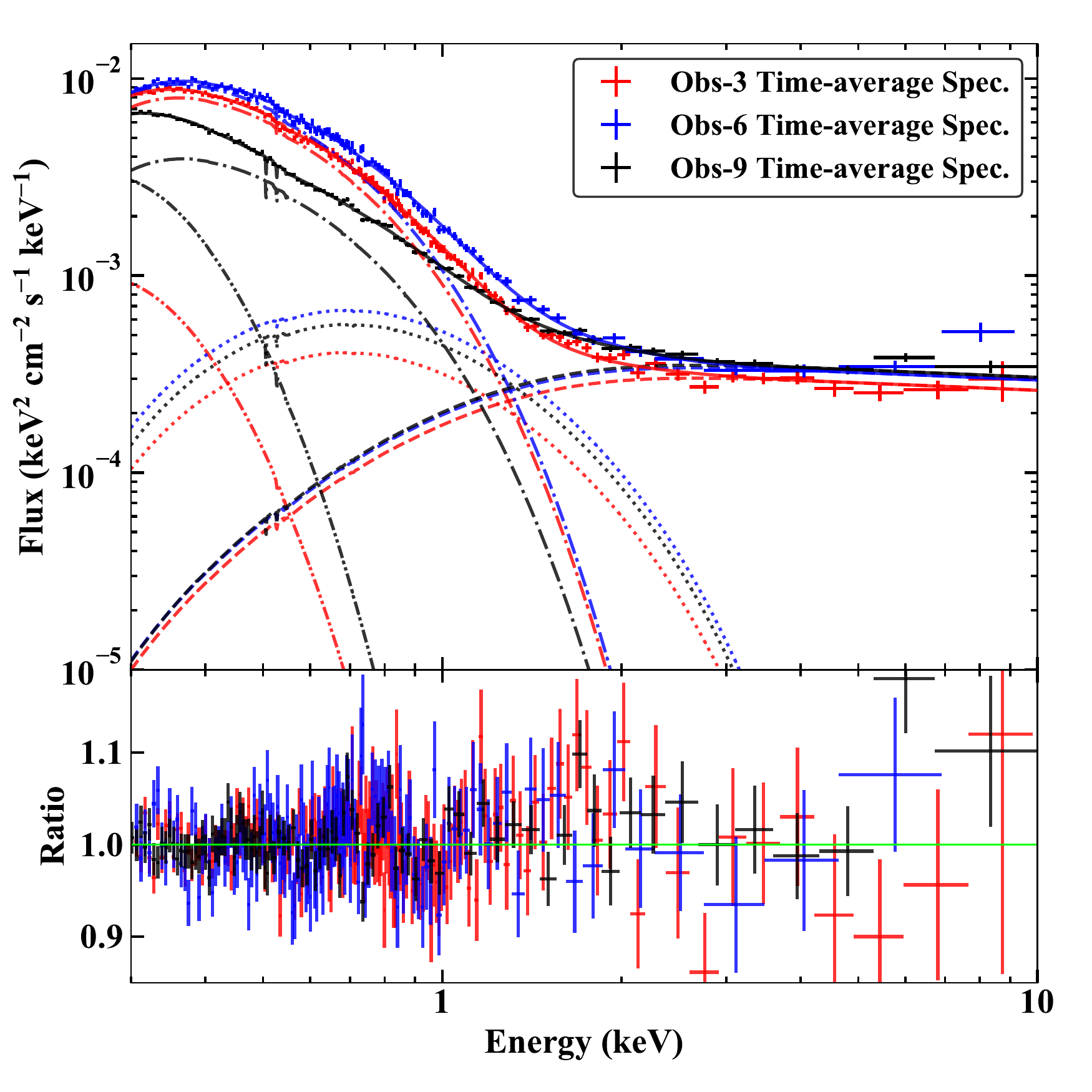}
\caption{Applying the best-fit Model-2 of Obs-9 to the time-average spectra in Obs-3 and 6 when the QPO is not detected. The black, red and blue spectra are the time-averaged spectra from Obs-9, 3 and 6, respectively. The dash-dot lines represent the non-QPO {\tt compTT-1} component, which have caused the much stronger soft excesses in Obs-3 and 6.}
\label{fig-spec-others}
\end{figure}

Figure~\ref{fig-spec-others} shows the best-fit results for these two observations. The overall minimal $\chi^2$ for Obs-3 and 6 is 778.0 for 746 dof, indicating reasonably good fits to both spectra. The best-fit $N_{\rm H, host}$ is $0.97^{+0.84}_{-0.78}\times10^{20}$ cm$^{-2}$ for Obs-3 and $1.09^{+0.83}_{-0.11}\times10^{20}$ cm$^{-2}$ for Obs-6. If we tie the normalization of {\tt diskbb} between Obs-3 and Obs-6, the best-fit $\chi^2$ only increases by 1.0, equivalent to 1 $\sigma$ significance. If we set the normalizations of all the components to be the same between Obs-3 and 6, except $N_{\rm H, ins}$ and an overall normalization factor, then the best-fit $\chi^2$ increases by 10.8 for a reduction of 3 dof, equivalent to 2.5 $\sigma$ significance. Thus the shape difference between the two time-averaged spectra from Obs-3 and Obs-6 is not significant, mainly the overall flux in Obs-6 is a factor of 1.27 $\pm$ 0.02 higher than in Obs-3.

The most remarkable difference is the enhancement of {\tt compTT-1} in the two non-QPO observations, as also shown in Figure~\ref{fig-spec-others}. In comparison with the time-averaged spectra in Obs-9, {\tt compTT-1} is a factor of 1.97$^{+0.05}_{-0.10}$ stronger in Obs-3, and a factor of 2.35$^{+0.09}_{-0.13}$ stronger in Obs-6. Interestingly, {\tt compTT-1} is also the only spectral component that is not required by the QPO's variability spectra in Obs-2 and 9. Therefore, {\tt compTT-1} is very likely to be the main driver for the anti-correlation between the intensity of the soft excess and the QPO's detectability as reported in Paper-I.

However, we also note that the increase of the flux of {\tt compTT-1} in Obs-3 and 6 is not enough to dilute the QPO signal across the entire 0.3-10 keV band. Thus the disappearance of the QPO is not a simple flux-dilution effect. In other words, if the QPO were present at the same strength in {\tt comptt} as in Obs-9, then there would be sufficient S/N to detect it in Obs-3 and 6. Therefore, the enhancement of {\tt compTT-1} must instead indicate a physical change in the accretion flow into a state which does not produce a QPO.

\begin{table}
\centering
\caption{The best-fit parameters of two broadband SEDs for \rej1034\ in Obs-9. `f' indicates that this parameter is fixed. For linked parameters, we put the parameter names in the table instead of values. Default values are adopted for all the model parameters not listed here.}
\begin{tabular}{lcccc}
\hline
Comp. & Par. & SED-1 & SED-2 & Unit \\
\hline
{\tt TBabs} & $N_{\rm H, gal}$ & 1.36 (f) & 1.36 (f) & $10^{20}$ cm$^{-2}$ \\
{\tt zTBabs} & $N_{\rm H, host}$ & 2.64 $^{+0.65}_{-0.97}$ & 0.59 $^{+0.42}_{-0.43}$ & $10^{20}$ cm$^{-2}$ \\
{\tt redden} & $E_{\rm B-V, gal}$ & 1.34  (f) & 1.34 (f) & $10^{-2}$ \\
{\tt zredden} & $E_{\rm B-V, host}$ & \multicolumn{2}{c}{$=N_{\rm H, host}\times1.7\times10^{-22}$} & \\
{\tt agnsed} & $M_{\rm BH}$ & 1.68 $^{+0.05}_{-0.25}$ & 11.83 $^{+1.64}_{-4.09}$ & $10^{6}M_{\odot}$ \\
{\tt agnsed} & log($\dot{m}$) & 0.43 $^{+0.08}_{-0.07}$ & -0.70 $^{+0.19}_{-0.06}$ & \\
{\tt agnsed} & $a_{*}$ & 0.26 $^{+0.04}_{-0.24}$ & 0.97 $^{+0.01}_{-0.07}$ & \\
{\tt agnsed} & $kT_{\rm e, warm}$ & 0.21 $^{+0.03}_{-0.02}$ & 0.23 $^{+0.02}_{-0.02}$ & keV \\
{\tt agnsed} & $kT_{\rm e, hot}$ & 100 (f) & 100 (f) & keV \\
{\tt agnsed} & $\Gamma_{\rm hot}$ & 2.19 $^{+0.04}_{-0.05}$ & 2.16 $^{+0.04}_{-0.04}$ & \\
{\tt agnsed} & $\Gamma_{\rm warm}$ & 3.34 $^{+0.16}_{-0.14}$ & 3.43 $^{+0.12}_{-0.13}$ & \\
{\tt agnsed} & $R_{\rm hot}$ & 6.0 $^{+1.0}_{-0.9}$ & 2.08 $^{+0.85}_{-0.36}$ & $R_{\rm g}$ \\
{\tt agnsed} & $R_{\rm warm}$ & 10.2 $^{+2.3}_{-0.3}$ & 131.8 $^{+143.5}_{-60.0}$ & $R_{\rm g}$ \\
{\tt hostgal} & norm & 0.33 $^{+0.03}_{-0.03}$ & 0.28 $^{+0.03}_{-0.02}$ & \\
{\tt const} &  & 1.04 $^{+0.04}_{-0.04}$ & 1.05 $^{+0.03}_{-0.04}$ & \\
\hline
$\chi^2/dof$ &  & 634.3/596 & 633.8/596 &  \\
\hline
\end{tabular}
\label{tab-sedfit}
\end{table}

\section{Multi-wavelength Properties}
In this section we explore how the QPO properties depend on the wider multi-wavelength properties of \rej1034. 
\subsection{Broadband Spectral Energy Distribution \& Modelling}
\label{sec-sed}
In steady state models the accretion flow has constant mass accretion rate at all radii. Thus the optical/UV emission from the outer accretion disc can be used to determine the accretion rate through the X-ray emitting inner regions once the black hole mass is known (e.g. Davis \& Laor 2011, Done et al. 2012), modulo corrections for spin and inclination. Since there is very little host galaxy absorption, or even much warm absorption, it seems most likely that \rej1034 is close to face on. The host galaxy is similarly face on (with a bar, see {\it HST} image\footnote{ http://archive.stsci.edu/cgi-bin/mastpreview?mission=hst\&dataid=JD9512010})

To build the spectral energy distribution (SED) extending down to the optical/UV, we use the UVOT data from the 
simultaneous \swift\ ToO observation. This provides 
multi-band photometry (UVW2, UVM2, UVW1, U, B, V) to support the \xmm/OM data which was taken in the UVW1 fast mode only to search for fast variability. These data, together with the \xmm/EPIC and \nustar\ spectra, allow us to build the best simultaneous broadband SED of \rej1034. 

We fit these multi-wavelength data with the 
{\tt agnsed} model (Kubota \& Done 2018) in {\sc xspec}. This model is based on the physical concept described above, that the mass accretion rate is constant with radius and the radial emissivity corresponds to that of a thin disc (Novikov-Thorne). Then the emission mechanism changes from blackbody to warm Comptonisation 
at $R_{\rm warm}$, and from warm Comptonisation to hot Comptonisation at $R_{\rm hot}$. {\tt agnsed} is an refinement of the {\tt optxagnf} model (Done et al. 2012), as it
uses the more sophisticated passive disc scenario for the warm Comptonisation (Petrucci et al. 2018), and calculates the temperature of seed photons of the hot Comptonisation internally. The model only includes a single temperature warm Comptonisation region, rather than the two-temperature model preferred by all the spectral-timing results, but we focus first on the overall energetics rather than the detailed soft X-ray shape.

We fix the electron temperature of the hot corona at its default value of 100 keV\footnote{This parameter has very little effect on the SED fitting. We also try to fix it at 200 keV, and the resultant difference in the $\chi^2$ is only 0.1, and there are almost no changes in the best-fit parameters.}. The outer radius of the disc is chosen to be the self-gravity radius calculated inside the model following Laor \& Netzer (1989). The inclination angle is taken to be $30\degr$. Since it is known that the host galaxy star light contributes significantly to the optical emission of \rej1034\ (Bian \& Huang 2010; Czerny et al. 2016), we use the spectral template of an Sb galaxy from Polletta et al. (2007) to model it, which is incorporated as the local {\tt hostgal} model. Galactic absorption is considered in the same way as in previous X-ray analysis. The Galactic dust reddening{\footnote{https://irsa.ipac.caltech.edu/applications/DUST}} is modelled by the {\tt zredden} model with $E_{\rm B-V, gal}$ being fixed at 0.0134 for the light-of-sight of \rej1034\ (Schlegel, Finkbeiner \& Davis 1998). Dust reddening from the host galaxy ($E_{\rm B-V, host}$) is assumed to be $1.7\times10^{-22}~N_{\rm H, host}$ (Bessell 1991). Based on the cosmology model noted in Section~\ref{sec-introduction}, we adopt a co-moving distance of 175.2 Mpc for \rej1034\ (Wright 2006). A free constant is included in the model to account for potential calibration differences between the normalizations of \xmm/OM and \swift/UVOT.

After performing the SED fitting in {\sc xspec}, we find
a range of local minima in $\chi^2$ depending on the black hole mass/spin. These span from low mass, low spin solutions, where the standard disc extends down into the lowest energy end of the 
\xmm/EPIC bandpass, and the warm Comptonisation only fills in between this and the hot Comptonisation
(SED-1), up to the highest mass, highest spin solution where most of the UV and soft X-ray emission is produced by the warm Comptonisation, with the standard disc component only contributing to the optical/UV emission (SED-2). 

SED-1 and SED-2 provide comparably good fits to all the data (see Table~\ref{tab-sedfit}), but the black hole mass in SED-1
is consistent with the values derived from independent mass estimators such as the H$\beta$ line velocity width (Czerny et al. 2016). We therefore prefer SED-1, with a lower black hole mass of 
1.68 $^{+0.05}_{-0.25}\times10^{6}M_{\odot}$. The 
black hole spin is also low at $a_{*}=0.26^{+0.04}_{-0.24}$.
We note that {\tt agnsed} does not include the combined effects of red and blue-shifts expected from the fast orbital velocities of gas under the influence of strong gravity, but these effects are not large for low spin solutions (see e.g. Done et al. 2013).

The mass accretion rate through the outer disc ($\dot{m}$) is high at 2.69$^{+0.54}_{-0.40}$. This is marginally super-Eddington, though not sufficiently extreme for advection of radiation to produce a large change in the predicted SED (Kubota \& Done 2019). Nonetheless, as
an additional check, we replace the {\tt agnsed} model by the {\tt agnslim} model (Kubota \& Done 2019). {\tt agnslim} has a similar set of parameters, but it also takes into account the suppressed radiative efficiency in the inner disc region due to the strong advection and/or disk wind operating in the super-Eddington state. The best-fit results of {\tt agnslim} are very similar to {\tt agnsed}, with {\tt agnslim} having a slightly larger $\chi^2$ of 636.9 for 597 dof. This result confirms that \rej1034\ is only a slightly super-Eddington NLS1, so the additional effects of super-Eddington accretion state are not significant.

The decomposition of SED-1 in the X-ray band is similar to the best-fit Model-1 in that there is a standard disc component at the lowest energies. But we have previously shown that Model-2, with two warm Comptonisation components, gives a better solution for the X-ray spectral variability of \rej1034. Therefore, we refit the unfolded X-ray spectrum of SED-1 with the components of Model-2. Unsurprisingly, a good fit can be achieved because the unfolded spectra from Model-1 and Model-2 are very similar (see Figure~\ref{fig-spec-rmscov1}a and Figure~\ref{fig-spec-rmscov2}a). The reason of adopting this approach is to retain energy conservation between the disc and corona in SED-1. Figure~\ref{fig-sed} shows the best-fit SED, where the warm and hot Comptonisation components in SED-1 have now been replaced by the three Comptonisation components in Model-2. We present this result as the by far best physical SED decomposition for \rej1034. The optical emission is dominated by the star-light from the host galaxy, the UV bump is dominated by the emission from the outer accretion disc, and the X-ray emission originates from the inner disc and the three Comptonisation regions.

\subsection{Exploring the UV Variability}
\begin{figure}
\centering
\includegraphics[trim=0.05in 0.3in 0.0in 0.0in, clip=1, scale=0.58]{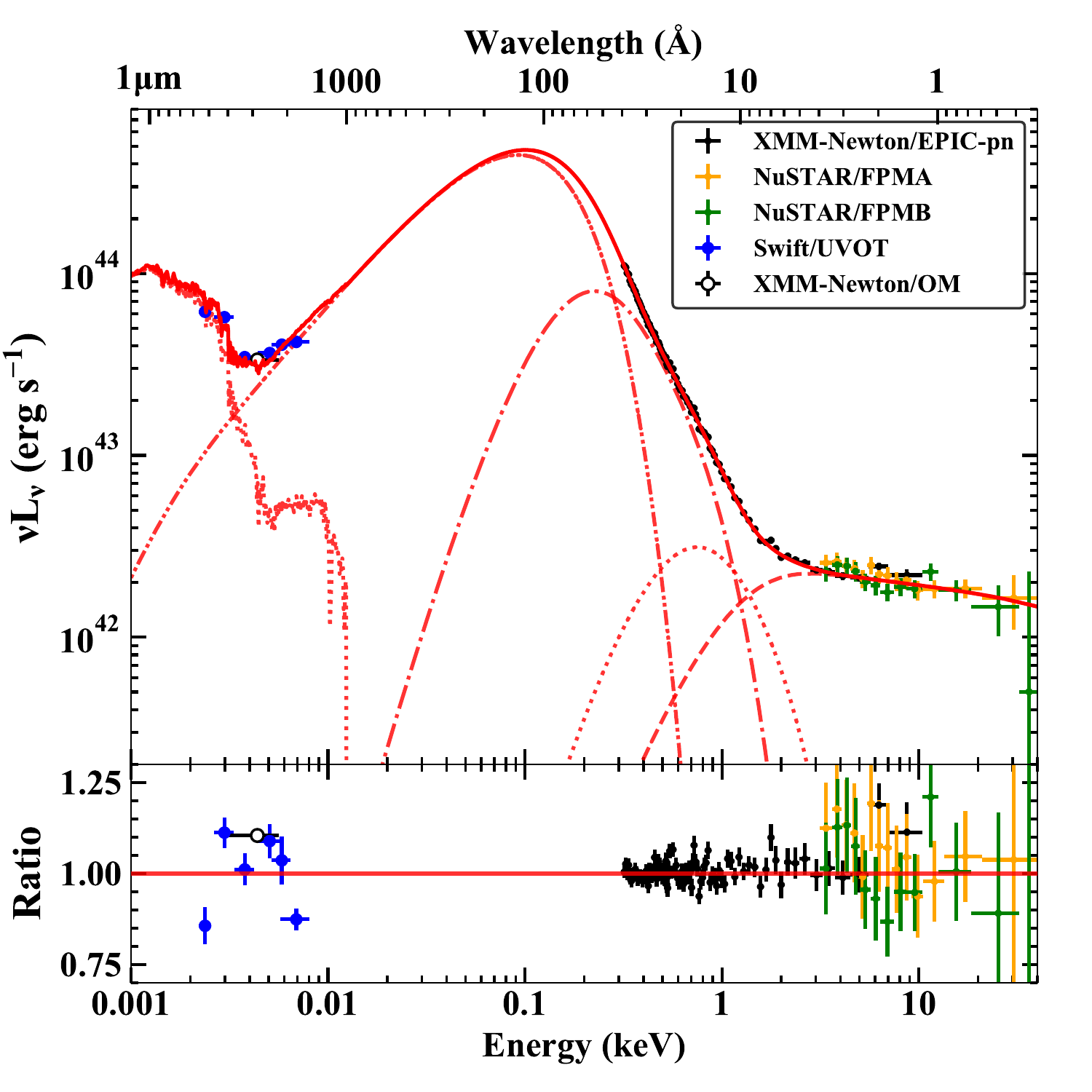}
\caption{Upper panel: the best-fit broadband SED of \rej1034\ in Obs-9. The multi-wavelength data comprise the \swift\ UVOT and \xmm\ OM data in optical/UV, the \xmm\ EPIC-pn spectrum in 0.3-10 keV and the \nustar\ FPMA/FPMB spectra in 3.0-40 keV. This SED is based on the best-fit SED-1 model, but the two Comptonisation components have been replaced by the three Comptonisation components in Model-2. The red solid line shows the entire best-fit SED. The red dotted lines in the optical/IR shows the best-fit host galaxy component. The red dash-dot-dot, dash-dot, dotted, and dash lines represent the best-fit accretion disc component in SED-1, the {\tt compTT-1}, {\tt compTT-2} and {\tt nthComp} components in Model-2, respectively. Lower panel: ratios between the data and the best-fit SED.}
\label{fig-sed}
\end{figure}

\begin{figure}
\centering
\includegraphics[trim=0.1in 0.3in 0.0in 0.0in, clip=1, scale=0.57]{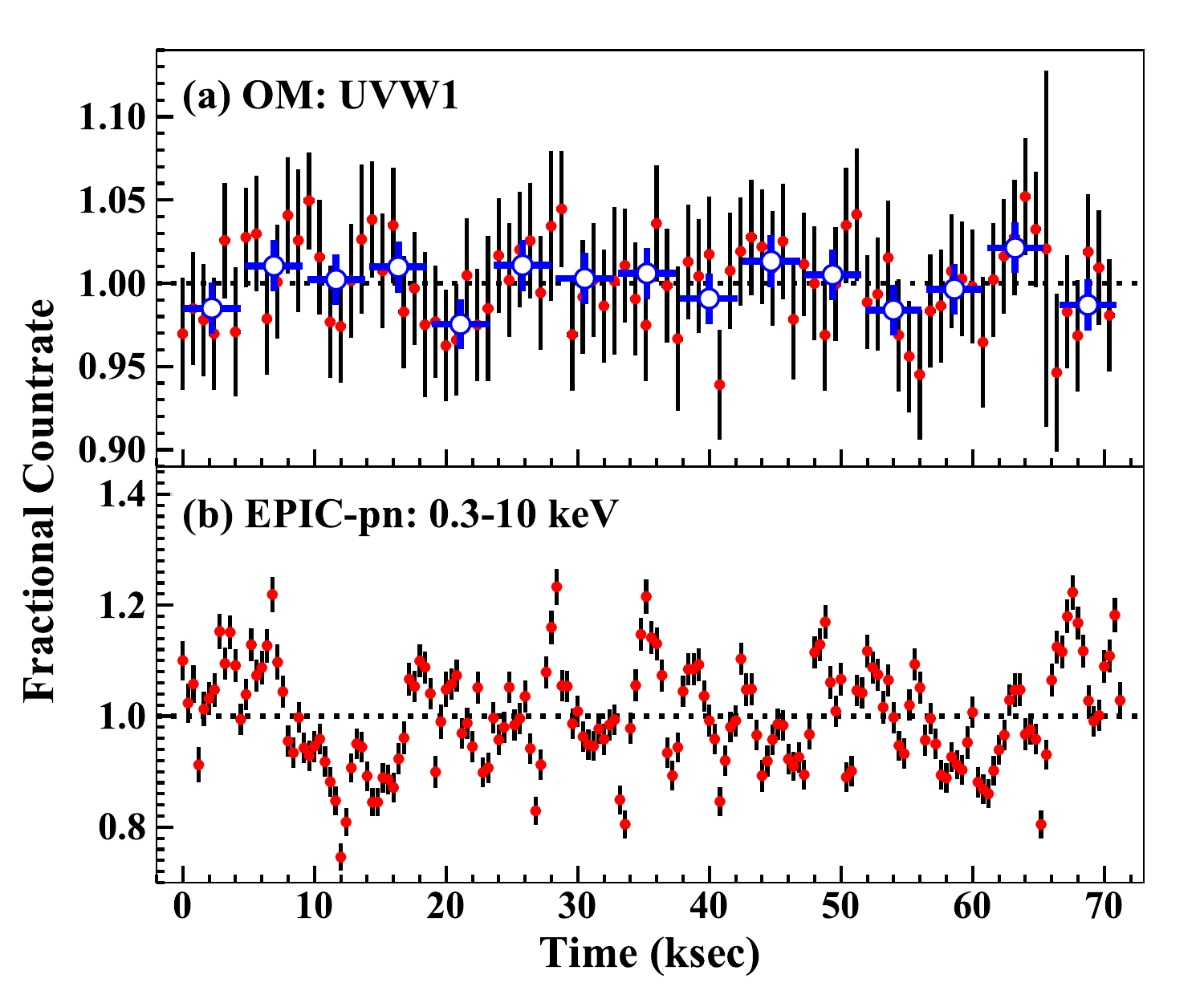}
\caption{Panel-a: the OM fast-mode light curve (800 s binned) in the UVW1 filter in Obs-9, as shown by the red points. The blue points show photometric fluxes derived from the simultaneous imaging-mode exposures. Panel-b: the simultaneous EPIC-pn light curve (400 s binned) in 0.3-10 keV plotted for comparison. There is no significant correlation between the two light curves (Pearson's correlation coefficient is -0.05). The intrinsic rms of the UVW1 light curve is also consistent with zero.}
\label{fig-omlc}
\end{figure}

As shown in Figure~\ref{fig-sed}, the UVW1 band of OM is dominated by the accretion disc emission of \rej1034, and so it tracks the mass accretion rate through the outer disc. We can explore the short-term UV variability by extracting the light curve in the UVW1 filter in Obs-9, which is shown in Figure~\ref{fig-omlc}a. No intrinsic rms can be found in this UV band, nor does there exist any UV/X-ray correlation, as indicated by the Pearson's correlation coefficient of -0.05. We also examine the cross-correlation function between the UV and X-ray light curves, and find no significant lagged correlation. Therefore, we can conclude that the UV variability is much weaker than the X-ray variability. This is very similar to that observed in several other super-Eddington NLS1s such as \rxj04, whose optical/UV emission is also dominated by the disc emission, with no intrinsic UV variability (Jin et al. 2017b). But this is in contrast to some AGN with much lower Eddington ratios such as NGC 5548, where significant variability is observed in the UV/optical band, which is also correlated with the X-ray variability because of the X-ray reprocessing (e.g. Edelson et al. 2015; Gardner \& Done 2017). 

\begin{figure}
\centering
\includegraphics[trim=0.1in 0.3in 0.0in 0.0in, clip=1, scale=0.57]{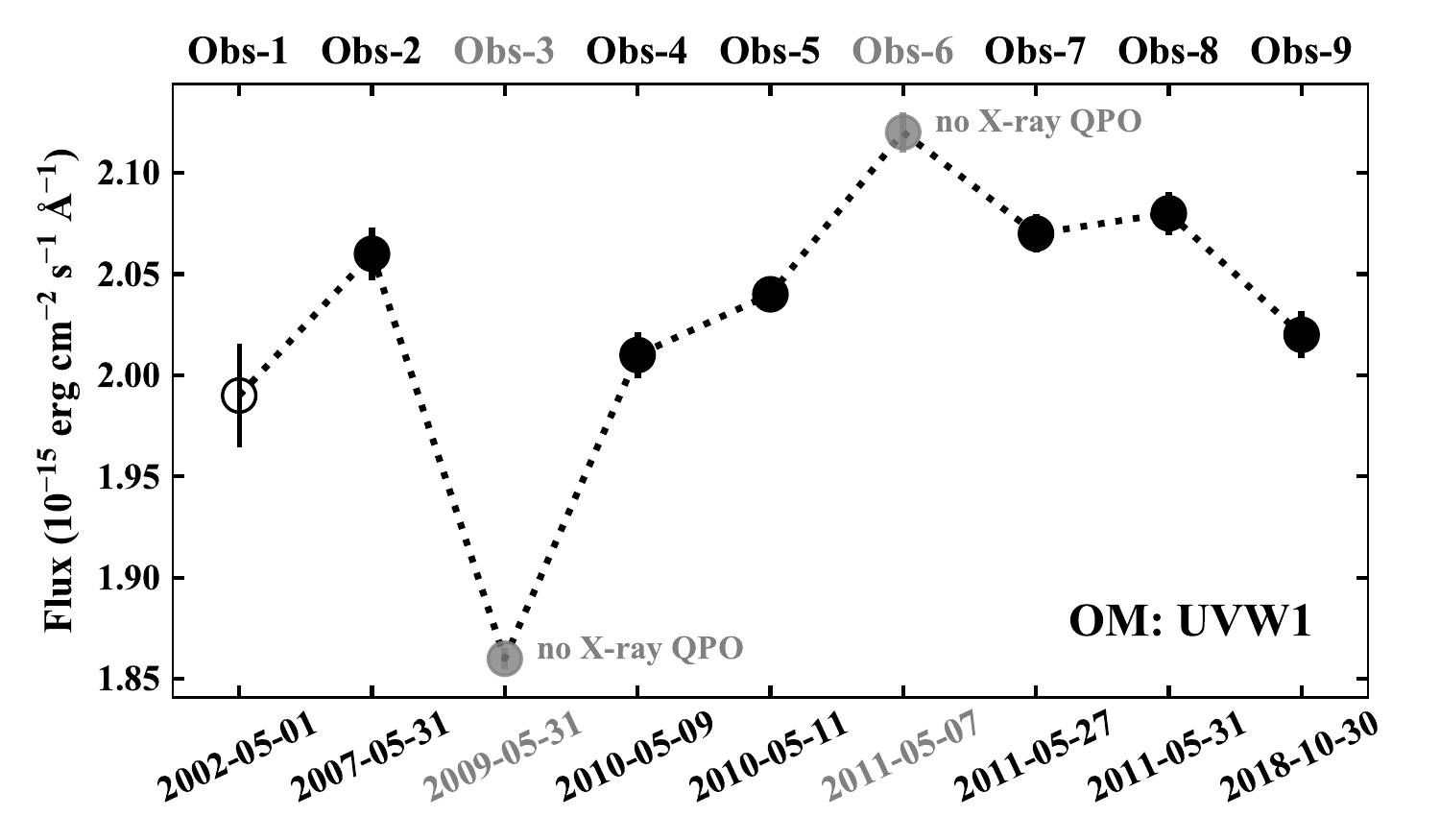}
\caption{The long-term UV light curve of \rej1034\ as observed by \xmm/OM through the UVW1 filter. Six observations show the X-ray QPO (black points), two observations do not show any detectable X-ray QPO (grey points), one observation does not have sufficient data quality for the QPO analysis (open circle point). No correlation is found between the UV flux and the QPO's detectability.}
\label{fig-uvlc}
\end{figure}

Since all the 9 \xmm\ observations of \rej1034\ have simultaneous ultraviolet observations in UVW1, we can construct a long-term light curve in UVW1. This begins with the first X-ray observation in 2002, through to 
the latest observation in 2018, as is shown in Figure~\ref{fig-uvlc}. This light curve shows the variation of $\dot{m}$ though the outer disc of \rej1034\ during the past 16 years. Specifically, the two observations showing no QPO signal, Obs-3 and Obs-6, have the lowest and highest $\dot{m}$, respectively. Therefore, we can exclude $\dot{m}$ as the direct trigger of the X-ray QPO, although a delayed influence cannot be ruled out due to the poor sampling of this long-term light curve.

\section{Analogy of Spectral-timing Properties between \rej1034\ and \grs1915}
\grs1915\ is a famous Galactic black hole binary (BHB) with a super-Eddington luminosity and a low-temperature Comptonisation component (Done, Wardzi\'{n}ski \& Gierli\'{n}ski 2004; Middleton et al. 2006). During its disc-dominated spectral state, a high-frequency QPO at $\sim$67 Hz is sometimes observed, with a quality factor of $\sim$20 (Morgan, Remillard \& Greiner 1997). The peak frequency of this QPO varies between 65 and 72 Hz, and its rms amplitude increases from 1 per cent at a few keV to 11 per cent at 40 keV (Belloni \& Altamirano 2013; Belloni et al. 2019). Previous studies have shown that the QPO frequency seen in \rej1034\ is
consistent with a simple mass scaling factor of the 67 Hz QPO in \grs1915, and that their X-ray spectra are similarly dominated by a soft component (Middleton et al. 2009; Middleton \& Done 2010; Czerny et al. 2016). These two QPOs also have similar quality factors and small variation of peak frequency (Paper-I). 

However, a key failure in this analogy is the soft X-ray lag seen in the QPO spectrum of Obs-2 (e.g. Zoghbi \& Fabian 2011), which is the opposite to the soft X-ray lead seen in \grs1915\ (M\'{e}ndez et al. 2013).
However, this issue is now resolved, because we showed in Paper-I that the new data clearly show a soft X-ray lead for the stronger QPO detection in Obs-9.
However, another difference is that so-far no harmonics have been found for the QPO in \rej1034\ (Alston et al. 2014; Paper-I), whereas they have been detected in \grs1915, but only in some observations (e.g. Belloni, M{\'e}ndez \& S{\'a}nchez-Fern{\'a}ndez 2001; Strohmayer 2001; Remillard et al. 2002). Thus this particular difference may not be a very serious issue.

Here we use the stronger QPO in Obs-9 to explore its similarity to the 67~Hz QPO of \grs1915\ in more detail, with a direct comparison of the spectral-timing properties between the two sources. We take the rms spectrum of the 67 Hz QPO from Fig. 6 in Belloni \& Altamirano (2013), and take its phase lag spectrum from Fig. 4 in M\'{e}ndez et al. (2013). Both of these two studies use the \rxte\ data of \grs1915\ taken on 21th Oct. 2003 (OBSID: 80701-01-28-00, 01, 02). The frequency of the QPO in Obs-9 is $2.82\times10^{-4}$ Hz, which is a factor of $2.38\times 10^5$ lower than the 67 Hz QPO of \grs1915. Assuming the black hole mass of \grs1915\ is $12.4~M_\odot$ (Reid et al. 2014), the black hole mass of \rej1034\ is estimated to be $3\times 10^6~M_\odot$.

\begin{figure*}
\centering
\begin{tabular}{cc}
\includegraphics[trim=0.1in 0.0in 0.0in 0.1in, clip=1, scale=0.55]{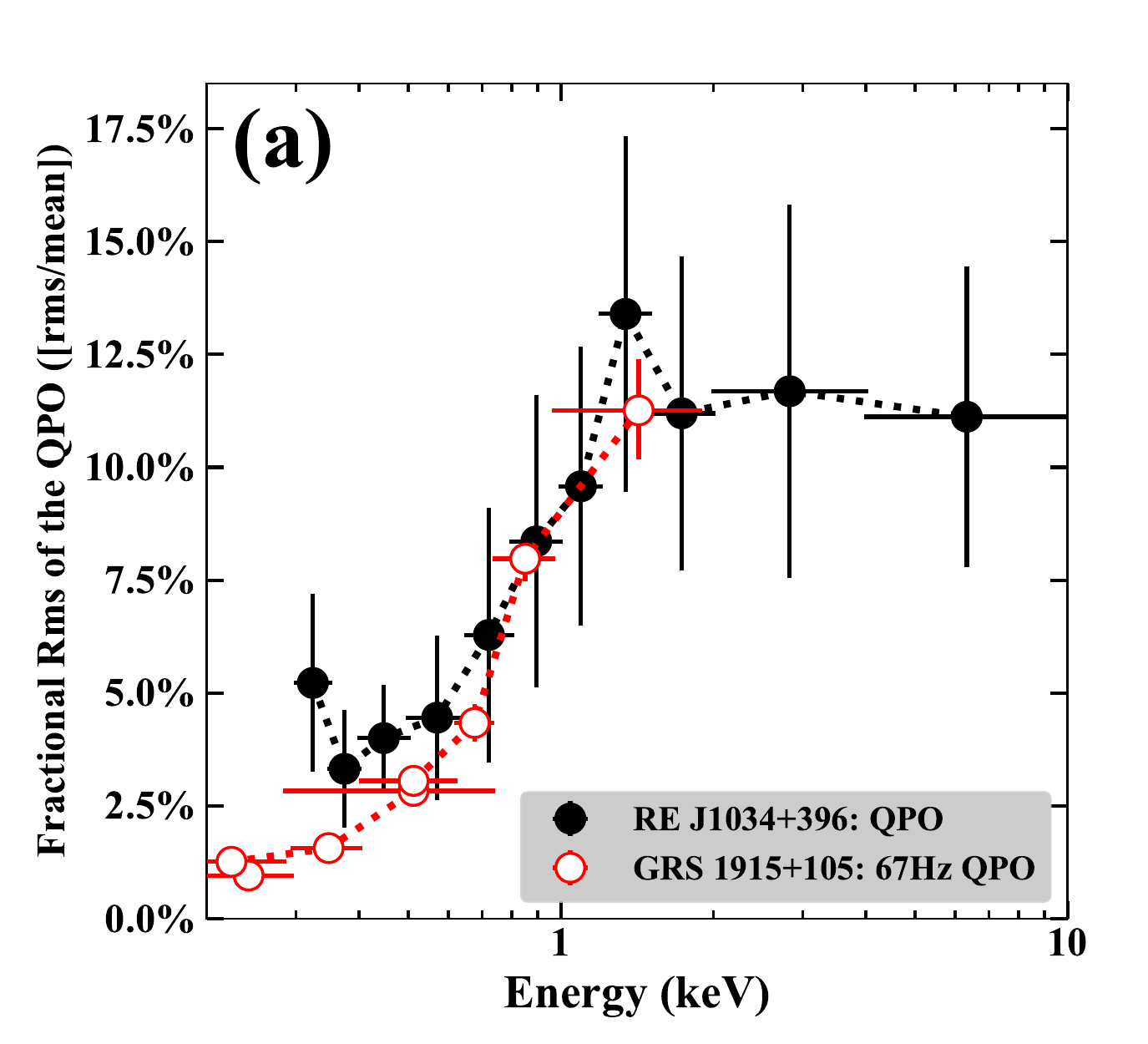} &
\includegraphics[trim=0.2in 0.0in 0.0in 0.1in, clip=1, scale=0.55]{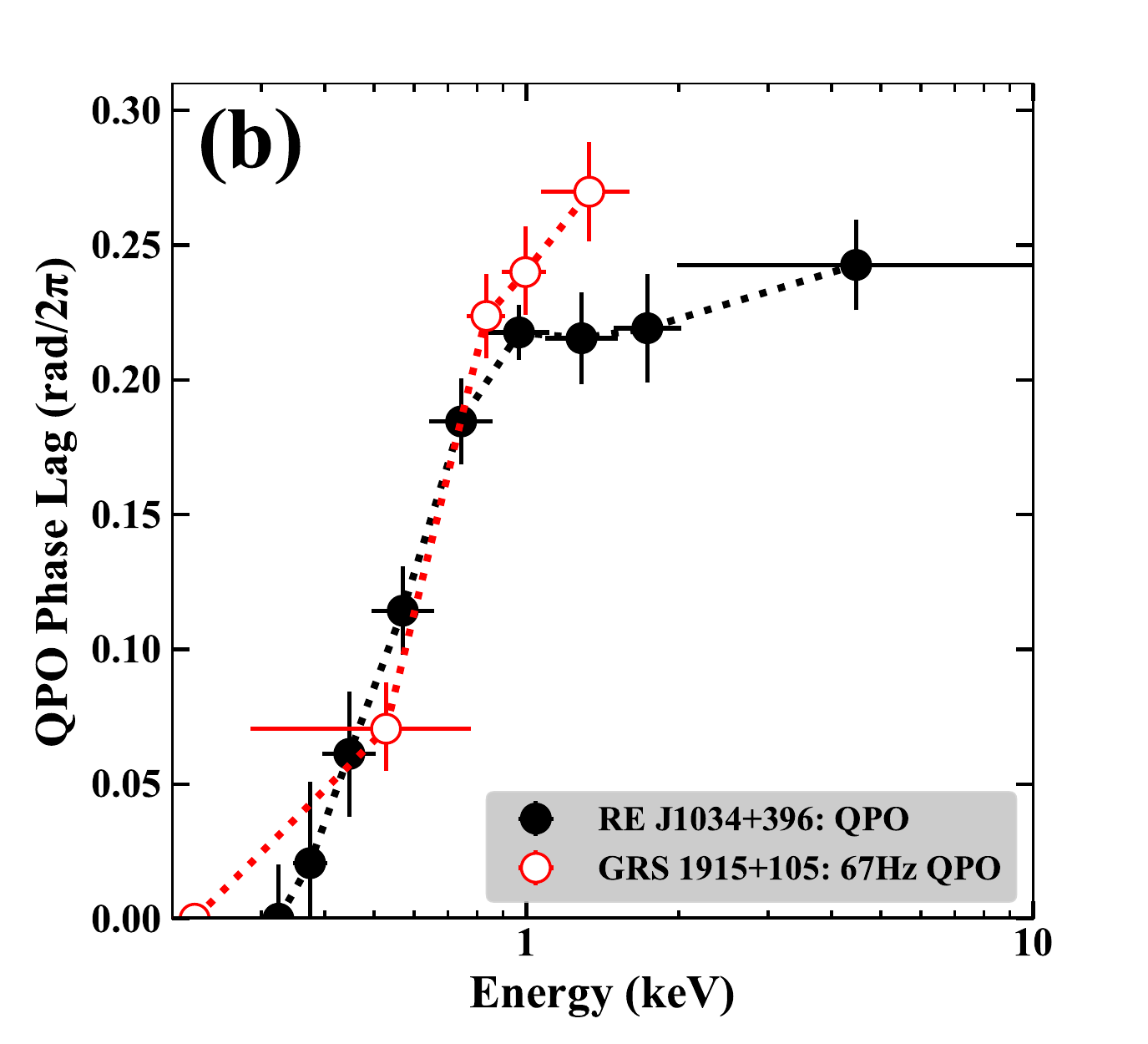} \\
\includegraphics[trim=-0.2in 0.1in 0.0in 0.1in, clip=1, scale=0.55]{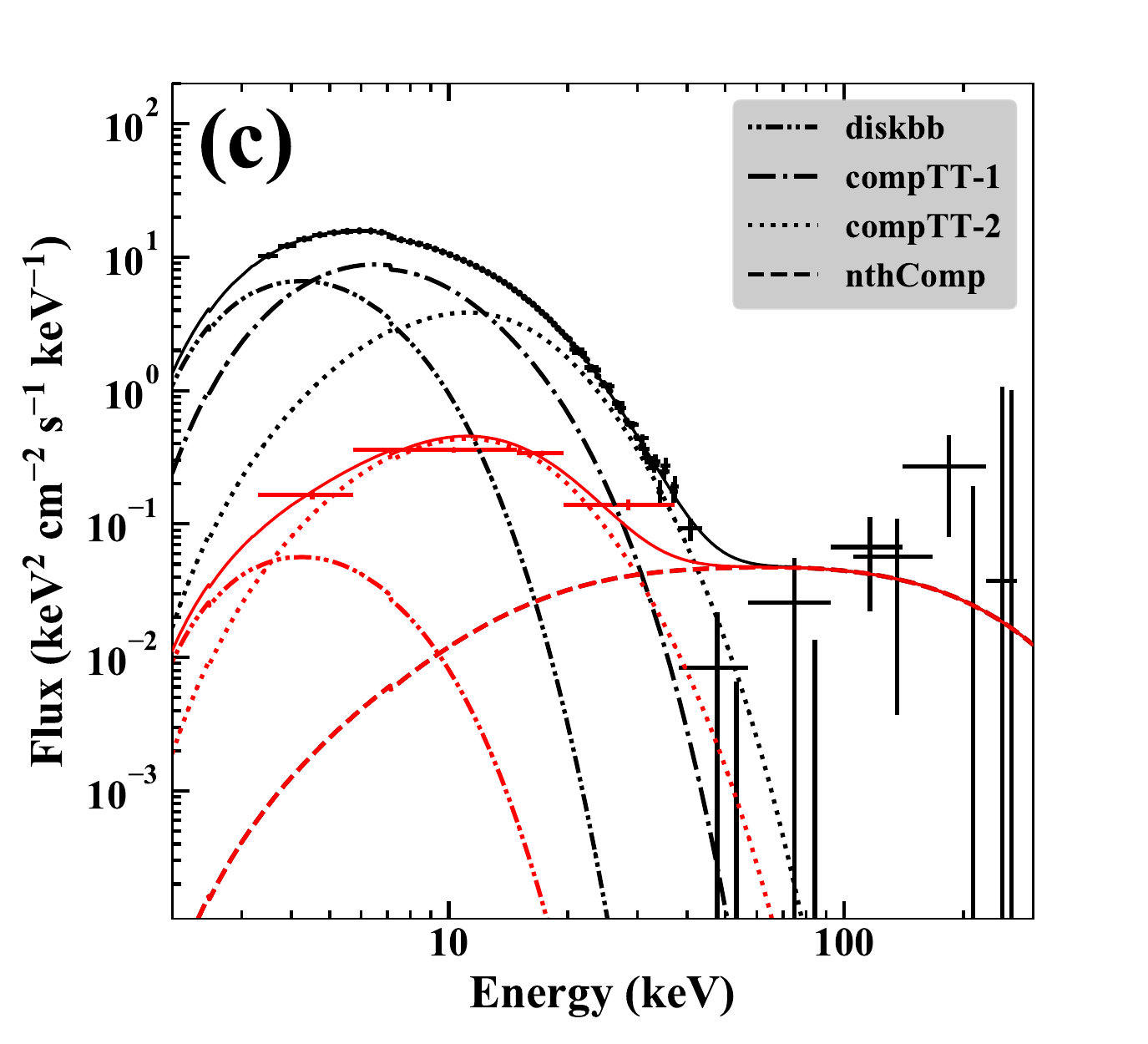} &
\includegraphics[trim=0.2in 0.1in 0.0in 0.1in, clip=1, scale=0.55]{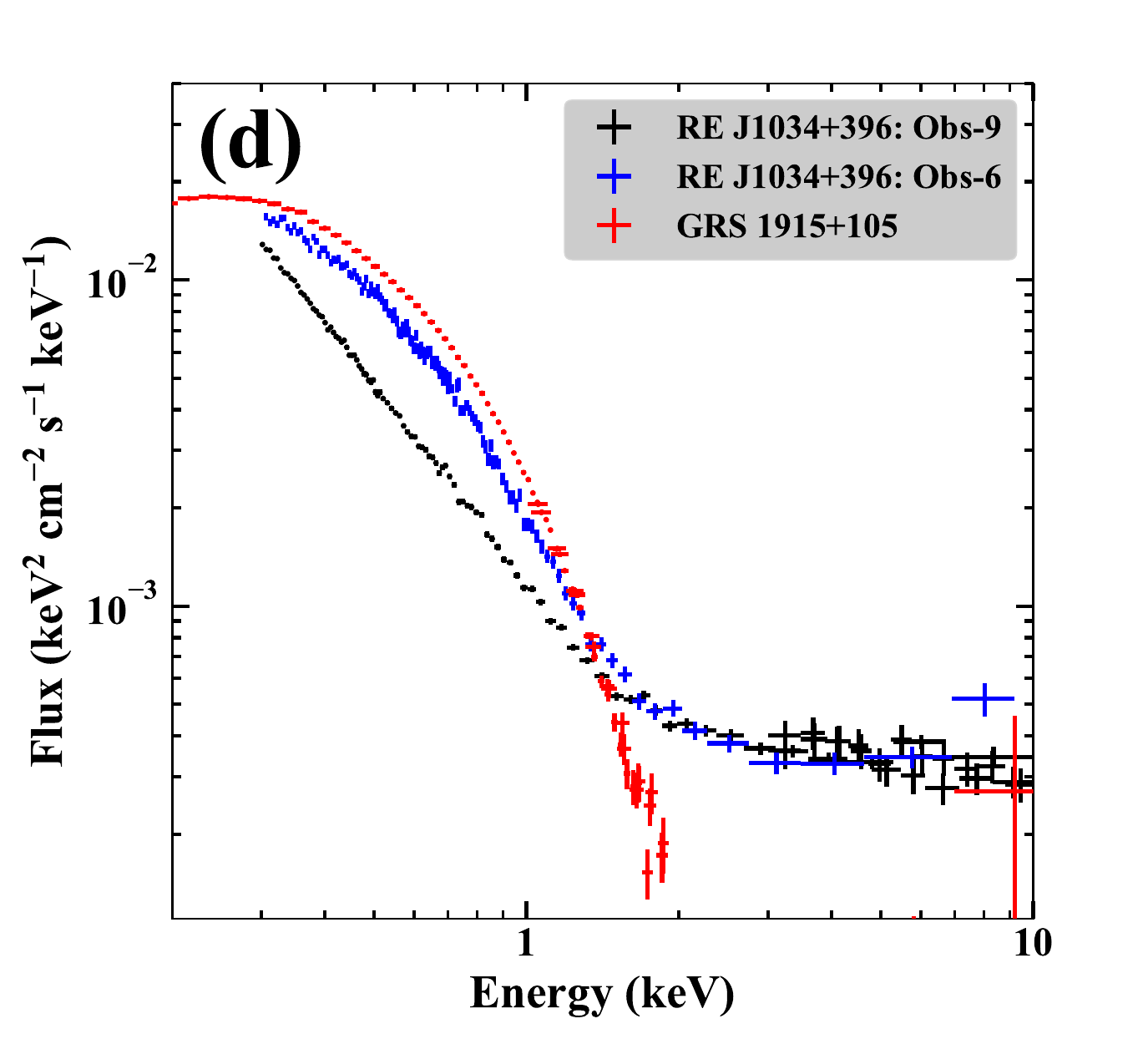} \\
\end{tabular}
\caption{Comparison of spectral-timing properties between \rej1034\ and \grs1915. Panels-a and b: comparisons for the QPO's rms and phase lag spectra between \rej1034\ in Obs-9 and \grs1915\ (67 Hz QPO, shifted down by a factor of 20 in energy). The rms and phase lag spectra of the 67 Hz QPO of \grs1915\ are obtained from Fig.4a in M\'{e}ndez et al. (2013) and Fig.6 in Belloni et al. (2013), separately. Panel-c: applying a similar model to the \rxte\ spectra of \grs1915\ during its 67 Hz QPO observations (OBSID: 80701-01-28-00, 01, 02). The rms spectrum of the 67 Hz QPO is shown in red. Panel-d: comparison of unfolded and unabsorbed spectra between \rej1034\ in Obs-9 (with a QPO) and Obs-6 (without a QPO), and \grs1915\ being shifted down by a factor of 20 in energy and a factor of 1000 in flux.}
\label{fig-rgs1915}
\end{figure*}

Since the disc temperature scales as $T^4\propto \dot{m}/M$, if these two black hole systems have similar Eddington ratios, then their disc temperatures should be a factor $\sim 20$ different (see also Middleton et al. 2009). The \rxte\ spectra of \grs1915\ from 3-40~keV, with the energy shifted down by a factor of 20, serendipitously matches very well to the \xmm\ energy range for \rej1034. We apply this shifting factor, and compare the rms variability and phase lag from these two objects directly. Figure~\ref{fig-rgs1915} Panels-a and b show that the QPO of \rej1034\ and the 67 Hz QPO of \grs1915\ are indeed very similar in terms of both rms and phase lag spectra.

We extract the time-averaged spectra of \grs1915\ from the above \rxte\ datasets, and create an absolute rms spectrum for the 67 Hz QPO. These spectra are fitted with the four spectral components in the Model-2 configuration. Figure~\ref{fig-rgs1915}c shows the best-fit results. Note that the {\tt nthComp} component is poorly constrained due to the poor S/N at high energies, so we simply fix its slope at 2.0, and set its normalization as the same between the time-averaged spectrum and the rms spectrum. Since the two sources have very different black hole masses, it is not correct to directly compare their model parameters. But generally speaking, this spectral decomposition of \grs1915\ is similar to \rej1034\ (see Figure~\ref{fig-spec-rmscov2}a). The most noticeable point is that the {\tt compTT-1} component does not appear in the 67 Hz QPO's rms spectrum, either.

Finally, we compare the unfolded, unabsorbed and time-averaged spectra between the two sources in a phenomenological way, which is shown in Figure~\ref{fig-rgs1915}d, where again we shift the energy of \grs1915\ down by a factor of 20 in order to match the energy range of \rej1034. Then the disc luminosity should roughly scale down by a factor of $20^4$. But the flux also scales with $D^{-2}$ where $D$ is the source distance. Adopting the luminosity distance of $D_{\rm R} = 182.6$ Mpc for \rej1034\ and $D_{\rm G} = 8.6 $ kpc for \grs1915\ (Reid et al. 2014), the shifting factor for the flux is $20^4 \times (D_{\rm G}/D_{\rm R})^2\simeq3.5\times 10^{-4}$. We can also adopt another factor of $\sim 3$ to account for the much larger inclination angle of \grs1915\ than \rej1034, so the final shifting factor for the flux is chosen to be 1000. 

Surprisingly, the spectra shown in Figure~\ref{fig-rgs1915}d are somewhat different: the
soft excess of \rej1034\ is weaker than \grs1915, but its hard X-ray tail is much stronger. In particular the spectrum of \grs1915\ has the convex curvature which is more like that seen in the non-QPO observations of \rej1034 (blue points).

However, this might also be related to the different inclination angles between \grs1915\ and \rej1034. \grs1915\ is at a moderate/high inclination angle, where the Doppler boost from the high velocity inner disc offsets its gravitational redshift, whereas \rej1034\ is much more likely to be face on, so gravitational redshifts dominate. We note that the {\tt agnsed} model used for SED-1 does not include these effects. There is also the possibility of differences in black hole spin but this is rather uncertain in both objects. As we discuss above, we prefer the low mass which is also low spin solution of SED-1 for \rej1034, but we note that the stronger redshifts expected for a face on disc could allow a higher spin solution, while in \grs1915\ there are different spin values claimed for different assumed electron temperature of the {\tt nthcomp} component (Middleton et al. 2006; Remillard \& McClintock 2006).

\section{Discussion}
\label{sec-discussion}
%\subsection{Origin of the QPO in \rej1034}
%\label{sec-mechanism}
The soft X-ray excess is a ubiquitous feature of the majority of the bright AGN population, but an X-ray QPO is a rather rare phenomenon, detected in less than 10 AGN so far, and only in \rej1034\ with both high significance and repeatability. 
Thus \rej1034\ provides us with a unique laboratory in which to investigate the relationship between the soft excess and the presence of a QPO. It is clear from Paper-I and previous studies that the QPO is directly related to the shape of the soft X-ray spectrum in \rej1034, where the QPO is quenched when the soft X-ray spectrum is stronger, and has a more convex (thermal-looking) shape.

Our latest Obs-9 data with the strongest and most coherent AGN QPO ever seen, allows us to explore the various energy-dependent properties of this QPO and the stochastic variability both at low and high frequencies (although the latter has more limited statistics).
Our analysis favors a model in which the QPO traces the hottest part of the soft X-ray excess emission, while the stochastic variability links to a cooler component. We also show that the previous datasets from \rej1034\ can also be well fitted by this model, although in Obs-9 the QPO is also very strong at the lowest energies (which we interpret as part of the {\tt diskbb} component), which is not the case in Obs-2.

The QPO in \rej1034\ is so long lasting that it must be a fundamental frequency of the system itself 
rather than just some transient `hot spot' or occultation event. It is probably analogous to the 
67 Hz QPO observed in \grs1915\ as the frequency scales with the black hole mass, and here we show that the 67 Hz QPO is likewise associated with the hottest part of the strong soft X-ray emission in \grs1915. 
It seems that there is a region at the inner edge of the accretion flow which is radially distinct enough that it exhibits particular spectral and variability behaviours.

The {\tt agnsed} spectral model has distinct radial regions, with the high energy 
hot Comptonisation from $R_{\rm isco}-R_{\rm hot}$ and the warm Comptonisation(s) 
from $R_{\rm hot}-R_{\rm warm}$. In this model, the warm Comptonisation is associated with the optically thick, geometrically thin(ish) accretion disc, while the hot component is part of an optically thin(ish), geometrically thick accretion flow. Coherent oscillations are easier to excite and maintain when the radial region is small. This occurs when the source is close to the Eddington limit for the hot Comptonisation, and maybe for the warm Comptonisation region as well (Kubota \& Done 2018). Both \rej1034\ and \grs1915\ show that it is the hottest part of the warm Comptonisation region which is oscillating most strongly, so we are observing a small radial region at the edge of a geometrically thin(ish) disc. 

There are multiple oscillation modes which can exist for a narrow, thin disc ring. The simplest are from purely geometric displacements such as a vertical or radial offset about an equilibrium position centred on 
some radius $r_0$, with width $dr_0$, on the equatorial plane at $z_0=0$. However, pressure forces within the disc should be important, which can be incorporated in models of slender (or non-slender) tori. These have an internal vertical structure, with elliptical cross-section disc centred on $r_0, z_0=0$, but now with height $dz_0$ at $r_0$, tapering down to $0$ at $r_0\pm dr_0$.  
These tori have the geometric displacement modes in radius and vertical height as before, but also have modes which change the vertical structure. The {\it breathing} mode changes $dz_0$ without changing $dr_0$, the {\it plus} mode is an anti-correlated increase in $dz_0$ with decreasing $dr_0$, and the {\it X} mode is an anti-correlated increase in $z_0$ at $r_0+dr_0$ with a decrease at $r_0-dr_0$. Intrinsically, the proper area of the torus does not change for a vertical or {\it X} mode, whereas it does for the {\it radial}, {\it plus} and {\it breathing} modes. 

Vincent et al. (2014) showed the power generated by the changing position and/or shape of these tori, assuming constant emissivity per unit (proper) area, when seen through a fully general relativistic (GR) spacetime. The change in proper area of the {\it radial}, {\it plus} and {\it breathing} modes mean that these have more intrinsic power, but the GR effects also amplify any changes in vertical extent at high inclination. Thus for face on inclinations, the {\it radial}, {\it plus} and {\it breathing} modes have most power, whereas at $85^\circ$ this changes to the {\it plus} mode and {\it breathing} modes (the vertical mode does not include any intrinsic area change and the {\it radial} mode has no vertical extent for the GR to amplify).

The similarity of the (phase) lags in the QPO in both \rej1034\ and \grs1915\ argues for this being the same type of oscillation in each case, and 
while \grs1915\ is at moderately high inclination, \rej1034\ is most probably face on. Thus the only mode which has high power at both high and low inclination is the {\it plus} mode, so we favour the association of the 
QPO on both objects with the {\it plus} mode. 

However, none of these simulations allow for a real change in heating/cooling rate, and hence the real change in temperature and emissivity which will accompany such oscillations. For example, the expansion/compression of the torus in the {\it breathing} and {\it plus} modes could also give rise to intrinsic changes in cooling/heating rates.  This would allow both the {\it breathing} and {\it plus} modes to be strong at all inclinations.

\section{Conclusions}
\label{sec-conclusion}
This paper builds on and extends the results of Paper-I for the re-detection of the QPO in Obs-9. Here we use these data to examine the QPO in more detail, using the combined spectral-timing techniques as well as the full multi-wavelength data in the simultaneous \xmm, \nustar\ and \swift\ observations. These results provide new insights on the relationship between the soft excess and the QPO, and reinforce the QPO's analogy with the 67 Hz QPO detected in the Galactic micro-quasar \grs1915. We summarize the main results below.

\begin{itemize}
\item we present the detailed energy dependence of the QPO's timing properties, including its peak frequency, rms, phase lag and coherence in Obs-9. Steep increases of rms and lag (i.e. soft X-ray lead) are observed from 0.3 to 1 keV, and both of them flatten towards harder X-rays. The maximal rms is 12.4\%, and the maximal lag is 861 s (equivalent to a 0.24 phase lag).

\item we extract the rms, coherence and lag spectra for the LF and HF bands outside the QPO frequency bin in Obs-9. Both HF and LF rms spectra increase from 0.3 to 10 keV, but an extra hump is observed at LF below 1 keV. No significant covariance is found between the soft and hard X-rays for these stochastic variabilities. A tentative time lag of $\sim$1000 is noticed below 0.4 keV at the frequency band below $10^{-4}$ Hz.

\item we fit the \xmm\ and \nustar\ time-averaged spectra and the five types of variability spectra simultaneously. We find that in order to produce good fits to all the spectra, four spectral components are needed, including a disc component ({\tt diskbb}), two warm Comptonisation components ({\tt compTT-1}, {\tt compTT-2}) and a hot Comptonisation component ({\tt nthComp}). Our results indicate that {\tt compTT-1} exhibits strong stochastic variability, while the QPO is contained in the hotter, less luminous {\tt compTT-2}. This model can also fit the QPO lag spectrum very well, with {\tt diskbb} leading {\tt compTT-2} by 679 s, which in turn leading {\tt nthComp} by 180 s. 

\item the best-fit spectral components for Obs-9 also produce good fits to the spectra from Obs-2, 3 and 6, without changing the shape of each component. The main difference in Obs-2 is that {\tt compTT-2} also exhibits some LF stochastic variability, which is also correlated with {\tt compTT-1}, and so the QPO may have been more severely affected, which explains why the QPO is less coherent in Obs-2 than in Obs-9. For the non-QPO observations of Obs-3 and 6, the main difference is that {\tt compTT-1} is much stronger, so it may have `killed' the QPO signal internally.

\item we construct the optical to hard X-ray broadband SED of \rej1034\ with the state-of-the-art models. We find that the QPO's detectability is not directly correlated with the mass accretion rate through the outer disc as traced by the UV flux. No short-term UV variability is detected by \xmm/OM.

\item using the simultaneous \xmm\ and \nustar\ spectra, we constrain the underlying reflection component to be less than 20\% in the hard X-rays, and there is no significant detection of the iron K$\alpha$ line,

\item we show that the rms and phase lag spectra between the QPO of \rej1034\ and the 67 Hz QPO of \grs1915\ are very similar. The four spectral components of \rej1034\ can also decompose the time-averaged spectra and QPO's rms spectrum of \grs1915\ in a similar way after shifting the energy scale for the mass difference, although the curvature of the soft excess is slightly different. This may be due to the difference in inclination between these two sources.

\end{itemize}

We emphasize the importance of having more data to verify the association of the QPO with the highest temperature part of the soft X-ray excess, as well as more sensitive high energy observations to better constrain the QPO in the 2-10~keV range, and especially any iron line/reflection signature. Our speculation associating the QPO with the {\it plus} mode of an overheated inner edge of the disc predicts that the warm Comptonisation region pulsation is seen as seed photons by the hot Compton region, so this seed photon modulation should
propagate into the hot Comptonisation on the light travel time. Since we measure this at 180 s, this implies a distance of $\sim (18-12)~R_{\rm g}$ for a black hole mass of $(2-3)~\times 10^6~M_\odot$. 

However, the faintness of the source above 2~keV means that while deeper \xmm\ observations will help, a full picture may only emerge from future studies with the large effective area of missions such as {\it Athena}. The next generation X-ray all sky monitor such as {\it Einstein Probe} can also observe \rej1034\ regularly, and thus can monitor its long-term spectral evolution, verify its anti-correlation with the QPO's detectability, and even help to constrain the duty cycle of the QPO's presence.

\section*{Acknowledgements}
We thank the referee for providing useful comments and suggestions to improve the paper. We thank the \swift\ team for approving and conducting the target-of-opportunity observations.
CJ thanks Jiren Liu for helpful discussions on the line features.
CD thanks Frederic Vincent and Chris Fragile for useful discussions about the torus modes, and especially acknowledges the beautiful movies from Frederic Vincent of the ray traced simulations.
CJ acknowledges the National Natural Science Foundation of China through grant 11873054, as well as the support by the Strategic Pioneer Program on Space Science, Chinese Academy of Sciences through grant XDA15052100.
CD and MJW acknowledge the Science and Technology Facilities
Council (STFC) through grant ST/P000541/1 for support.

This work is based on observations conducted by \xmm, an ESA
science mission with instruments and contributions directly funded by
ESA Member States and the USA (NASA).
This work also made use of data from the \nustar\ mission, a project led by the California Institute of Technology, managed by the Jet Propulsion Laboratory, and funded by the National Aeronautics and Space Administration.
This research has made use of the NASA/IPAC Extragalactic Database (NED) which is operated by the Jet Propulsion Laboratory, California Institute of Technology, under contract with the National Aeronautics and Space Administration.

%%%%%%%%%%%%%%%%%%%%%%%%%%%%%%%%%%%%%%%%%%%%%%%%%%
\section*{Data Availability}
The data underlying this article are all publicly available in the High Energy Astrophysics Science Archive Research Center (HEASARC) at https://heasarc.gsfc.nasa.gov, as well as the \xmm\ Science Archive (XSA) at https://www.cosmos.esa.int/web/xmm-newton/xsa.



%\bibliographystyle{mnras}

\begin{thebibliography}{99}
\bibitem[Alston et al.(2014)]{2014MNRAS.439.1548A} Alston W. N., Done C., Vaughan S., 2014, \mnras, 439, 1548 

\bibitem[\protect\citeauthoryear{Ar{\'e}valo \& Uttley}{2006}]{2006MNRAS.367..801A} Ar{\'e}valo P., Uttley P., 2006, MNRAS, 367, 801

\bibitem[\protect\citeauthoryear{Arnaud}{1996}]{1996ASPC..101...17A} Arnaud K. A., 1996, ASPC, 101, 17 

\bibitem[\protect\citeauthoryear{Arnaud, et al.}{1985}]{1985MNRAS.217..105A} Arnaud K.~A., et al., 1985, MNRAS, 217, 105

\bibitem[\protect\citeauthoryear{Ballantyne, Ross \& Fabian}{2001}]{2001MNRAS.327...10B} Ballantyne D.~R., Ross R.~R., Fabian A.~C., 2001, MNRAS, 327, 10

\bibitem[\protect\citeauthoryear{Belloni \& Altamirano}{2013}]{2013MNRAS.432...10B} Belloni T.~M., Altamirano D., 2013, MNRAS, 432, 10

\bibitem[\protect\citeauthoryear{Belloni \& Hasinger}{1990}]{1990A&A...230..103B} Belloni T., Hasinger G., 1990, A\&A, 230, 103

\bibitem[\protect\citeauthoryear{Belloni, M{\'e}ndez \& S{\'a}nchez-Fern{\'a}ndez}{2001}]{2001A&A...372..551B} Belloni T., M{\'e}ndez M., S{\'a}nchez-Fern{\'a}ndez C., 2001, A\&A, 372, 551

\bibitem[\protect\citeauthoryear{Belloni, et al.}{2019}]{2019MNRAS.489.1037B} Belloni T.~M., Bhattacharya D., Caccese P., Bhalerao V., Vadawale S., Yadav J.~S., 2019, MNRAS, 489, 1037

\bibitem[Bessell(1991)]{1991A&A...242L..17B} Bessell M. S., 1991, A\&A, 242, L17 

\bibitem[\protect\citeauthoryear{Bian \& Huang}{2010}]{2010MNRAS.401..507B} Bian W. H., Huang K., 2010, MNRAS, 401, 507

\bibitem[\protect\citeauthoryear{Blackburn}{1995}]{1995ASPC...77..367B} Blackburn J.~K., 1995, ASPC, 77, 367, ASPC...77

\bibitem[Boller et al. (1996)]{Boller96} Boller T., Brandt W. N., Fink H., A\&A, 305, 53

\bibitem[Boroson(2002)]{2002ApJ...565...78B} Boroson T. A., 2002, ApJ, 565, 78

\bibitem[\protect\citeauthoryear{Crummy et al.}{2006}]{Crummy06}Crummy J., Fabian A. C., Gallo L., Ross R. R., 2006, ApJ, 365, 1067

\bibitem[\protect\citeauthoryear{Czerny, et al.}{2016}]{2016A&A...594A.102C} Czerny B., et al., 2016, A\&A, 594, A102

\bibitem[Davis \& Laor(2011)]{2011ApJ...728...98D} Davis S. W., Laor A., 2011, ApJ, 728, 98 

\bibitem[Done et al. (2012)]{2012MNRAS.420.1848D} Done C., Davis S. W., Jin C., Blaes O., Ward, M., 2012, MNRAS, 420, 1848

\bibitem[\protect\citeauthoryear{Done \& Gierli{\'n}ski}{2006}]{2006MNRAS.367..659D} Done C., Gierli{\'n}ski M., 2006, MNRAS, 367, 659

\bibitem[Done \& Jin(2016)]{2016MNRAS.460.1716D} Done C., Jin, C., 2016, \mnras, 460, 1716 

\bibitem[\protect\citeauthoryear{Done, et al.}{2013}]{2013MNRAS.434.1955D} Done C., Jin C., Middleton M., Ward M., 2013, MNRAS, 434, 1955

\bibitem[\protect\citeauthoryear{Done, Wardzi{\'n}ski \& Gierli{\'n}ski}{2004}]{2004MNRAS.349..393D} Done C., Wardzi{\'n}ski G., Gierli{\'n}ski M., 2004, MNRAS, 349, 393

\bibitem[\protect\citeauthoryear{Edelson, et al.}{2015}]{2015ApJ...806..129E} Edelson R., et al., 2015, ApJ, 806, 129

\bibitem[\protect\citeauthoryear{Emmanoulopoulos, McHardy \& Papadakis}{2011}]{2011MNRAS.416L..94E} Emmanoulopoulos D., McHardy I.~M., Papadakis I.~E., 2011, MNRAS, 416, L94

\bibitem[Fabian et al.(2009)]{2009Natur.459..540F} Fabian A. C., et al., 2009, Nature, 459, 540

\bibitem[\protect\citeauthoryear{Foreman-Mackey, et al.}{2013}]{2013PASP..125..306F} Foreman-Mackey D., Hogg D.~W., Lang D., Goodman J., 2013, PASP, 125, 306

\bibitem[Gallo(2006)]{2006MNRAS.368..479G} Gallo L. C., 2006, \mnras, 368, 479 

\bibitem[\protect\citeauthoryear{Gardner \& Done}{2017}]{2017MNRAS.470.3591G} Gardner E., Done C., 2017, MNRAS, 470, 3591

\bibitem[\protect\citeauthoryear{Gardner \& Done}{2014}]{2014MNRAS.442.2456G} Gardner E., Done C., 2014, MNRAS, 442, 2456

\bibitem[\protect\citeauthoryear{Gierli{\'n}ski \& Done}{2004}]{2004MNRAS.349L...7G} Gierli{\'n}ski M., Done C., 2004, MNRAS, 349, L7

\bibitem[Gierli{\'n}ski et al.(2008)]{2008Natur.455..369G} Gierli\'{n}ski M., Middleton M., Ward M., Done C., 2008, \nat, 455, 369 

\bibitem[\protect\citeauthoryear{Harrison, et al.}{2013}]{2013ApJ...770..103H} Harrison F.~A., et al., 2013, ApJ, 770, 103

\bibitem[\protect\citeauthoryear{Hu, et al.}{2011}]{2011ApJ...740...67H} Hu C.-P., Chou Y., Wu M. C., Yang T. C., Su Y. H., 2011, ApJ, 740, 67

\bibitem[\protect\citeauthoryear{Huang, et al.}{1998}]{1998RSPSA.454..903H} Huang N. E., et al., 1998, RSPSA, 454, 903

\bibitem[Jansen et al.(2001)]{2001A&A...365L...1J} Jansen F., Lumb D., Altieri B., et al., 2001, \aap, 365, L1 

\bibitem[\protect\citeauthoryear{Jin, Done \& Ward}{2020}]{2020MNRAS.495.3538J} Jin C., Done C., Ward M., 2020, MNRAS, 495, 3538 (Paper-I)

\bibitem[Jin et al.(2017a)]{JDW2017a} Jin C., Done C., Ward M., 2017a, \mnras, 468, 3663

\bibitem[Jin et al.(2017b)]{JDW2017b} Jin C., Done C., Ward M., 2017b, \mnras, 471, 706

\bibitem[Jin et al.(2016)]{2016MNRAS.455..691J} Jin C., Done C., Ward M., 2016, \mnras, 455, 691 

\bibitem[Jin et al.(2013)]{2013MNRAS.436.3173J} Jin C., Done C., Middleton M., Ward M., 2013, MNRAS, 436, 3173

\bibitem[Jin et al. (2012a)]{2012MNRAS.420.1825J} Jin C., Ward M., Done C., Gelbord J., 2012a, MNRAS, 420, 1825

\bibitem[Jin, Ward \& Done (2012b)]{2012MNRAS.422.3268J} Jin C., Ward M., Done C., 2012b, MNRAS, 422, 3268

\bibitem[Jin, Ward \& Done (2012c)]{2012MNRAS.425..907J} Jin C., Ward M., Done C., 2012c, MNRAS, 425, 907

\bibitem[\protect\citeauthoryear{Jin et al.}{2009}]{Jin09} Jin C., Done C., Ward M., Gierli\'{n}ski M., Mullaney J., 2009, MNRAS, 398, L16

\bibitem[\protect\citeauthoryear{Kara, et al.}{2016}]{2016MNRAS.462..511K} Kara E., Alston W.~N., Fabian A.~C., Cackett E.~M., Uttley P., Reynolds C.~S., Zoghbi A., 2016, MNRAS, 462, 511

\bibitem[\protect\citeauthoryear{Kara, et al.}{2017}]{2017MNRAS.468.3489K} Kara E., Garc{\'\i}a J.~A., Lohfink A., Fabian A.~C., Reynolds C.~S., Tombesi F., Wilkins D.~R., 2017, MNRAS, 468, 3489

\bibitem[\protect\citeauthoryear{Kolehmainen, Done \& D{\'\i}az Trigo}{2011}]{2011MNRAS.416..311K} Kolehmainen M., Done C., D{\'\i}az Trigo M., 2011, MNRAS, 416, 311

\bibitem[\protect\citeauthoryear{Kotov, Churazov \& Gilfanov}{2001}]{2001MNRAS.327..799K} Kotov O., Churazov E., Gilfanov M., 2001, MNRAS, 327, 799

\bibitem[\protect\citeauthoryear{Kubota \& Done}{2019}]{2019MNRAS.489..524K} Kubota A., Done C., 2019, MNRAS, 489, 524

\bibitem[\protect\citeauthoryear{Kubota \& Done}{2018}]{2018MNRAS.480.1247K} Kubota A., Done C., 2018, MNRAS, 480, 1247

\bibitem[\protect\citeauthoryear{Kubota \& Done}{2004}]{2004MNRAS.353..980K} Kubota A., Done C., 2004, MNRAS, 353, 980

\bibitem[\protect\citeauthoryear{Kubota, Makishima \& Ebisawa}{2001}]{2001ApJ...560L.147K} Kubota A., Makishima K., Ebisawa K., 2001, ApJL, 560, L147

\bibitem[Laor(1991)]{1991ApJ...376...90L} Laor A., 1991, ApJ, 376, 90

\bibitem[\protect\citeauthoryear{Laor \& Netzer}{1989}]{1989MNRAS.238..897L} Laor A., Netzer H., 1989, MNRAS, 238, 897

\bibitem[\protect\citeauthoryear{Maitra \& Miller}{2010}]{2010ApJ...718..551M} Maitra D., Miller J. M., 2010, ApJ, 718, 551

\bibitem[\protect\citeauthoryear{M{\'e}ndez, et al.}{2013}]{2013MNRAS.435.2132M} M{\'e}ndez M., Altamirano D., Belloni T., Sanna A., 2013, MNRAS, 435, 2132

\bibitem[\protect\citeauthoryear{Middleton \& Done}{2010}]{2010MNRAS.403....9M} Middleton M., Done C., 2010, MNRAS, 403, 9

\bibitem[\protect\citeauthoryear{Middleton et al.}{2009}]{Middleton09} Middleton M., Done C., Ward M., Gierli\'{n}ski M., Schurch N., 2009, MNRAS, 394,250

\bibitem[\protect\citeauthoryear{Middleton, et al.}{2006}]{2006MNRAS.373.1004M} Middleton M., Done C., Gierli{\'n}ski M., Davis S.~W., 2006, MNRAS, 373, 1004

\bibitem[Middleton et al.(2011)]{2011MNRAS.417..250M} Middleton M., Uttley P., Done C., 2011, \mnras, 417, 250 

\bibitem[\protect\citeauthoryear{Morgan, Remillard \& Greiner}{1997}]{1997ApJ...482..993M} Morgan E.~H., Remillard R.~A., Greiner J., 1997, ApJ, 482, 993

\bibitem[\protect\citeauthoryear{Nowak et al.}{1999}]{1999ApJ...510..874N} Nowak M. A., Vaughan B. A., Wilms J., Dove J. B., Begelman M. C., 1999, ApJ, 510, 874

\bibitem[\protect\citeauthoryear{Page, et al.}{2004}]{2004MNRAS.352..523P} Page K.~L., Schartel N., Turner M.~J.~L., O'Brien P.~T., 2004, MNRAS, 352, 523

\bibitem[\protect\citeauthoryear{Parker, Miller \& Fabian}{2018}]{2018MNRAS.474.1538P} Parker M.~L., Miller J.~M., Fabian A.~C., 2018, MNRAS, 474, 1538

\bibitem[\protect\citeauthoryear{Petrucci, et al.}{2018}]{2018A&A...611A..59P} Petrucci P.-O., et al., 2018, A\&A, 611, A59

\bibitem[Polletta et al.(2007)]{2007ApJ...663...81P} Polletta M., et al., 2007, ApJ, 663, 81 

\bibitem[Porquet et al.(2004)]{2004A&A...422...85P} Porquet D., Reeves J. N., O'Brien P., Brinkmann W., 2004, A\&A, 422, 85 

\bibitem[\protect\citeauthoryear{Puchnarewicz, et al.}{1995}]{1995MNRAS.276...20P} Puchnarewicz E. M., Mason K. O., Siemiginowska A., Pounds K. A., 1995, MNRAS, 276, 20

\bibitem[\protect\citeauthoryear{Reid, et al.}{2014}]{2014ApJ...796....2R} Reid M.~J., McClintock J.~E., Steiner J.~F., Steeghs D., Remillard R.~A., Dhawan V., Narayan R., 2014, ApJ, 796, 2

\bibitem[\protect\citeauthoryear{Remillard \& McClintock}{2006}]{2006ARA&A..44...49R} Remillard R. A., McClintock J. E., 2006, ARA\&A, 44, 49

\bibitem[\protect\citeauthoryear{Remillard, et al.}{2002}]{2002ApJ...580.1030R} Remillard R.~A., Muno M.~P., McClintock J.~E., Orosz J.~A., 2002, ApJ, 580, 1030

\bibitem[\protect\citeauthoryear{Ross \& Fabian}{1993}]{1993MNRAS.261...74R} Ross R.~R., Fabian A.~C., 1993, MNRAS, 261, 74

\bibitem[\protect\citeauthoryear{Schlegel, Finkbeiner \& Davis}{1998}]{1998ApJ...500..525S} Schlegel D.~J., Finkbeiner D.~P., Davis M., 1998, ApJ, 500, 525

\bibitem[\protect\citeauthoryear{Strohmayer}{2001}]{2001ApJ...554L.169S} Strohmayer T.~E., 2001, ApJL, 554, L169

\bibitem[Titarchuk(1994)]{1994ApJ...434..570T} Titarchuk L., 1994, ApJ,  434, 570 

\bibitem[\protect\citeauthoryear{Turner \& Pounds}{1988}]{1988MNRAS.232..463T} Turner T.~J., Pounds K.~A., 1988, MNRAS, 232, 463

\bibitem[Uttley et al.(2014)]{2014A&ARv..22...72U} Uttley P., Cackett E. M., Fabian A. C., Kara E., Wilkins D. R., 2014, A\&ARv, 22, 72

\bibitem[\protect\citeauthoryear{Vaughan \& Nowak}{1997}]{1997ApJ...474L..43V} Vaughan B. A., Nowak M. A., 1997, ApJL, 474, L43

\bibitem[Verner et al.(1996)]{1996ApJ...465..487V} Verner D. A., Ferland G. J., Korista K. T., Yakovlev D. G., 1996, \apj, 465, 487

\bibitem[\protect\citeauthoryear{Vincent et al.}{2014}]{2014A&A...563A.109V} Vincent F.~H., Mazur G.~P., Straub O., Abramowicz M.~A., Klu{\'z}niak W., T{\"o}r{\"o}k G., Bakala P., 2014, A\&A, 563, A109

\bibitem[\protect\citeauthoryear{Wilkinson \& Uttley}{2009}]{Wilkinson09}Wilkinson T., Uttley P., 2009, MNRAS, 397, 666

\bibitem[Wilms et al.(2000)]{2000ApJ...542..914W} Wilms J., Allen A., McCray R., 2000, \apj, 542, 914 

\bibitem[Wright(2006)]{2006PASP..118.1711W} Wright E. L., 2006, \pasp, 118, 1711 

\bibitem[Wu \& Huang(2009)]{Wu&Huang2009} Wu, Z., Huang, N. E. 2009, AADA,1,1

%\bibitem[\protect\citeauthoryear{You, Bursa \& %{\.Z}ycki}{2018}]{2018ApJ...858...82Y} You B., Bursa M., {\.Z}ycki P.~T., %2018, ApJ, 858, 82

\bibitem[Zdziarski et al.(1996)]{1996MNRAS.283..193Z} Zdziarski A. A., Johnson W. N., Magdziarz P., 1996, MNRAS, 283, 193 

\bibitem[\protect\citeauthoryear{Zoghbi, et al.}{2013}]{2013ApJ...767..121Z} Zoghbi A., Reynolds C., Cackett E.~M., Miniutti G., Kara E., Fabian A.~C., 2013, ApJ, 767, 121

\end{thebibliography}


%%%%%%%%%%%%%%%%%%%%%%%%%%%%%%%%%%%%%%%%%%%%%%%%%%

%%%%%%%%%%%%%%%%% APPENDICES %%%%%%%%%%%%%%%%%%%%%

\appendix

\section{Searching for the Soft X-ray Absorption Lines}
\label{sec-lines}
The soft X-ray spectrum of \rej1034\ was reported to contain a few absorption lines, which were considered to be produced by the partially ionized oxygen, neon and iron elements (Middleton, Uttley \& Done 2011). It was also reported that an extra absorption feature appears during the QPO trough at 0.86 keV, which is interpreted as the O {\sc viii} K-edge at its theoretical energy of 0.87 keV (i.e. 0.84 keV in the observer's frame, Maitra \& Miller 2010). This absorption feature was proposed to be evidence for the existence of a warm absorber, whose variability is correlated with the QPO. The location of this warm absorber was estimated to be $\sim$15 $R_{\rm g}$ from the central black hole, and the QPO can be explained as the periodic obscuration by this ionized material. However, Middleton, Uttley \& Done (2011) analyzed the RGS spectrum of Obs-2, and argued that the small radius of this absorber is in contradiction with the low turbulence and outflow velocities as measured from the RGS spectra.

Since the absorption lines are claimed to be related to the QPO, we also search for soft X-ray absorption features in the RGS spectra whose spectral resolution is higher than EPIC cameras. Since the significance of a line feature is directly related to the underlying continuum, we use the spectral components in the best-fit Model-2 (see Section~\ref{sec-model2}) to fit the continua of RGS spectra, and use narrow Gaussian components to fit potential absorption lines.

\begin{figure}
\begin{tabular}{c}
\includegraphics[trim=0.25in 0.2in 0.0in 0.0in, clip=1, scale=0.49]{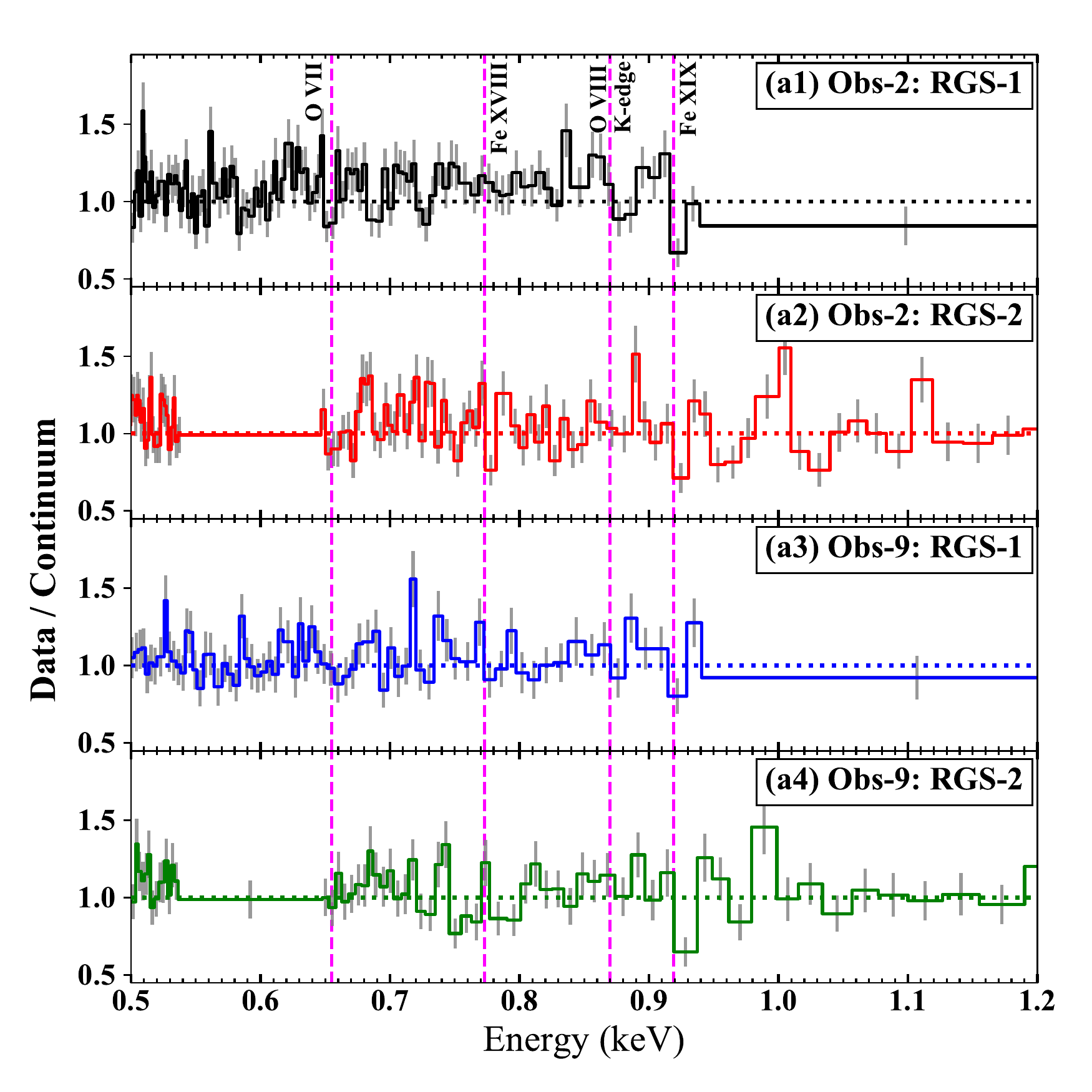}\\
\end{tabular}
\caption{The RGS spectra of \rej1034\ observed in Obs-2 and 9. These are shown as the ratio between the data and the best-fit continuum model. The positions of some important absorption lines are marked by the vertical dash lines at their energies in the AGN's rest-frame. Note that the O {\sc viii} K-edge is not detected.}
\label{fig-rgs-spec}
\end{figure}

Figure~\ref{fig-rgs-spec} shows the RGS spectra relative to their best-fit continua for both Obs-2 and 9. The rest frame energies of various potential features are marked, but plainly there is nothing which is highly significant. A feature consistent with 0.923 keV can be seen in all spectra. However, its significance is not straightforward to be determined precisely due to the `look elsewhere' effect, from the number of independent energies searched. Nonetheless, since this has been claimed before, we use a standard $\Delta\chi^2=9$ for a $3\sigma$ significance. This gives
3.3$\sigma$ in Obs-2 and 4.3$\sigma$ in Obs-9. If we assume this line is a blue-shifted O {\sc viii} K-edge, then the velocity along the line-of-sight would be as large as $\sim0.048c$ ($c$ is the speed of light). Thus this line would appear broad either because of strong turbulence or the velocity distribution of the outflowing material, which is not observed in the spectrum. Therefore, we rule out the origin of O {\sc viii} K-edge, and attribute this line to the closer Fe {\sc xix} (0.919 keV) line. We can also identify other absorption lines such as O {\sc viii} and Fe {\sc xviii} in both Obs-2 and 9, but their statistical significances are all less than 3$\sigma$ (see Table~\ref{tab-line-rgs}).

In order to check if these absorption lines correlate with the QPO phase (Maitra \& Miller 2010) and the continuum flux, we perform QPO phase-resolved spectroscopy and 0.3-10 keV flux-resolved spectroscopy using the EPIC-pn data in Obs-2 and 9, with a similar method as described in Maitra \& Miller (2010). Our approach is to use the intrinsic mode function (IMF) at the QPO's timescale to identify local peaks and troughs of the QPO, which is derived by applying the Ensemble Empirical Mode Decomposition (EEMD) method to the 0.3-10 keV light curve (Huang et al. 1998; Wu \& Huang 2009; Hu et al. 2011; Paper-I). We define the time and flux at every QPO extremum as [$T_{\rm min}$, $F_{\rm min}$] for every QPO trough, and [$T_{\rm max}$, $F_{\rm max}$] for every QPO peak. Then each of the interval between the consecutive pair of extrema, i.e. [$T_{\rm min}$,$T_{\rm max}$] or [$T_{\rm max}$,$T_{\rm min}$], is divided into three sub-intervals. We use the factor $\mu$ to define the flux within the interval as $F_{\rm min}$+$\mu$($F_{\rm max}$-$F_{\rm min}$), thus the value of $\mu$ is within (0,1). The three sub-intervals are defined for $\mu=$ (0,0.3), (0.3-0.7), (0.7,1), and are referred to as low phase (LP), intermediate phase (IP) and high phase (HP), respectively. For the flux-resolved spectra, we define four flux ranges, which are 0-5.0, 5.0-5.7. 5.7-6.5 and $>$ 6.5 counts s$^{-1}$ for Obs-2 and 0-5.0, 5.0-5.7. 5.7-6.5 and $>$ 6.5 counts s$^{-1}$ for Obs-9, and these flux ranges are signified by Flux-1 to 4. Similarly, the spectral components in Model-2 are used to fit the continuum in these spectra. Potential absorption lines are fitted with Gaussian components.

However, we cannot detect a broad O {\sc viii} K-edge in any of the spectra derived above in either Obs-2 or 9. This is most likely because that comparing to Maitra \& Miller (2010), our continuum model is more sophisticated, and the spectral quality of the new Obs-9 dataset is better. The narrow Fe {\sc xix} absorption line can be detected. Table~\ref{tab-line-pn} compares the significance of Fe {\sc xix} absorption line in different spectra. Despite the low significance of detection, we notice that the Fe {\sc xix} line appears most significant in the QPO-LP spectra in both Obs-2 and 9, and so it is possible that this line is correlated with the QPO phase. If this correlation is cause by the response of ionization state to the changing illumination, as proposed by Maitra \& Miller (2010), then it should show an even stronger correlation with the flux state. However, the significances of this line in the flux-resolve spectra are all much smaller than in the QPO-LP spectrum, thus the illumination scenario can be ruled out, unless the obscuring material only `sees' the variation of QPO but not the continuum flux, which is difficult to explain. However, this potential correlation between the Fe {\sc xix} line and QPO phase is still too vague, but we encourage to explore it in more detail with future observations.

\begin{table}
\centering
\caption{Soft X-ray absorption lines detected in the RGS spectrum. The listed parameters are for the best-fit Gaussian profile. Negative line velocities indicate outflows. The last column indicates the detection significance.}
\begin{tabular}{lccccc}
\hline
Absorption Line & Obs & $E_{\rm line,res}$ & $v_{\rm line}$ & EW & Sig. \\
 & & (keV) & ($c$) & (eV) & ($\sigma$) \\
\hline
Fe {\sc xix} (0.919 keV) & Obs-2 & 0.923 & 0.004 & -3.9 & 3.3 \\
 & Obs-9 & 0.924 & 0.006 & -5.4 & 4.3 \\
Fe {\sc xviii} (0.773 keV) & Obs-2 & 0.780 & 0.009 & -0.8 & 1.5 \\
 & Obs-9 & 0.785 & 0.016 & -1.1 & 1.1 \\
O {\sc viii} (0.655 keV)& Obs-2 & 0.656 & 0.002 & -1.2 & 3.1 \\
 & Obs-9 & 0.655 & 0.001 & -0.6 & 1.7 \\
\hline
\end{tabular}
\label{tab-line-rgs}
\end{table}

\begin{table}
\centering
\caption{Best-fit parameters by fitting a Gaussian profile to the Fe {\sc xix} (rest frame: 0.919 keV) absorption line in the QPO phase-resoled spectra and 0.3-10 keV flux-resolved spectra. QPO-HP, MP and LF indicate the high-phase, medium-phase and low-phase. Flux-1 to 4 indicate the sequence from the highest flux to the lowest.}
\begin{tabular}{ccccc} 
\hline
Obs & Spectrum & $E_{\rm line, res}$ & EW & Significance  \\
 & & (keV) & (eV) & ($\sigma$)  \\
\hline
Obs-2 & QPO-HP & 0.920 ($v_{\rm line}=0.001c$)  & -0.0 & 0.0  \\
Obs-2 & QPO-MP & tied to QPO-HP & -1.0 & 0.3  \\
Obs-2 & QPO-LP & tied  to QPO-HP & -5.1 & 1.8  \\
Obs-9 & QPO-HP & 0.938 ($v_{\rm line}=0.02c$) & -0.0 & 0.0  \\
Obs-9 & QPO-MP & tied  to QPO-HP & -0.4 & 0.1  \\
Obs-9& QPO-LP & tied  to QPO-HP & -8.0 & 3.1  \\
\hline
Obs-2 & Flux-1 & 0.971 ($v_{\rm line}=0.06c$) & -2.8 & 0.6  \\
Obs-2 & Flux-2 & tied to Flux-1& -5.4 & 1.7  \\
Obs-2 & Flux-3 & tied to Flux-1 & -0.0 & 0.0  \\
Obs-2 & Flux-4 & tied to Flux-1 & -0.0 & 0.0  \\
Obs-9 & Flux-1 & 0.954 ($v_{\rm line}=0.04c$) & -5.4 & 1.6  \\
Obs-9 & Flux-2 & tied to Flux-1 & -2.4 & 0.9  \\
Obs-9 & Flux-3 & tied to Flux-1 & -3.1 & 1.3  \\
Obs-9 & Flux-4 & tied to Flux-1 & -2.5 & 0.6  \\
\hline
\end{tabular}
\label{tab-line-pn}
\end{table}

\section{Effect of pile-up in the variability spectra of Obs-2}
\label{sec-pileup}
A potential issue associated with Obs-2 is the pile-up effect, because EPIC-pn was in the full-frame mode, and the pile-up threshold is below the count-rate of \rej1034. In paper-I we have determined that in order to safely remove the pile-up effect, the central 10 arcsec of the point-spread-function (PSF) should be excised. This inner radius is slightly larger than the 7.5 arcsec radius which was adopted by some previous works (Geirli\'{n}ski et al. 2008; Alston et al. 2014). However, the excision of the core of PSF will cause a decrease of the S/N because of the loss of source photons, thus there is often a compromise between S/N and the pile-up level.

\begin{figure*}
\centering
\begin{tabular}{ccc}
\includegraphics[trim=0.2in 0.2in 0.0in 0.1in, clip=1, scale=0.41]{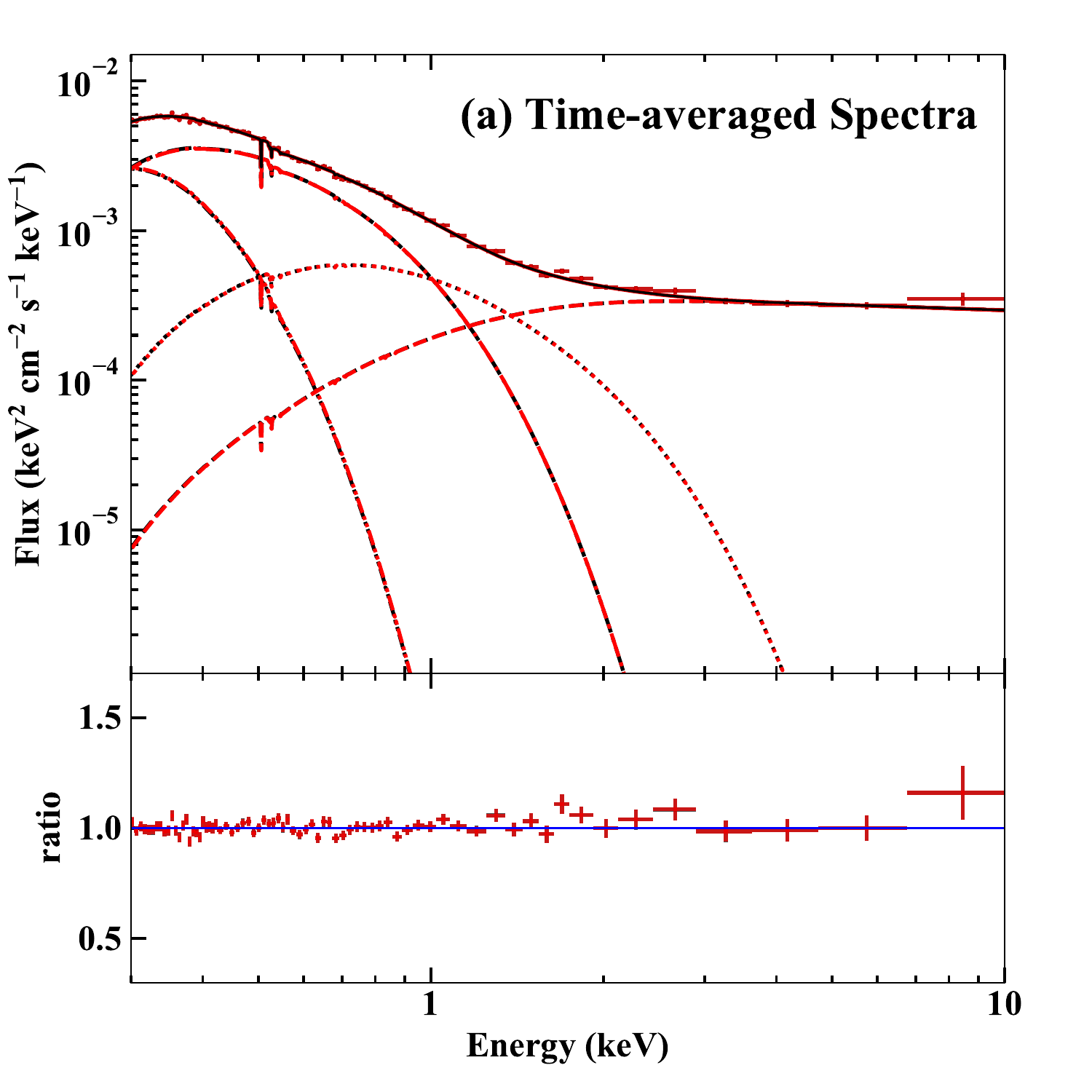} &
\includegraphics[trim=0.7in 0.2in 0.0in 0.1in, clip=1, scale=0.41]{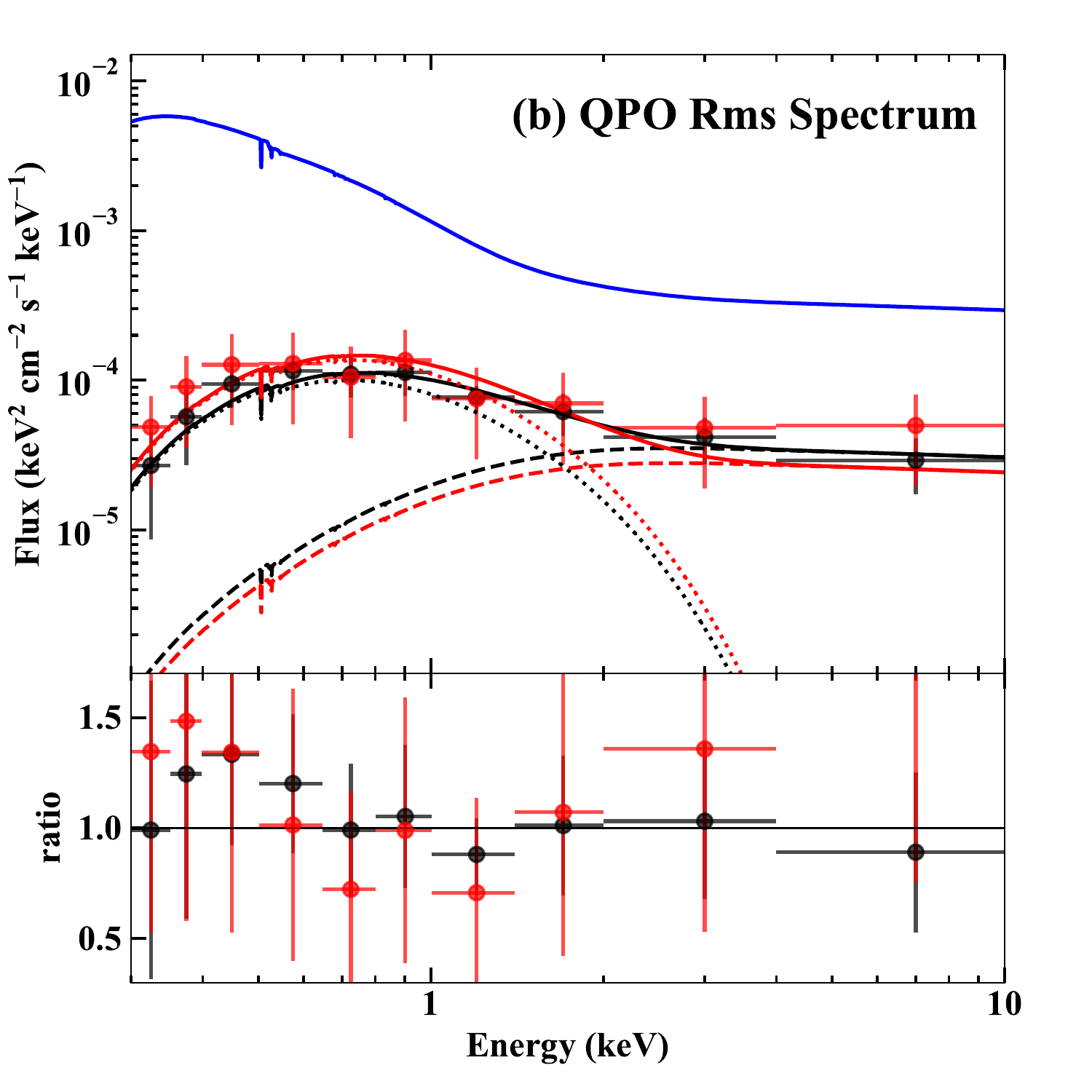} &
\includegraphics[trim=0.75in 0.2in 0.0in 0.1in, clip=1, scale=0.41]{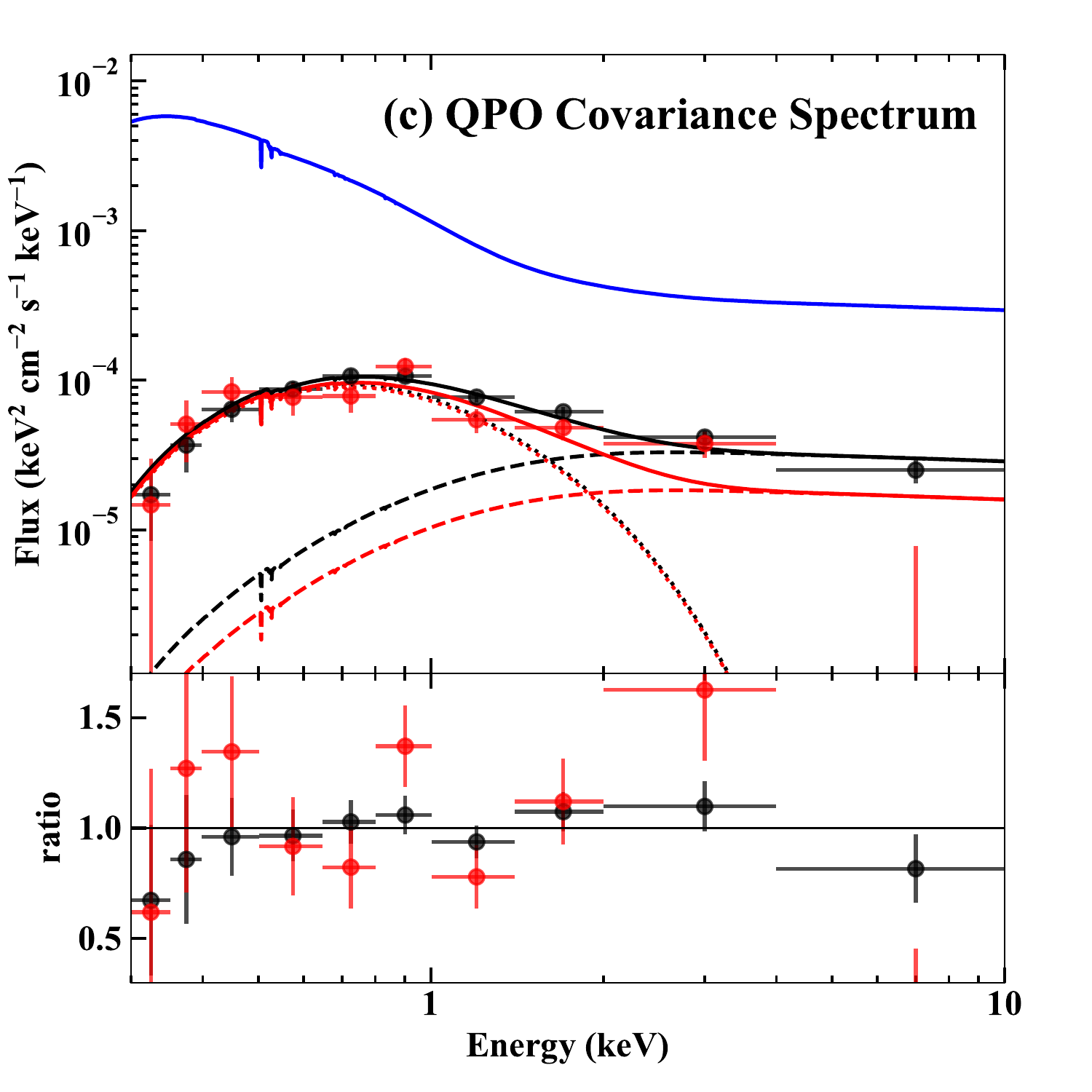} \\
\includegraphics[trim=0.2in 0.2in 0.0in 0.1in, clip=1, scale=0.41]{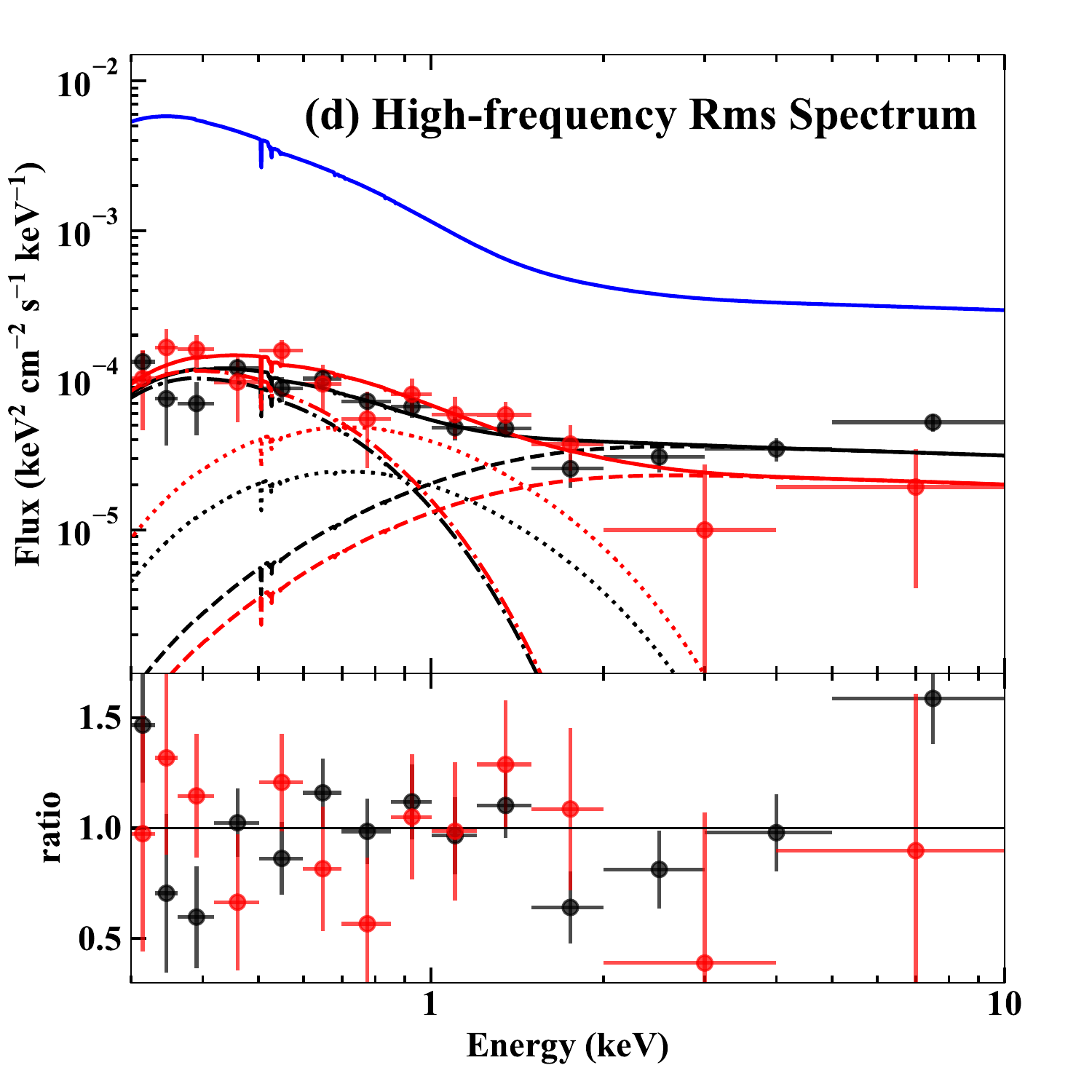} &
\includegraphics[trim=0.7in 0.2in 0.0in 0.1in, clip=1, scale=0.41]{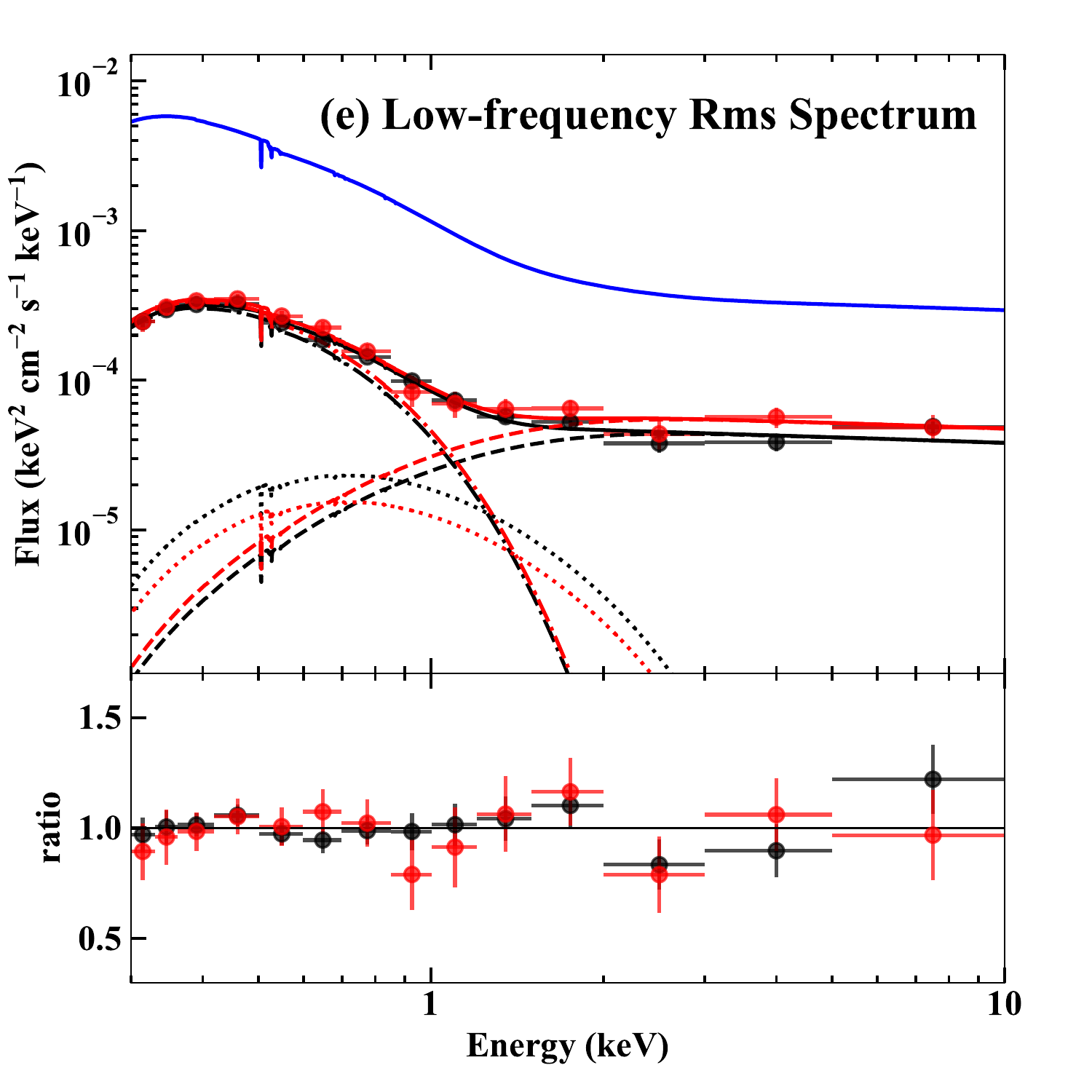} &
\includegraphics[trim=0.75in 0.2in 0.0in 0.1in, clip=1, scale=0.41]{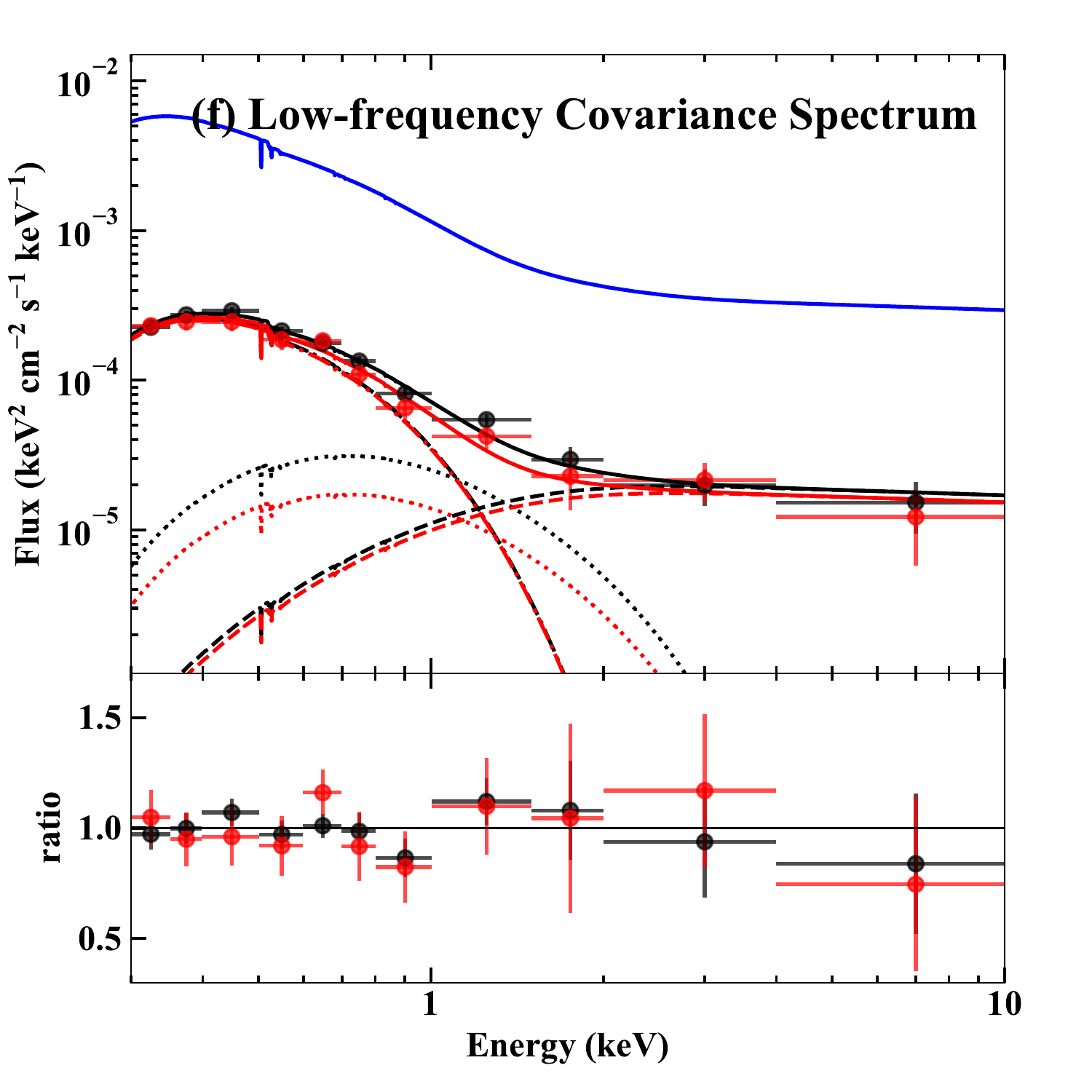} \\
\end{tabular}
\caption{Comparing the variability spectra of \rej1034\ from the Obs-2 dataset using different source extraction regions. The black spectra are the same as those plotted in Figure~\ref{fig-spec-rmscov2-obs2}. For the red spectra, the central circular region of 10 arcsec radius is excised to remove the pile-up effect. No significant differences are found between the two sets of variability spectra, except that the red spectra have lower S/N due to the loss of source photons.}
\label{fig-spec-rmscov2-obs2b}
\end{figure*}

In Section~\ref{sec-lagspec-obs2} we applied the best-fit Model-2 of Obs-9 to the same set of variability spectra from Obs-2. To maximize the S/N for the model test, we did not excise the core of PSF for the variability spectra. Here we show the results after excising the central 10 arcsec of the PSF. Figure~\ref{fig-spec-rmscov2-obs2b} compares the same types of variability spectra before and after removing the core of PSF. Note that the time-averaged spectrum is always extracted outside the pile-up region, so the red and black time-averaged spectra are exactly the same. It is clear that all the variability spectra are still consistent, and the spectral decompositions are not affected significantly. In this case, the main influence of removing the pile-up region is the decrease of S/N. Similarly, Zoghbi \& Fabian (2011) showed that pile-up does not affect the time lag measurement for the dataset of Obs-2. Therefore, we can conclude that the pile-up in Obs-2 does not affect our results in this work.

\section{MCMC Analysis for the Best-fit Model-2}
\label{sec-mcmc-model2}
\begin{figure*}
\centering
\includegraphics[trim=0.15in 0.2in 0.0in 0.1in, clip=1, scale=0.45]{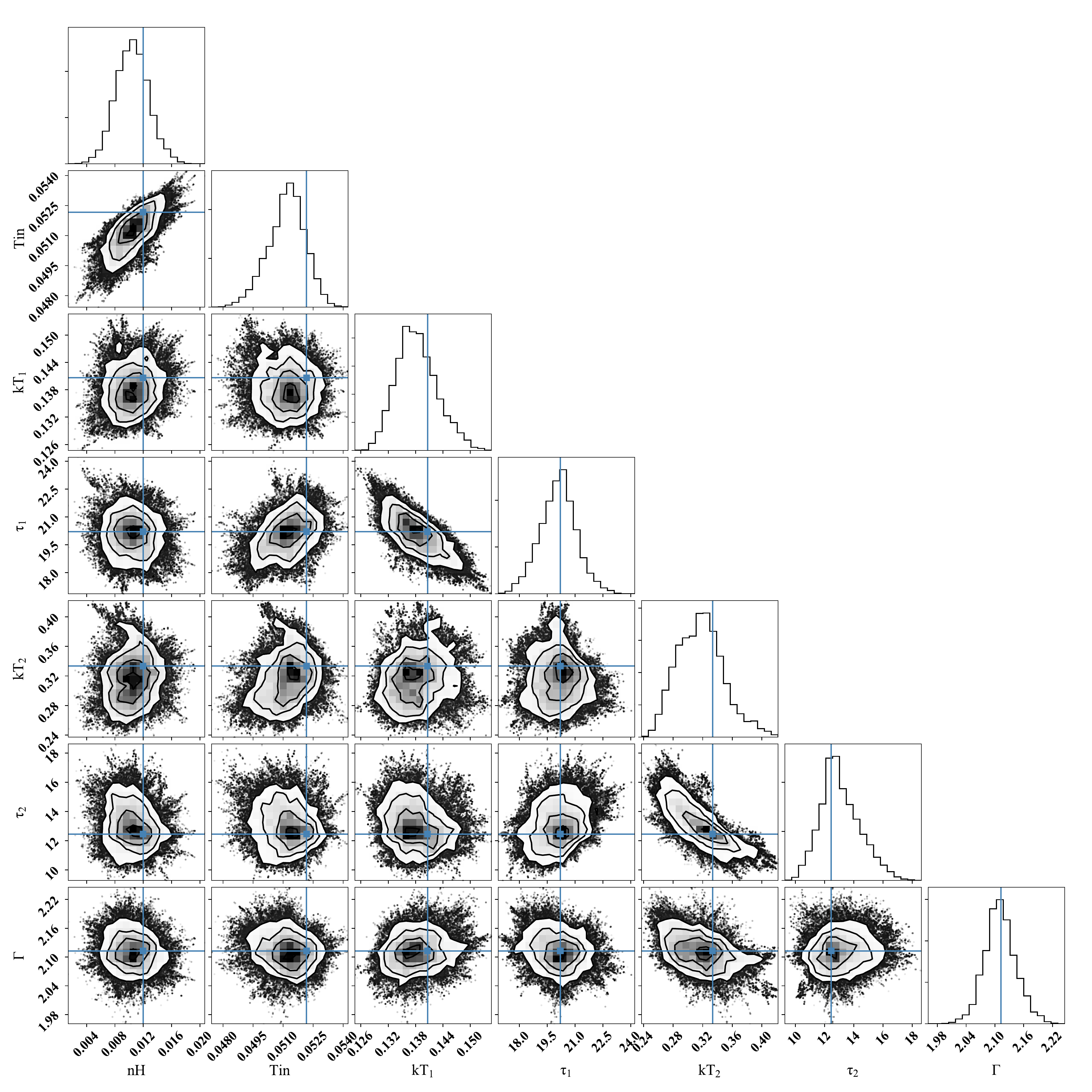}
\caption{Results of MCMC sampling for the parameter space of the best-fit Model-2 in Table~\ref{tab-varspecfit}. $\tau_1$ and $kT_1$ are the optical depth and electron temperature for {\tt compTT-1}, while $\tau_1$ and $kT_1$ are for {\tt compTT-2}. The blue solid lines indicate the best-fit values. All the physical parameters of Model-2 are well constrained.}
\label{fig-mcmc-model2}
\end{figure*}

The time-averaged spectrum is often degenerated to different models and spectral components. In this work, we have shown that by fitting various variability spectra simultaneously, the degeneracy can be greatly suppressed, thereby allowing additional spectral components to be identified, such as the two soft X-ray Comptonisation components in Model-2. As a further demonstration of the strong model constraints, we perform MCMC sampling of the multi-dimensional parameter space of the best-fit Model-2 in Table~\ref{tab-varspecfit}. The resultant covariances between different physical parameters are plotted in Figure~\ref{fig-mcmc-model2}. A weak correlation is found between the inner disc temperature $T_{\rm in}$ and the intrinsic absorption column $N_{\rm H}$. Weak anti-correlations are found between the optical depth $\tau$ and electron temperature $kT$ of both {\tt CompTT-1} and {\tt CompTT-1}. But there is no major parameter degeneracy, and all the parameters are well constrained. Therefore, our MCMC test confirms that the spectral decomposition of Model-2 is statistically robust.

%%%%%%%%%%%%%%%%%%%%%%%%%%%%%%%%%%%%%%%%%%%%%%%%%%

% Don't change these lines
\bsp	% typesetting comment
\label{lastpage}
\end{document}